\documentclass[aps,prappl,twocolumn,superscriptaddress]{revtex4-2}
 \usepackage{graphicx}
\usepackage{physics}
\usepackage{xcolor}
\usepackage{float}

\usepackage{cancel}

\usepackage[breaklinks=true,colorlinks,citecolor=blue,linkcolor=blue,urlcolor=blue]{hyperref}

\draft 

\begin{document}

\title{Ultrastrong Magnon-Photon Coupling Achieved by Magnetic Films in Contact with Superconducting Resonators} 

\author{Alberto Ghirri}
\email[mail to: ]{alberto.ghirri@nano.cnr.it}
\affiliation{Istituto Nanoscienze - CNR, Centro S3, via G. Campi 213/A, 41125, Modena, Italy}
\author{Claudio Bonizzoni}
\affiliation{Dipartimento di Scienze Fisiche, Informatiche e Matematiche Universit\`a di Modena e Reggio Emilia, via G. Campi 213/A, 41125, Modena, Italy}
\affiliation{Istituto Nanoscienze - CNR, Centro S3, via G. Campi 213/A, 41125, Modena, Italy}
\author{Maksut Maksutoglu}
\affiliation{Institute of Nanotechnology, Gebze Technical University, 41400, Gebze, Kocaeli, Turkey}
\author{Alberto Mercurio}
\affiliation{Dipartimento di Scienze Matematiche e Informatiche, Scienze Fisiche e  Scienze della Terra, Universit\`{a} di Messina, I-98166 Messina, Italy}
\author{Omar Di Stefano}
\affiliation{Dipartimento di Scienze Matematiche e Informatiche, Scienze Fisiche e  Scienze della Terra, Universit\`{a} di Messina, I-98166 Messina, Italy}
\author{Salvatore Savasta}
\affiliation{Dipartimento di Scienze Matematiche e Informatiche, Scienze Fisiche e  Scienze della Terra, Universit\`{a} di Messina, I-98166 Messina, Italy}
\author{Marco Affronte}
\affiliation{Dipartimento di Scienze Fisiche, Informatiche e Matematiche Universit\`a di Modena e Reggio Emilia, via G. Campi 213/A, 41125, Modena, Italy}
\affiliation{Istituto Nanoscienze - CNR, Centro S3, via G. Campi 213/A, 41125, Modena, Italy}

\date{\today}

\begin{abstract}
{Coherent coupling between spin wave excitations (magnons) and microwave photons in a cavity may disclose new paths to unconventional phenomena as well as for novel applications. Here, we present a systematic investigation on YIG (Yttrium Iron Garnet) films on top of coplanar waveguide resonators made of superconducting YBCO. We first show that spin wave excitations with frequency higher than the Kittel mode can be excited by putting in direct contact a 5~$\mu$m thick YIG film with the YBCO coplanar resonator (cavity frequency $\omega_c/2 \pi = 8.65$~GHz). With this configuration, we obtain very large values of the collective coupling strength $\lambda/2 \pi \approx 2$~GHz and cooperativity $C=5 \times 10^4$. Transmission spectra are analyzed by a modified Hopfield model for which we provide an exact solution that allows us to well reproduce spectra by introducing a limited number of free parameters. It turns out that the coupling of the dominant magnon mode with photons exceeds 0.2 times the cavity frequency, thus demonstrating the achievement of the ultrastrong coupling regime with this architecture. Our analysis also shows a vanishing contribution of the diamagnetic term which is a peculiarity of pure spin systems.}
\end{abstract}

\maketitle 

\section{Introduction}

The interplay between magnetic excitations and electromagnetic radiation has recently assumed a pivotal role in many fields of research such as magnonics, spintronics, magneto-opto-mechanics and information processing for its potentialities in the development of hybrid systems and devices \cite{PirroNatRevMater21, Lachance-QuirionApplPhysExpr19,  RameshtiPhysRep22, HybMagnonics}. The control of magnon-photon coupling and cooperativity is one of the keys for enabling the exploitation of unique functionalities related to the coherent dynamics in these systems. Novel applications, including memory devices \cite{ZhangNatCommun15}, coherent spin pumping \cite{BaiPRL17}, haloscopes for axion detection \cite{CresciniEurPhysJC18}, microwave-optical transducers \cite{HisatomiPRB16} and coherent microwave sources \cite{YaoPRL23} have been already tested. An open issue is the realization of all-on-chip devices for their efficient integration in microwave circuits \cite{HybMagnonics}.

In the prototypical case of a ferromagnetic sample embedded in a microwave resonator, spin waves couple with resonant electromagnetic modes \cite{RobertsJAP62} and the system can be modelled by combining Maxwell and Landau-Lifshitz-Gilbert (LLG) equations \cite{Auld1963, GuerevichMelkov}. This classical description works remarkably well in the case of the ferrimagnetic Yttrium Iron Garnet (YIG) \cite{RameshtiPhysRep22}, that has been studied in detail showing a combination of several optimal features, including the exceptionally low damping of magnetization precession \cite{MaierFlaigPRB17}. 

\begin{figure}[ht]
\begin{center}
\includegraphics[width=\linewidth]{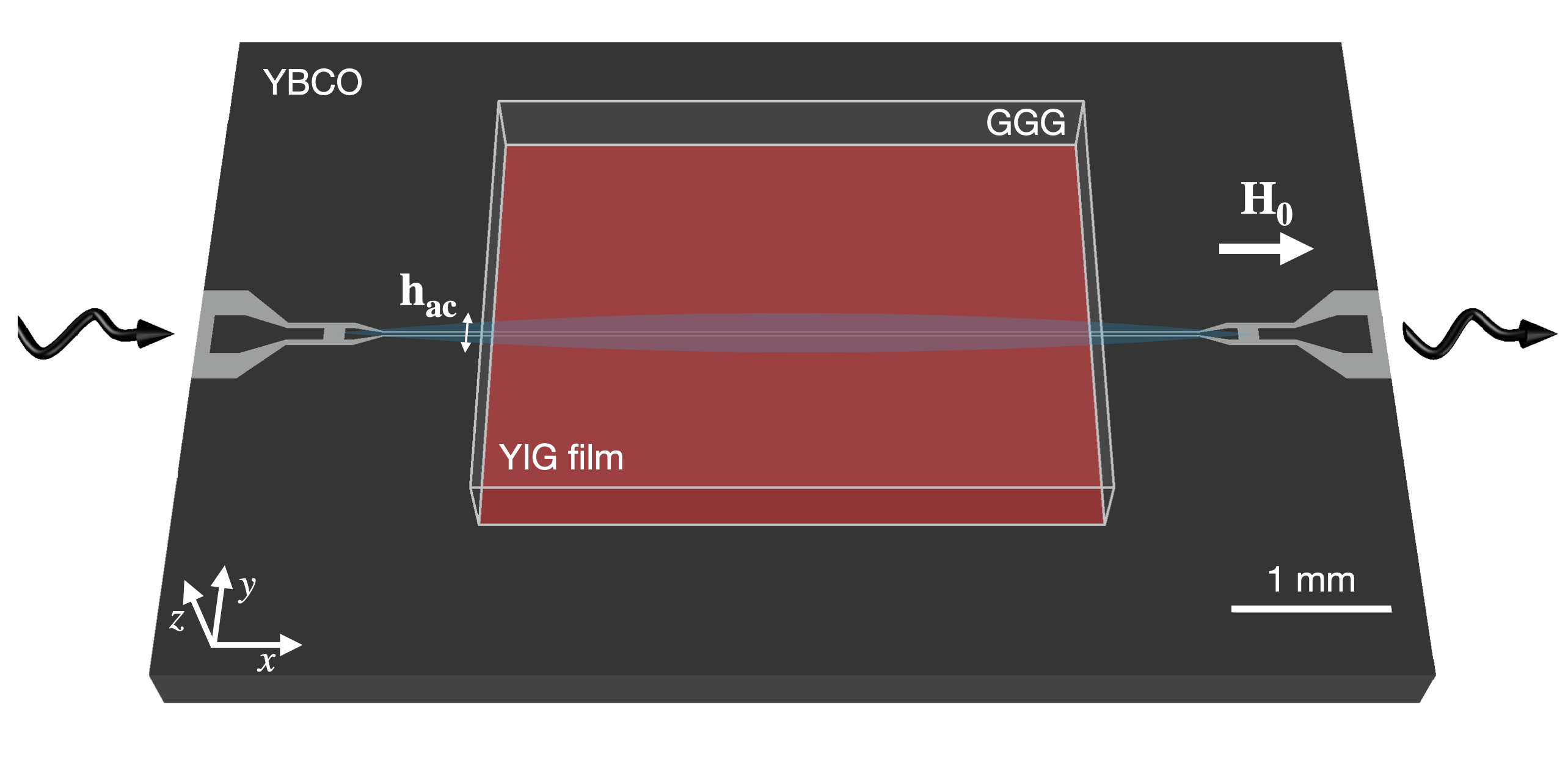}
\caption{Sketch of the YBCO/sapphire CPW resonator with the YIG/GGG film positioned above.}
\label{Fig-sketch}
\end{center}
\end{figure}

In the quantum regime, spin waves are collective (bosonic) excitations (magnons) that may coherently interact with cavity photons \cite{ImamogluPRL2009, SoykalPRL2010}. The Dicke model and its evolution in the Hopfield version, which also includes the diamagnetic term, have been developed to describe a large variety of hybrid light-matter quantum systems \cite{Kockum2019,FornDiazRM19}. In the strong coupling regime, the rotating wave approximation (RWA) and the (Jaynes) Tavis-Cummings Hamiltonian have been mainly used to interpret experimental data \cite{RameshtiPhysRep22}. However, there is a general trend now to push these studies beyond  conventional coupling regimes and more specifically to reach coupling strengths, $\lambda$, being a non-negligible fraction of the cavity frequency, $\omega_c$. Under these conditions novel effects related to processes that do not conserve the number of excitations in the system have been predicted \cite{Kockum2019,FornDiazRMP19}. 
 For $\lambda /  \omega_c \geq$0.1, we refer to the ultrastrong coupling (USC) regime, in which counter-rotating terms, neglected in RWA, must be considered. In this context, the description of the fundamental interaction between the electromagnetic field and the magnetic system, including the diamagnetic term which plays a key role in the superradiant phase transition \cite{nataf2010nogo, mazza2019superradiant, andolina2020condensation, zueco2021condensation}, still need to be clarified and tested on real magnetic materials.

\begin{figure}[th]
\centering
\includegraphics[width=\linewidth]{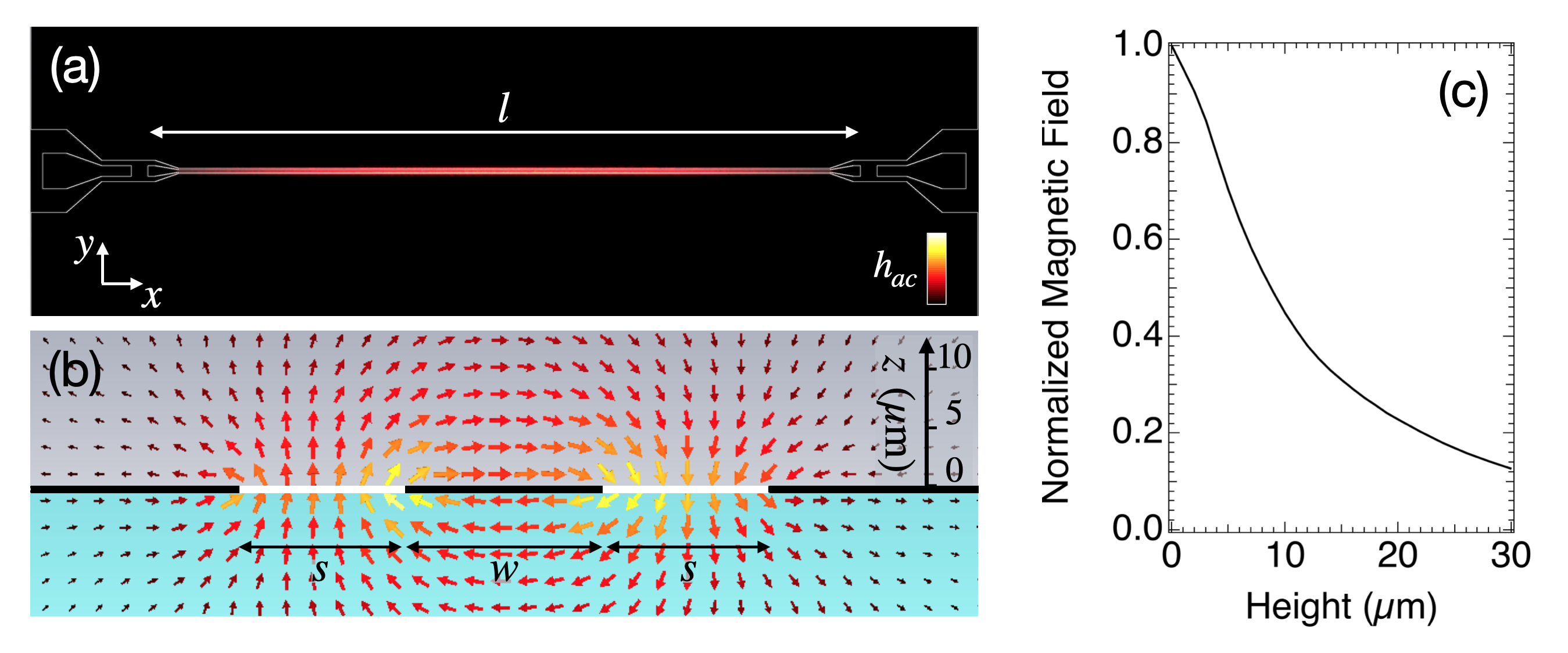}
\caption{Calculated profile of the oscillating field (magnetic component) for the fundamental mode of the bare CPW resonator. (a) Top view of the distribution $h_{ac}(x,y, z=0)$ showing the magnetic antinode in the middle of the resonator. (b) Profile of $\mathbf{h_{ac}}$ (phase = 0) in the plane perpendicular to the $x$-direction. The $x$ component of $\mathbf{h_{ac}}$ is vanishingly small. (c) Mean value of $h_{ac}(z)$ plotted as a function of the $z$ height.}
\label{fig-EMsim}
\end{figure}

\begin{figure*}[th]
\centering
\includegraphics[width=\textwidth]{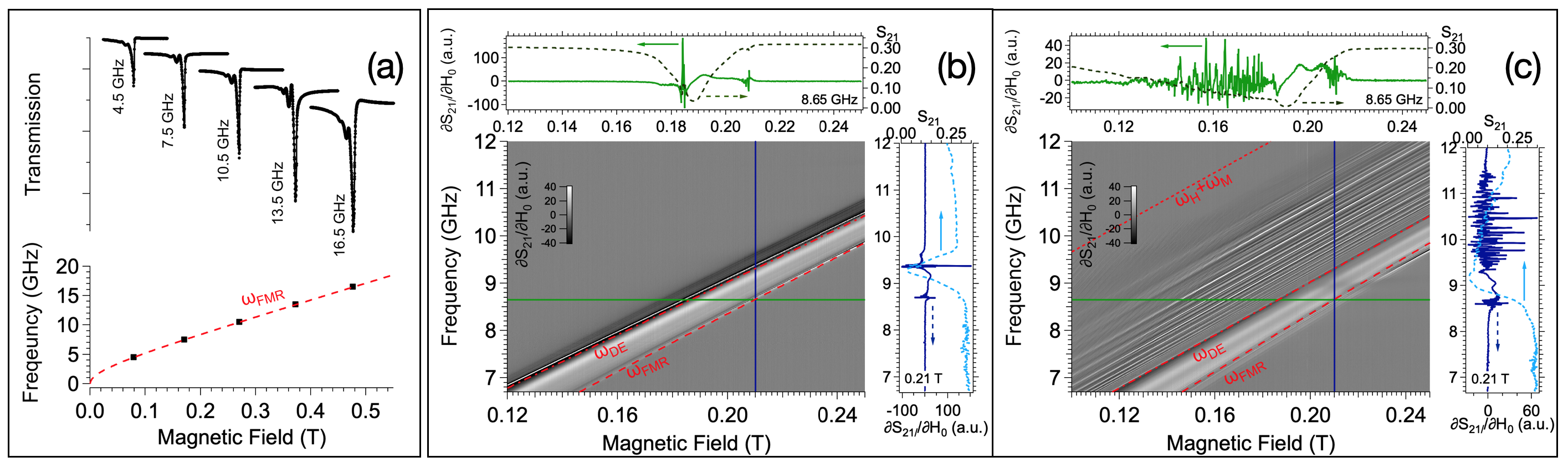}
\caption{Broadband characterization of the YIG film. (a) Ag/alumina microstrip, case $\#$A. Top: field-swept transmission spectra (temperature $T=50$~K), taken at different continuous wave frequencies, as indicated. Bottom: plot of the excitation frequencies as a function of the measured resonance field; the dashed line shows the fit with Eq.~\ref{Kittel}. (b) YBCO/sapphire CPW line, case $\#$B ($T=30$~K). Center:  Spectral map showing $\partial S_{21}(\omega, H_0)/\partial{H_0}$. Top: Plot of $S_{21}$ and $\partial S_{21}/\partial{H_0}$ as function of $H_{0}$ for $\omega/2\pi=8.65$~GHz. Right: Plot of $S_{21}$ and $\partial S_{21}/\partial{H_0}$ as function of $\omega$ for $H_0=0.21$~T. (c) Same as (b) with YIG in contact with the CPW line, case $\#$C.}
\label{fig-broadband}
\end{figure*}

Few experimental reports show magnetic systems achieving the USC regime with microwave resonators. The USC regime was first reported in the case of mm-size YIG crystals in 3D cavities \cite{PhysRevApplied.2.054002,BourhillPRB16, KostylevAPL2016, RameshtiPRB15, BourcinPRB23} and few other magnetic materials \cite{Flower_2019, PhysRevLett.123.117204, PhysRevApplied.15.044018, PhysRevB.101.214414}.
In view of the realization of scalable architectures, small magnets coupled to superconducting planar resonant geometries have been recently designed and observed to achieve the (ultra-)strong coupling regime \cite{HueblPRL13, morris2017, HouPRL2019, UstinovScAdv2021, PhysRevLett.128.047701, UstinovPRA2021, MacedoPRAppl21, RinconPRAppl23}. In spite of these encouraging results, the optimization of these hybrid systems for the USC is still largely unexplored. For instance, the geometry of the hybrid superconductor-magnet system needs to be optimized in order to maximize their mutual coupling, while magnetic materials may be preferably insulating in order to minimize the damping. 
Moreover, the vicinity to a superconducting material, which has to be resilient to magnetic fields \cite{GhirriAPL15} to allow for ferromagnetic resonance (FMR) experiments, may significantly affect the spectrum of magnetic excitations through several mechanisms \cite{UstinovScAdv2021, NiedzielskiPRAppl23}.

In this article, we address the problem of reaching the USC regime in coupled superconductor/ferrimagnet hybrid architectures. By means of a series of systematic experiments carried out with YIG films and planar transmission line resonators, we demonstrate the achievement of high coupling rates by positioning the magnetic film in direct contact with the superconducting resonator. In our geometry, the excitation of spin waves take place at the superconductor/ferrimagnet interface, where the amplitude of the microwave field is maximum. The optimized magnon-photon coupling results in collective coupling strengths as large as 0.2 times the cavity frequency. Data analysis, carried out with a modified Hopfield model for which we provide an exact solution, also evidences vanishingly small diamagnetic coupling for magnon excitations in YIG. 

The article is organised as follows: we start presenting the experimental methods in Section~\ref{sect-exp} whist in Section~\ref{sect-EM_simulations}  we report numerical simulations of the profile of the oscillating field which allow us to estimate volume and coupling strength involved in the spin-photon process. Experimental results are presented in progressive way by considering the broadband characterization first in Section~\ref{sect-broadband}, then transmission spectra on resonators in Section~\ref{sect-coupling}. Modelling and data analysis are reported in Section~\ref{sect-hamiltonian}, then discussion and concluding remarks in the latest sections.

\section{Experimental methods}
\label{sect-exp}

We prepared planar resonators with different sizes and geometries, including meanders with multiple resonance frequencies and inverse anapole resonators with focused electromagnetic radiation \cite{bonizzoni_coupling_2022}, to study the coupling with different YIG samples. Here we report on two types of planar devices: microstrip and coplanar waveguide (CPW) resonator (Fig.~\ref{Fig-sketch}). For each of them, the corresponding broadband transmission line, having the same dimensions and geometry (except for the presence of the input and output coupling gaps defining the half-wavelength resonator), has been fabricated to investigate the spin wave excitation spectrum in the different cases.  Microstrip lines having $500~\mu$m wide central strips were obtained by wet etching of $\mathrm{Ag/Al_2O_3}$ films \cite{GhirriAdvQTechnol20}. Superconducting coplanar waveguide (CPW) lines (Fig.~\ref{Fig-sketch}) were fabricated from commercial  YBa$_2$Cu$_3$O$_7$ (YBCO) films deposited on sapphire. Etching was carried out by Ar plasma in a reactive ion etching (RIE) chamber. The central conductor had length $l\approx 6$~mm, width $w=(17 \pm 1)~\mu$m and separation $s=(14 \pm 1)~\mu$m from the lateral ground planes. Further details are given in \cite{supplementary}. 

The experiments were carried out at low temperature in applied magnetic field ($\mathbf{H_0}$), which was oriented in the plane of the film along $x$ (Fig.~\ref{Fig-sketch}). Frequency-swept spectra were acquired using a Vector Network Analyzer with typical incident power $P_{inc}=-8$~dBm. We studied YIG films grown by liquid-phase epitaxy on $500~\mu$m thick gadolinium gallium garnet (GGG) substrates. Unless specified, the film has rectangular shape with thickness of $5~\mu$m and in-plane sizes of $4 \times 3~\mathrm{mm}^2$. The sample was mounted in diverse ways on the aforementioned microwave transmission lines and resonators. We anticipate that different spin-wave spectra and, ultimately, different magnon-photon coupling strengths were obtained in different cases. In the following we thus consider three experimental configurations: YIG film glued on metallic microstrip (case $\#$A);  YIG film glued on YBCO CPW line (case $\#$B); YIG film in contact with the YBCO CPW line (case $\#$C). In $\#$A and $\#$B the estimated film-resonator separation was on the order of $10~\mu$m. In case $\#$C we obtained a good contact between YBCO and YIG surfaces thanks to a polytetrafluoroethylene (PTFE) screw that gently pushed the GGG substrate against the YBCO surface from the backside.

\section{Electromagnetic simulations}
\label{sect-EM_simulations}

Finite-element electromagnetic simulations were carried out using commercial software to evaluate amplitude and distribution of the resonator fields. The maps calculated for the CPW resonator (Fig.~\ref{fig-EMsim}(a,b)) show the expected in-plane and out-of-plane profiles of the microwave magnetic field, $\mathbf{h_{ac}}$. 
The resonator field is mostly confined around the central line between the lateral ground planes, in a region of approximate width $w+2 s=45~\mathrm{\mu m}$ \cite{supplementary}. To evaluate the $z$ dependence of the resonator field, we calculated the in-plane averaged value of $h_{ac}$ as a function of the $z$ height (Fig.~\ref{fig-EMsim}(c)). The decay of $h_{ac} (z)$ is quasi-exponential and for $z=10~\mu \mathrm{m}$ it results in a reduction of a factor $\approx 2$ with respect to the maximum value.

We can exploit these electromagnetic simulations to estimate the amplitude of the vacuum fluctuation $b_{vac}$ and, consequently, of the spin-photon coupling strength ($g_s$) expected with this particular CPW resonator. By rescaling the calculated magnetic field to the single photon power level \cite{supplementary}, we obtain $b_{vac} \approx 3$~nT. This value is in perfect agreement with the value derived from $b_{vac}\approx \mu_0 \omega_c/(4 w) \sqrt{h/Z_0}$ \cite{TosiAIPAdv14, HueblPRL13}, being $h=6.626 \times 10^{-34}$~J~s the Planck constant and $Z_0= 58~\mathrm{\Omega}$ the nominal impedance of the CPW resonator. The spin-photon coupling strength then results $g_s=\gamma b_{vac}/4=21$~Hz, where $\gamma$=28.02~GHz/T is the electron's gyromagnetic ratio. For comparison, the value of the spin-photon coupling expected for the fundamental mode of the microstrip resonator is less than 1~Hz.

\begin{figure*}[ht]
\centering
\includegraphics[width=\textwidth]{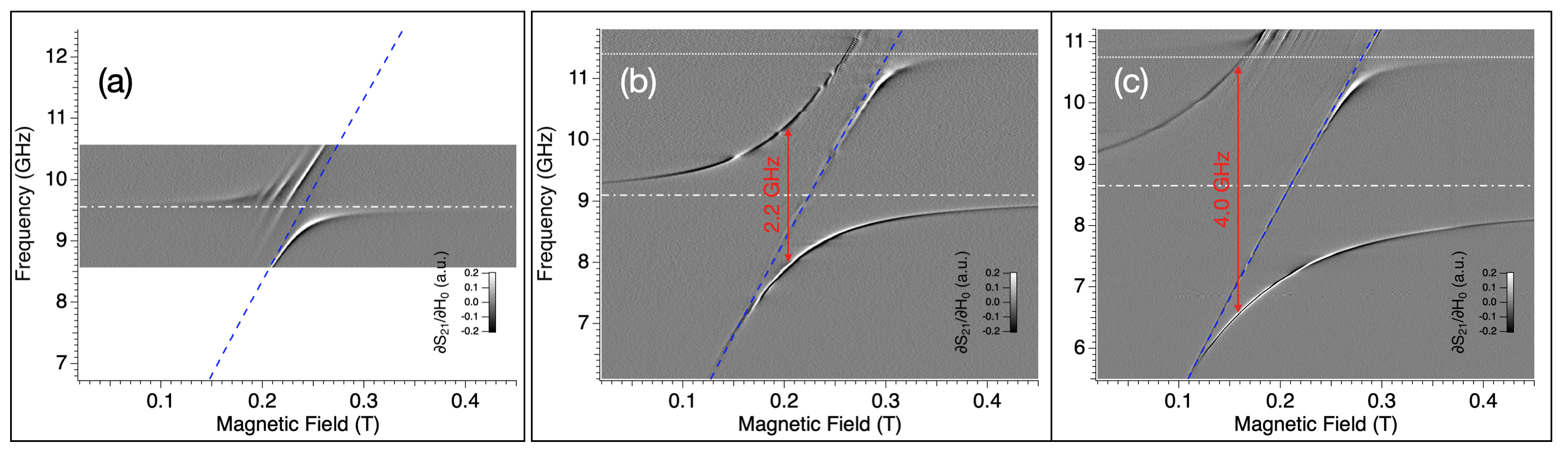}
\caption{Coupling of the YIG film with different planar resonators ($P_{inc}=-8$~dBm). (a) Microstrip resonator, case $\#$A ($T=50$~K). (b) CPW resonator, case $\#$B ($T=30$~K). (c) YIG in contact with the YBCO CPW resonator, case $\#$C ($T=30$~K). The blue dashed lines display $\omega_{FMR}(H_0)$ (Eq.~\ref{Kittel}) while the white dash-dot lines indicate the resonators' frequencies. The dotted lines correspond to the frequency of a broad mode due to the metal box hosting the CPW lines.}
\label{fig-coupling}
\end{figure*}

\section{Broadband transmission spectroscopy characterization}
\label{sect-broadband}

We first present a series of transmission ($S_{21}$) spectra acquired with broadband transmission lines by sweeping the microwave frequency ($\omega$) at steady values of the external magnetic field $H_0$ (Fig~\ref{fig-broadband}). In the case of the microstrip ($\#$A), the main resonance dip follows the Kittel relation (Fig.~\ref{fig-broadband}(a)), which, for in-plane magnetized films, reduces to \cite{KittelPR48}: 
\begin{eqnarray}
\omega_{FMR}=\sqrt{\omega_H(\omega_H+\omega_M)},
\label{Kittel}
\end{eqnarray}
where $\omega_H/2\pi=\gamma\mu_0 H_0$ and $\omega_M/2\pi=\gamma\mu_0 M_s$, being $\mu_0=4 \pi \times 10^{-7}$~H/m the vacuum permeability and $\mu_0M_s=0.245$~T the saturation magnetization of YIG as reported in the literature \cite{MaierFlaigPRB17, supplementary}.

For case $\#$B, the numerical derivative of the signal $\partial S_{21}(\omega, H_0)/\partial{H_0}$ is shown in Fig.~\ref{fig-broadband}(b). Additional spin wave resonance modes are well visible in this case. Such modes are commonly observed for YIG films \cite{Kittel, KajiwaraNature10} but their frequency specifically depends by the profile of the microwave field $\mathbf{h_{ac}}$. In particular, narrow CPW lines efficiently excite travelling spin waves with finite wavevector $0 < k \le 2 \pi/s$ \cite{MaksymovPhysE15}. Given the in-plane magnetization of the film ($\mathbf{H_0} ||x$) and the negligible $x$-component of $\mathbf{h_{ac}}$ (Fig.~\ref{fig-EMsim}), the dispersion of the Damon-Eshbach modes follows the characteristic expression \cite{GuerevichMelkov}:
\begin{eqnarray}
\omega_{DE}=\sqrt{\left(\omega_H+\frac{\omega_M}{2}\right)^2-\left(\frac{\omega_M}{2}\right)^2 e^{-2kd}},
\label{DE_dispersion}
\end{eqnarray}
where $k=k_s=2\pi/s=4.5 \times 10^5~\mathrm{rad~m^{-1}}$ (dash-dot line in Fig.~\ref{fig-broadband}(b)) \cite{Kennewell2007}. We note that $\omega_{DE}$ is near the maximum limit of Eq.~\ref{DE_dispersion}, $\omega_H(H_0)+\omega_M/2$, whilst the coupling between the  magnetization of YIG and GGG \cite{WangPRB20} is not evident from the measured spectra \cite{supplementary}.

Fig.~\ref{fig-broadband}(c) displays the spectrum obtained in case $\#$C. Here the transmission spectrum shows several additional spin wave modes whose origin can be related to the special boundary conditions imposed by the contact of the magnetic YIG with the superconducting YBCO planes, as observed in similar experiments \cite{TsutsumiIEEE97, OatesIEEE97}. Such modes are located in a wide absorption band between $\omega_{DE}(H_0)$ and $\omega_H(H_0)+\omega_M$ and their typical linewidth is $10 \lesssim \kappa_m \lesssim 40$~MHz.

\section{Enhanced Spin Photon Coupling and Cooperativity}
\label{sect-coupling}

Once we had identified the main features of the spin wave excitation spectrum, we performed experiments in a similar configuration, with the YIG film on half-wavelength resonators having dimensions and geometries equivalent to the ones of the broadband lines used in Fig.~\ref{fig-broadband} \cite{supplementary}. The $S_{21}$ map taken with the microstrip resonator (case $\#$A) is shown in (Fig.~\ref{fig-coupling}(a)). The lower branch has an oblique asymptote with the FMR mode (Eq.~\ref{Kittel}), while the upper branch is barely visible owing to multiple resonances with spin wave modes. This situation is common with other spin ensemble \cite{DinizPRA11, GhirriPRA16} or YIG \cite{RameshtiPhysRep22} systems in a resonator mode. From Fig.~\ref{fig-coupling}(a) the collective coupling strength can be roughly estimated to be of order of hundreds of MHz. 

The YBCO CPW resonator has, with respect to the microstrip, a smaller mode volume that results in a larger spin-photon coupling $g_s$ (Sect.~\ref{sect-EM_simulations}). A large anticrossing, having well defined polariton branches, is obtained in this case ($\#$B, Fig.~\ref{fig-coupling}(b)). Even larger couplings were obtained when the YIG film was put in direct contact with the YBCO resonator (case $\#$C): with this geometry the splitting raises to $2 \lambda /2 \pi \approx 4$~GHz (Fig.~\ref{fig-coupling}(c)), a factor $\approx 2$ larger than case $\#$B. Within the anticrossing gap, no additional modes are visible except for $\omega_{FMR}$. This suggests that this mode is weakly coupled to the CPW resonator. Additional spin wave modes are observed above $\approx 10$~GHz due to the presence of a spurious box mode (Fig.~\ref{fig-coupling}(b,c)). 

It is worth to estimate the cooperativity, that is commonly defined as $C=4 \lambda^2/( \kappa_m \kappa_c)$ \cite{FornDiazRMP19}. From our spectra we take $\kappa_m/2 \pi=40$~MHz and $\kappa_c/2 \pi=8$~MHz as decay rates of magnon (Sect.~\ref{sect-broadband}) and uncoupled cavity \cite{supplementary} modes. With these numbers, we achieve $C=5 \times 10^4$.  Considering the normalized parameter $U=\sqrt{(4 \lambda^2/ \kappa_m \kappa_c)(\lambda/\omega_c)}=108$, we infer that our results ranks among the best performing physical platforms for the reaching of the USC regime \cite{FornDiazRMP19}.

\section{System Hamiltonian}
\label{sect-hamiltonian}

We model our system by considering a quantized single-mode electromagnetic field (with $\omega_c$ cavity frequency) interacting with an ensemble of magnetic moments. We consider collective operators for the spin ensemble, the quantization of both spin excitations and the electromagnetic field which allows us to introduce the respective bosonic operators $\hat{a}$ and $\hat{b}$. Due to the vanishing orbital angular momentum of Fe$^{3+}$ in YIG \cite{GuerevichMelkov}, we expect a prominent Zeeman interaction of the type $\hat{\mathcal{H}}_{\rm Z} = -g_e \hat{\mathbf{\sigma}} \cdot \mu_B\hat{\mathbf{h}}$ for a single spin. Here $\comm{\hat \sigma_j}{ \hat \sigma_k} = i \epsilon_{jkl}\hat{\sigma}_l$ are the Pauli operators, $\hat{\mathbf{h}}$ is the magnetic field component of the cavity resonator while $\mu_B$ is the Bohr magneton and $g_e \approx 2$ in the case of a simple electron. However, to not exclude the possibility of having orbital angular momentum contributions in our hybrid system, we also consider this degree of freedom including a diamagnetic term, which comes from the usual minimal coupling replacement. 

The total Hamiltonian then reads ($\hbar = 1$)
\begin{eqnarray}
    \label{eq: cavity-spins}
    \hat{\mathcal{H}} &=& \omega_c \hat{a}^\dagger \hat{a} + \frac{\omega_b}{2} \sum_{j=1}^N \sigma_z^{(j)} + \frac{\lambda}{2 \sqrt{N}} \sum_{j=1}^N \sigma_x^{(j)} \left( \hat{a} + \hat{a}^\dagger \right) \nonumber \\
    &&+ \beta \left( \hat{a} + \hat{a}^\dagger \right)^2 ~,
\end{eqnarray}
where $\hat{a}$ ($\hat{a}^\dagger$) is the photon annihilation (creation) operator, $\omega_c$ is the cavity resonance frequency, $\omega_b$ is the resonance frequency of a single spin, $\lambda$ is the collective light-matter coupling, and $\beta$ is the coefficient of the diamagnetic term.

By using the collective spin operators $\hat{J}_z \equiv (1/2) \sum_{j=1}^N \hat{\sigma}_z^{(j)}$ and $\hat{J}_x = \hat{J}_+ + \hat{J}_- \equiv (1/2) \sum_{j=1}^N \hat{\sigma}_x^{(j)}$, we can apply the Holstein-Primakoff transformations \cite{holstein1940field}
\begin{equation}
    \label{eq: holstein-primakoff transformation}
    \hat{J}_z \to \hat{b}^\dagger \hat{b} - \frac{N}{2} ~, \quad \hat{J}_+ \to \hat{b}^\dagger \sqrt{N - \hat{b}^\dagger \hat{b}} ~ , \quad \hat{J}_- = \hat{J}_+^\dagger ~ ,
\end{equation}
where $\hat{b}$ and $\hat{b}^\dagger$ are the magnon annihilation and creation operators respectively, which obey to the standard bosonic commutation relations. In the thermodynamic limit (i.e. $N \to \infty$) we can approximate $\hat{J}_+ \approx \sqrt{N} \hat{b}^\dagger$. Then, by applying the Holstein-Primakoff transformation, we obtain:
\begin{eqnarray}
    \label{eq: final Hamiltonian}
    \hat{\mathcal{H}} &=& \omega_c \hat{a}^\dagger \hat{a} + \omega_b \hat{b}^\dagger \hat{b} + \lambda \left(\hat{b} + \hat{b}^\dagger \right) \left( \hat{a} + \hat{a}^\dagger \right) \nonumber \\
    && + \beta \left( \hat{a} + \hat{a}^\dagger \right)^2 ~ ,
\end{eqnarray}
that is the well-known Hopfield Hamiltonian \cite{hopfield1958theory}.

We consider the dependence of the magnon resonance frequency $\omega_b$ to the external magnetic field $H_0$ as described by $\omega_{b}=\sqrt{\omega_H(\omega_H+\omega_M)}+\Delta$, leaving as the sole free parameter the energy shift $\Delta$ characterizing high frequency magnons for the next step of our investigation. In our analysis we also leave as free parameters the cavity frequency $\omega_c$, the collective coupling $\lambda$ and the factor of the diamagnetic term $\beta$.

The Hamiltonian in Eq.~3 of the main text can be expressed in terms of two non-interacting harmonic oscillators $\hat{\mathcal{H}} = \Omega_- \hat{P}_-^\dagger \hat{P}^{}_- + \Omega^{}_+ \hat{P}_+^\dagger \hat{P}^{}_+$, where $\hat{P}_\pm$ are the polariton operators, which are linear combinations of light and matter operators $\hat{P}_\mu = c_1^{(\mu)} \hat{a} + c_2^{(\mu)} \hat{b} + c_3^{(\mu)} \hat{a}^\dagger + c_4^{(\mu)} \hat{b}^\dagger$, with $\mu = \pm$. To fit our parameters, we need first to find the polariton frequencies, and, being a proper bosonic excitation of the system, the operator $\hat{P}_\mu$ fulfills the equation of motion of the harmonic oscillator $[\hat{P}_\mu, \hat{\mathcal{H}}] = \Omega_\mu \hat{P}_\mu$. Since the polariton operator $\hat{P}_\mu$ is a linear combination of the light and matter operators, we need to calculate first the commutator of the latter with the Hamiltonian
\begin{eqnarray*}
    \comm{\hat{a}}{\hat{\mathcal{H}}} &=& \omega_c \hat{a} + \lambda \left( \hat{b} + \hat{b}^\dagger \right) \\
    \comm{\hat{b}}{\hat{\mathcal{H}}} &=& \omega_b \hat{b} + \lambda \left( \hat{a} + \hat{a}^\dagger \right) \\
    \comm{\hat{a}^\dagger}{\hat{\mathcal{H}}} &=& -\omega_c \hat{a}^\dagger - \lambda \left( \hat{b} + \hat{b}^\dagger \right) \\
    \comm{\hat{b}^\dagger}{\hat{\mathcal{H}}} &=& -\omega_b \hat{b}^\dagger - \lambda \left( \hat{a} + \hat{a}^\dagger \right) ~ ,
\end{eqnarray*}
and the polariton frequencies $\Omega_\mu$ are obtained by finding the positive eigenvalues of the following Hopfield matrix
\begin{equation}
    \label{eq: Hopfield matrix}
    \mathcal{M} = 
    \begin{pmatrix}
    \omega_c + 2 \beta & \lambda & -2 \beta & -\lambda\\
    \lambda & \omega_b & -\lambda & 0\\
    2 \beta & \lambda & -\omega_c - 2 \beta & -\lambda\\
    \lambda & 0 & -\lambda & -\omega_b
    \end{pmatrix} ~ .
\end{equation}
leading to:

\begin{equation}
    \label{eq: polariton frequencies}
    \Omega_\pm = \frac{1}{\sqrt{2}} \sqrt{\tilde{\omega}_c^2 + \omega_b^2 \pm \sqrt{\left(\tilde{\omega}_c^2 - \omega_b^2 \right)^2 + 16 \omega_c \omega_b \lambda^2}} ~,
\end{equation}
where $\tilde{\omega}_c = \sqrt{\omega_c (\omega_c + 4 \beta)}$.

The above equation fits well the peaks of the $S_{21}$ spectrum (case $\#$C). Figure \ref{fig:fit} shows the best fit result obtained with these parameters: $\omega_c /2 \pi= 8.65$ GHz, $\Delta /2 \pi= 2.05$~GHz, and $\lambda /2 \pi= 2.002$ GHz. For what concerns the diamagnetic parameter $\beta$, the only nontrivial result (i.e. nonzero result from fit) has been obtained by assuming a dependence on the magnon frequency: $\beta = \alpha / \sqrt{\omega_b}$, which can be justified by the fact that the presence of this term is dominant at low frequencies. With this assumption, we obtained $\alpha /\sqrt{2 \pi} = 3 \times 10^{-3} \ \text{GHz}^{\frac{3}{2}}$, corresponding to a value of the diamagnetic coefficient $\beta/2\pi \sim 10^{-3}$~GHz on resonance condition. The ratio between the collective coupling and the cavity frequency is approximately 0.23, fulfilling the criterion  $\lambda/\omega_c > 0.1$ for USC. The fit confirms that the influence of the diamagnetic term is almost negligible leading us to conclude that the system couples to the resonator mainly through the spins. Notice that its value is more than two orders of magnitude smaller than the standard diamagnetic term for electric dipolar interactions (Appendix~\ref{Appendix-diamagnetic_term}).

\section{Discussion}

The dispersion characteristics of the spin wave spectrum in an infinite ferromagnetic film have been reported by Kalinikos and Slavin~\cite{KalinikosJPhysC86}. Taking into account both dipole-dipole and exchange interactions, theory predicts the excitation of perpendicular standing spin waves (PSSW) modes even in the long wavenumber ($k_y d \ll 1$) limit. 
These modes are due to the broken translational invariance along the film thickness. Remarkably, in these conditions, the lowest mode shows a quasi-uniform  profile with a dispersion equation very similar to the Damon-Eshbach dipolar surface mode (Eq.~\ref{DE_dispersion}). Conversely, higher order PSSW modes display a nonuniform magnetization profile along $z$ \cite{KalinikosJPhysC86, DemokritovSpringer21}. 

Broadband spectroscopy data in Fig.~\ref{fig-broadband} evidence that the spin wave spectrum is influenced by two main factors: (i) the distribution of the exciting electromagnetic field and (ii) the boundary conditions at the ferrimagnet-superconductor interface. The effect of (i) emerges from the comparison between Fig.~\ref{fig-broadband}(a) and (b). In the case of the wide  microstrip line, the modes calculated with Eq.~\ref{Kittel} and Eq.~\ref{DE_dispersion}, the latter by considering $k=2 \pi/ w^{\prime}$ with $w^{\prime}=500~\mu \mathrm{m}$, are very near. They both follow the measured dispersion of lowest resonance mode in Fig.~\ref{fig-broadband}(a), showing that the magnetostatic approximation is valid in this case. Conversely, in the case of the narrow CPW line, the modes calculated with Eq.~\ref{Kittel} and Eq.~\ref{DE_dispersion} ($k=2 \pi/ w$) are different. Since the main absorption line is found for $\omega(H_0)\approx\omega_{DE}(H_0)$ in Fig.~\ref{fig-broadband}(b), according to the dipole-exchange theory \cite{KalinikosJPhysC86, DemokritovSpringer21}, we expect this mode shows a homogeneous magnetization profile along the film thickness.

The effect of (ii) is clearly evident from the direct comparison between Fig.~\ref{fig-broadband}(b) and (c), which shows the appearance of additional modes at $\omega>\omega_{DE}$ for superconductor and ferrimagnet in direct contact. The interplay between a magnetically ordered film in the vicinity of a superconductor is an interesting and open issue. In the first instance, we can assume that the Meissner effect (perfect diamagnetism) imposes the expulsion of the oscillating field at the interface. Intuitively this can be visualized as a superconducting plane reflecting the image of the magnetic excitations in YIG \cite{UstinovAFM2018, UstinovJAP2018}. Analysis of the portion of the spectrum can be attempted by simulations as suggested in \cite{UstinovJAP2018}, yet the dispersion law may depend to a large extent on the specific materials and geometry of the problem. We cannot exclude other mechanisms, like those related to vortex configuration and dynamics \cite{NiedzielskiPRAppl23}, anisotropy-induced surface pinning and other interfacial effects \cite{LeePRL20}, although these were reported for thin ferromagnetic films. Particular effects can also be induced on the superconductor and these may depend on the type of superconducting wave function and on its coherence length. Whilst our experiments represent a case study involving insulating YIG and the high critical temperature YBCO superconductor that merits further attention, this issue goes beyond the scope of our research that is focused on the ultrastrong coupling regime which, instead, is achieved also in different conditions and geometries.

In the experiments with YIG film and resonators (Fig.~\ref{fig-coupling}) the profile of the resonator field (Fig.~\ref{fig-EMsim}) may introduce additional quantization of the spin wave modes \cite{BayerSpringer06}. Consistently with Fig.~\ref{fig-broadband}(c)), we note that the fitted value of the frequency shift $\Delta /2 \pi= 2.05$~GHz is $\omega_{DE}<\Delta<\omega_M+\omega_H$, thus within the band of spin wave resonance modes observed for YIG and YBCO CPW line in direct contact. This frequency shift can be captured by finite-element electromagnetic simulations \cite{BourcinPRB23}, in which YIG film and resonator are respectively modelled as gyrotropic medium and perfect electric conductor (Appendix \ref{Appendix-EM_simulations}).

\begin{figure}
    \centering
    \includegraphics[width=\linewidth]{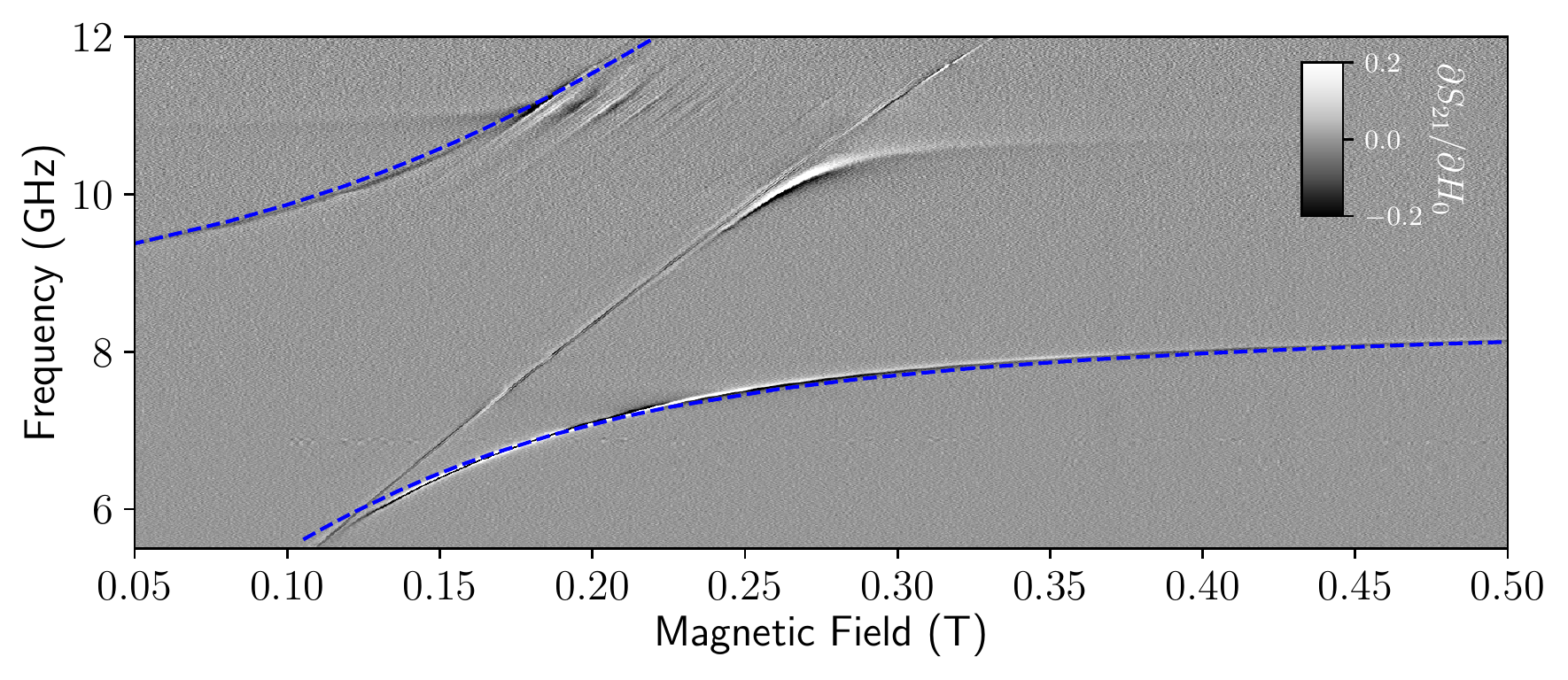}
    \caption{Best fit of the transmission spectrum obtained in case $\#$C with the YIG film pressed on top of the superconducting YBCO CPW. The obtained parameters are: $\omega_c/2 \pi = 8.65$ GHz, $\Delta/2 \pi = 2.05$ GHz, $\lambda/2 \pi = 2.002$ GHz, $\alpha/\sqrt{2 \pi} = 3 \times 10^{-3} \ \text{GHz}^{\frac{3}{2}}$.}
    \label{fig:fit}
\end{figure}

The collective coupling strength $\lambda/2 \pi=2.002$~GHz obtained by fitting the experimental data with Eq.~\ref{eq: polariton frequencies} (Fig.~\ref{fig:fit}) can be spelled out as $\lambda=g_s\sqrt{2 s_{\mathrm{Fe}} N}$, where $s_{\mathrm{Fe}}=5/2$ is the ground state spin of YIG and $N=1.8 \times 10^{15}$ is the total number of spins. The latter resulted much higher than the mean number of photons in the resonator \cite{SageJAP11, supplementary}. The spin-photon coupling is $g_s=21$~Hz as derived in Sect.~\ref{sect-EM_simulations}.
Considering the spin density $\rho=2 \times 10^{28}~\mathrm{m}^{-3}$ \cite{RameshtiPhysRep22} and the effective area of $4~\mathrm{mm} \times 45~\mu \mathrm{m}$ in which the microwave field overlaps the YIG film (Fig.~\ref{fig-EMsim}(a,b)), we can estimate an upper bound $d^{\prime} \approx 500~\mathrm{nm} \ll d$ for the thickness of the portion of YIG film coupled with the CPW resonator. This suggests that the magnon modes that are effectively coupled to the cavity one are located in close vicinity with the superconducting resonator. This observation is supported by similar experiments we performed with 20~$\mu$m thick YIG film, which provide results quite similar to those obtained with the 5~$\mu$m YIG film, thus confirming that the coupling is confined within few $\mu$m (Appendix~\ref{Appendix-film_thickness}). The ratio between the collective couplings in configuration $\#$C and $\#$B is $g_{s,C}/g_{s,B}\sqrt{N_C/N_B}\approx 2$ (Fig.~\ref{fig-coupling}(b,c)). From finite-element simulations, we estimated $g_{s,C}/g_{s,B} \approx 2$ due to the exponential decay of $b_{vac}$ with $z$ distance (Fig.~\ref{fig-EMsim}(c)). We therefore expect $N_C \approx N_B$. These observations indicate that in our experiments the achievement of the USC regime takes place mainly as a consequence of the optimized magnon-photon coupling strength.

The Hopfield model we used to analyze our spectra allows us to overcome the RWA approximation generally used in previous works with magnetic systems, thus providing a quantum description of the problem that can be applied to safely explore the USC regime. We stress that the analysis of the spectrum reported in Sect.~\ref{sect-hamiltonian} is quite robust since it relies on a minimal set of free parameters. Obviously, by introducing more degrees of freedom in the Hamiltonian the spectrum can also be fitted well. For instance, one may wonder whether more magnetic modes couple to cavity photons simultaneously. Results of simulations with three (and more) modes are reported in Appendix~\ref{Appendix-multimode_fit}. It always results that the best fit is obtained with one dominant mode, ultra-strongly coupled to the resonator's one, plus additional modes coupled more weakly. As a matter of fact, on the basis of the sole fitting of the polaritonic branches, one cannot exclude that additional magnetic excitations are involved in our and in similar experiments reported in the literature. However, the representation with only one effective mode allows putting stringent bounds to the other terms of the Hamiltonian. Specifically, this is the case for the diamagnetic term, that in our analysis results vanishingly small. This result is consistent to what is expected for a pure spin interaction, although the issue is still debated in the literature. For instance, we just mention that conclusions reported in Ref.~\cite{UstinovPRA2021} lead to a finite diamagnetic contribution yet, as those authors concluded, this may arise from surface plasmonic modes or by the different nature of magnetic material (permalloy). In our case, the absence of this diamagnetic term may be relevant for the observation of superradiant phase transition expected for $\lambda/\omega_c > 0.5$ (see discussion in Appendix~\ref{Appendix-diamagnetic_term}), making pure spin systems interesting and unique in this perspective.

As concerns a possible route for applications, the coexistence of superconductivity and magnetism is not trivial since the presence of a magnetic field can be detrimental to the superconductor. In our experiments, the use of high critical temperature YBCO resonators, resilient to high magnetic field \cite{GhirriAPL15}, indicates a good option for the realization of this kind of hybrid device. Our experiments also show that the contact between superconductor and magnet, and in particular the vanishing gap between the two, is critical to enhancing the coupling between magnetic and microwave modes (Fig.~\ref{fig-coupling}(b,c)). The direct growth of YIG on top of superconducting oxide is not straightforward but there can be different options to overcome this technical issue \cite{TremplerAPL20, KosenAPLMater19}. The availability of commercial YIG films of excellent quality allows the easy implementation of our experiment in different geometries.

\section{Conclusions}

In summary, our experimental and theoretical results show the achievement of the ultrastrong coupling with a YIG film positioned in direct contact with a superconducting CPW resonator. The obtained collective coupling strength $\lambda/2 \pi=2.002$~GHz is improved by at least one order of magnitude with respect to previous reports involving YIG films in 3D cavities \cite{ZhangJAP16} or bulk YIG crystals in planar resonators \cite{HueblPRL13, morris2017}. The estimated $\lambda/\omega_c$ ratio of 0.23 and the cooperativity $C=5 \times 10^4$ are among the largest reported so far for magnetic systems yet we believe that there are still margins to further increase them for instance by using planar resonators with more confined mode volumes and higher quality factors, or by means of YIG films detached from the GGG substrate \cite{TremplerAPL20, KosenAPLMater19} to reduce magnetic losses \cite{DanilovSovPhysJ89, MihalceanuPRB18}. The very small diamagnetic coefficient (compared to standard electric dipole interactions) observed in these systems makes them suitable for exploring superradiant phase transitions \cite{nataf2010nogo, mazza2019superradiant, andolina2020condensation, zueco2021condensation}.

\begin{acknowledgments}
We thank Prof.~Bulat Rameev for useful discussions. This work was partially supported by European Community through FET Open SUPERGALAX project (grant agreement No.~863313) and by NATO Science for Peace and Security Programme (NATO SPS Project No.~G5859). MM acknowledges TUBITAK-BIDEB for the support under the 2219 scholarship program. SS acknowledges the Army Research Office (ARO) (Grant No. W911NF1910065).
\end{acknowledgments}
\begin{appendix}
\renewcommand\thefigure{\thesection.\arabic{figure}} 
\setcounter{figure}{0}

\section{Additional finite-element electromagnetic simulations}
\label{Appendix-EM_simulations}

\begin{figure}[ht]
\centering
\includegraphics[width=\linewidth]{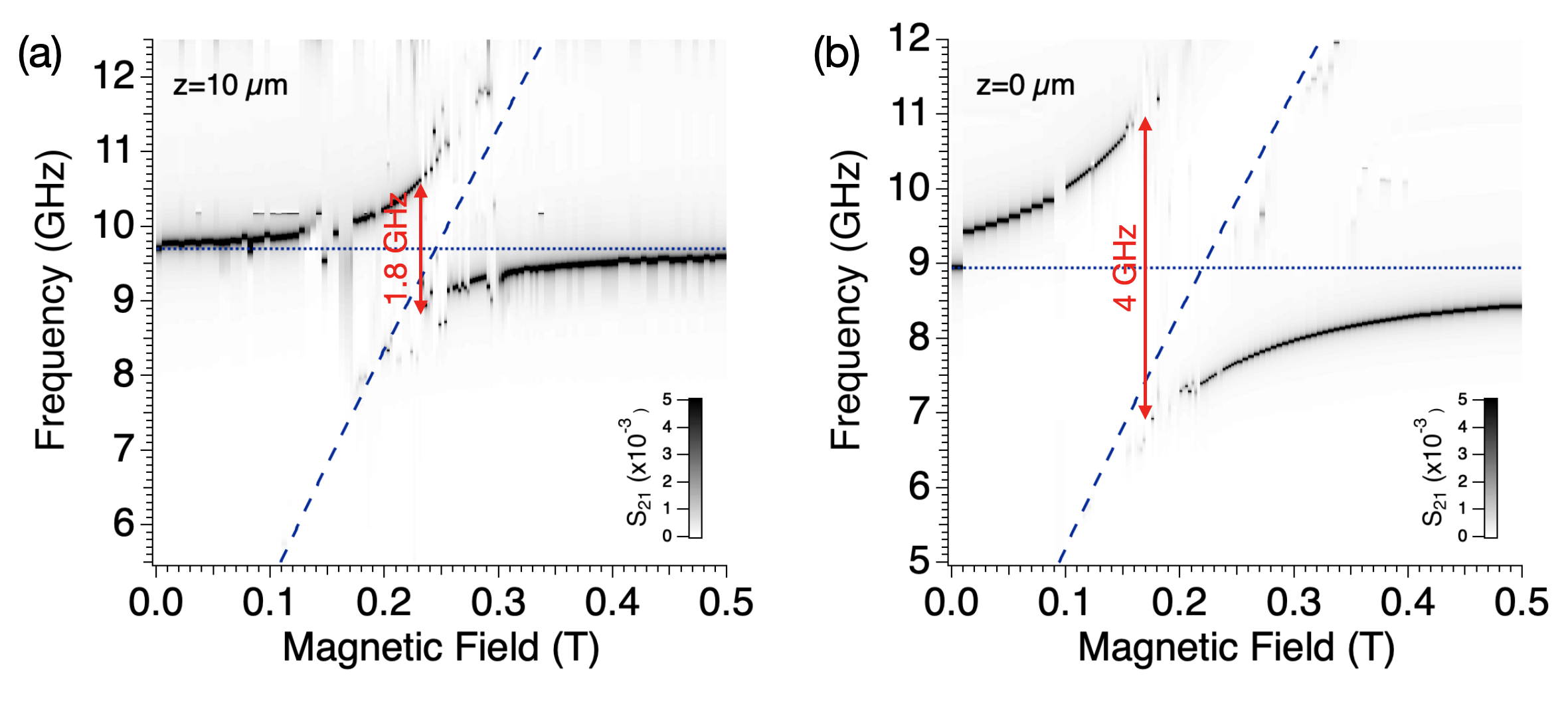}
\caption{Transmission spectral maps obtained by finite-element electromagnetic simulations. (a) The YIG film is lifted of 10~$\mu$m with respect to the surface of the CPW resonator. (b) The YIG film is in contact with the conducting surfaces of the resonator. Blue dashed lines indicate $\omega_{FMR}(H_0)$ while the dotted lines show $\omega_c$ in the two cases.}
\label{fig-App-EMsim_YIG_YIG}
\end{figure}

We carried out electromagnetic simulations with a commercial software (CST Microwave Studio) to evaluate the scattering parameters of the coupled system composed by CPW resonator and YIG film. The superconductor was modelled as a perfect electric conductor, whilst the magnetic film was included on the lower face of the GGG substrate \cite{ConnellyIEEE21} with a thickness $d=5~\mu$m.

Finite-element simulations were carried out by assuming that the precession of the magnetization in the ferrite can be described by the gyrotropic model, in which the permeability is modelled as a nonsymmetric Polder tensor with characteristic frequency dependence \cite{GuerevichMelkov}. The magnetic dispersion of YIG was defined by directly introducing the Larmor frequency, $\omega_H/2 \pi=\gamma\mu_0 H_0$, and the gyrotropic frequency, $\omega_M/2 \pi=\gamma\mu_0 M_s$, as input parameters of the simulator, with $\gamma$=28.02~GHz/T, $\mu_0=4 \pi \times 10^{-7}$~H/m and $\mu_0 M_s=0.245$~T. 

The simulation of frequency spectra was repeated for increasing values of the external magnetic field ($\mathbf{H_0}=H_0 \hat{x}$) to obtain the the spectral maps shown in Fig.~\ref{fig-App-EMsim_YIG_YIG}. The evolution of the coupled YIG-resonator modes displays the appearance of polaritonic branches. The simulations carried out with the YIG film lifted of 10~$\mu$m (panel (a)) or in contact (panel (b)) with the CPW resonator, essentially reproduce the experimental trend shown in Fig.~\ref{fig-coupling}(b,c), showing a good correspondence with the measured splittings of the polaritonic branches. In agreement with the measured spectra, the spectral maps in Fig.~\ref{fig-App-EMsim_YIG_YIG} show that in both cases the lower polariton branch converge towards $\omega_{FMR}(H_0)$ at low frequency. The increase of the spltting from panel (a) to (b) determines a shift of the anticrossing towards lower magnetic field, which is compatible with the frequency shift $\Delta$ introduced in Sect.~\ref{sect-hamiltonian}.

For the sake of completeness, we mention that finite element simulations fail to mimic the polariton lineshapes shown in \cite{supplementary}. A more detailed model is probably required to reproduce broadening effects, this however goes beyond the scope of our work.

\section{Effects of YIG films with different thickness}
\label{Appendix-film_thickness}

\begin{figure}[ht]
\centering
\includegraphics[width=\linewidth]{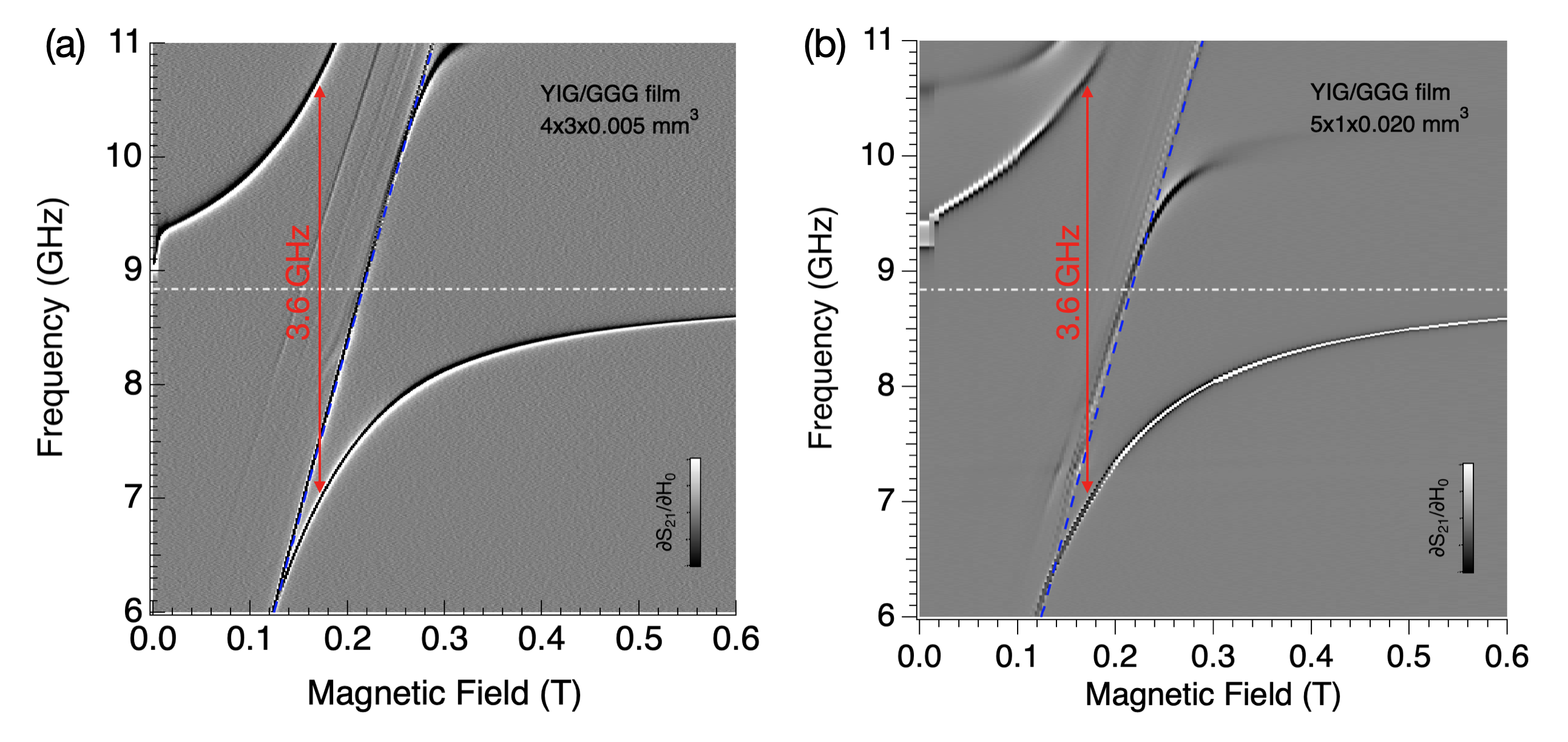}
\caption{Comparison between data obtained with YIG/GGG films of different thickness ($T=30$~K): (a) $5~\mu$m, (b) $20~\mu$m. The red arrows indicate the splitting $2 \lambda /2\pi \approx 3.6$~GHz. The blue dashed lines display $\omega_{FMR}(H_0)$. Both measurements were carried out by means of the same YBCO CPW resonator, horizontal dash-dot lines show the frequency of the fundamental mode in the two cases.}
\label{fig_app_film_thickness}
\end{figure}

Figure~\ref{fig_app_film_thickness} shows a direct comparison between transmission data taken with the same CPW resonator and films of different thickness and size. In these experiments, the sample was held on the resonator with a copper spring, this resulted in a lower coupling with respect to data reported in the main body of the article. Panel~(a) shows the spectral map taken with the YIG/GGG film having a thickness of $5~\mu$m and an area of $\approx 4 \times 3~\mathrm{mm}^2$. Panel~(b) shows data acquired on a YIG/GGG film having a thickness of $20~\mu$m and area of $\approx 5 \times 1~\mathrm{mm}^2$. The splitting indicated by the blue arrow, corresponding to $2 \lambda /2\pi \approx 3.6$~GHz, is comparable in the two cases.

\section{Multimode fit}
\label{Appendix-multimode_fit}

\begin{figure}[htb]
    \centering
    \includegraphics[width=0.8\linewidth]{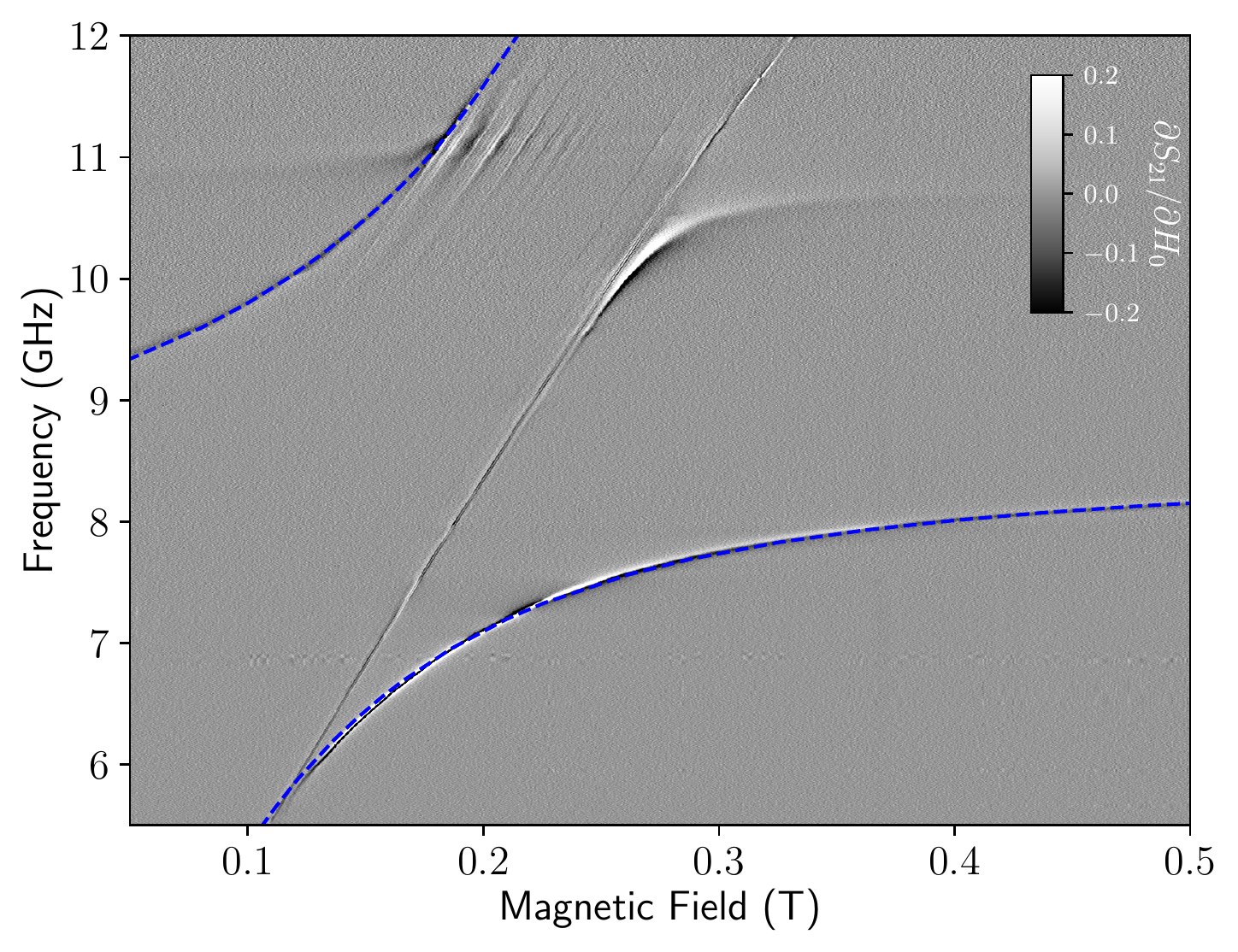}
    \caption{Best fit of the transmission spectrum obtained in case $\#$C with the YIG film pressed on top of the superconducting YBCO CPW by considering three magnonic modes instead of only one, are $\omega_c /2 \pi = 8.64$ GHz, $\Delta_0 /2 \pi = 0.23$ GHz, $\Delta_1 /2 \pi = 1.92$ GHz, $\Delta_2 /2 \pi = 3.17$ GHz, $\lambda_0  /2 \pi= 0.192$ GHz, $\lambda_1 /2 \pi = 1.93$ GHz, $\lambda_2 /2 \pi = 0.2$ GHz, $\alpha /\sqrt{2 \pi} = 0.035 \ \text{GHz}^{\frac{3}{2}}$. Among the three modes, only one is ultrastrongly coupled with the cavity field. Again, the factor $\alpha$ of the diamagnetic term is very small.}
    \label{fig:fit3modes}
\end{figure}

Despite the fitting analysis presented in the main body of the article involving one magnonic mode is statistically significant on its own, we extend the model by introducing more magnonic modes. In addition to the results shown in the main text, we present here the fit results considering three magnonic modes (Fig.~\ref{fig:fit3modes}). The method can be straightforwardly extended from the case of single mode, and the parameters obtained from the fit are $\omega_c /2 \pi= 8.64$ GHz, $\Delta_0/2 \pi = 0.23$ GHz, $\Delta_1/2 \pi = 1.92$ GHz, $\Delta_2/2 \pi = 3.17$ GHz, $\lambda_0/2 \pi = 0.192$ GHz, $\lambda_1/2 \pi = 1.93$ GHz, $\lambda_2/2 \pi = 0.2$ GHz, $\alpha/\sqrt{2 \pi} = 0.035 \ \text{GHz}^{\frac{3}{2}}$. It is worth noting that, among the three modes, only one results to be ultrastrongly coupled with the field of the resonator, corroborating the validity of considering only one magnonic mode. The obtained diamagnetic factor $\alpha$ is slightly larger than the single mode case. 

\section{On the diamagnetic term}
\label{Appendix-diamagnetic_term}

\begin{figure}[htb]
    \centering
    \includegraphics[width=\linewidth]{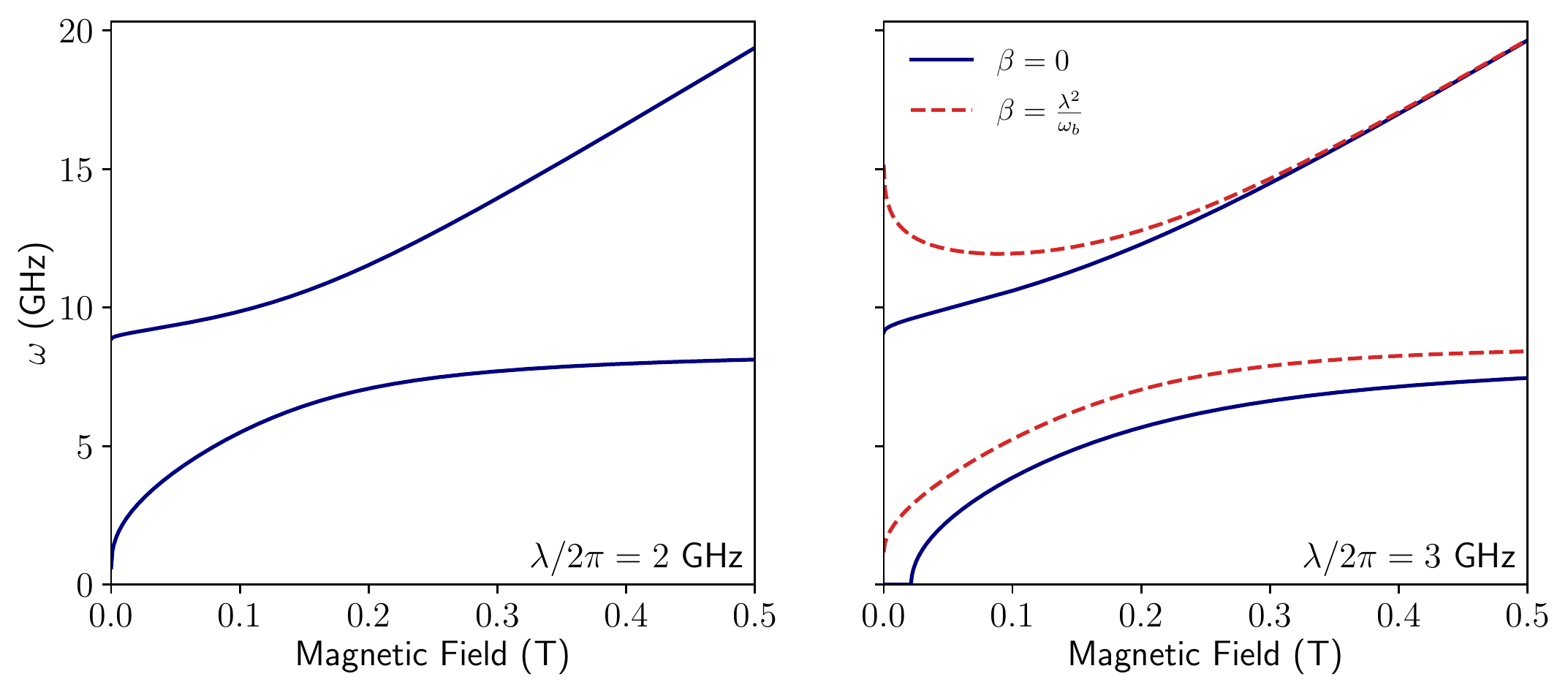}
    \caption{Left: polariton eigenfrequencies considering only one magnonic mode and using the same parameters of the fit in the main text, that are: $\omega_c/2 \pi = 8.65$ GHz, $\Delta/2 \pi = 2.05$ GHz, $\lambda/2 \pi = 2.002$ GHz, $\alpha/\sqrt{2 \pi} = 3 \times 10^{-3} \ \text{GHz}^{\frac{3}{2}}$. Here the fitted term of the diamagnetic term is relatively small. However, the coupling $\lambda$ is still too small to show the superradiant phase transition. Right: Polariton eigenfrequencies with the same parameters as before, except for $\lambda / 2 \pi = 3$ GHz, which is large enough to achieve the superradiant phase transition. The continuous blue lines correspond to the case of $\beta = 0$ (no diamagnetic term), clearly showing a critical point in the region of small magnetic fields. The red dashed lines correspond to $\beta = \lambda^2 / 
    \omega_b$ which avoids the superradiant phase transition.}
    \label{fig: supp - diamagnetic term}
\end{figure}

It is useful to compare the obtained diamagnetic terms with the {\em standard} one $\beta_{\rm std} = \lambda^2 / \omega_b$, which comes from the minimal coupling replacement \cite{nataf2010nogo, garziano2020gauge} for electric dipolar interactions. Notice that the expression for $\beta_{\rm std}$ is fixed by gauge-invariance requirements \cite{garziano2020gauge}. This constraint, preventing superradiance phase transitions, does not hold in the presence of magnetic interactions \cite{nataf2010nogo, mazza2019superradiant, andolina2020condensation, zueco2021condensation}. We calculate the ratio $\mathcal{B} \equiv \beta / \beta_{\rm std} = \alpha \sqrt{\omega_b} / \lambda^2$, which is $\mathcal{B} \approx 0.002$ and $\mathcal{B} \approx 0.027$ for the single mode and three modes cases, respectively. These results pave the way for the transition to a superradiant phase \cite{nataf2010nogo}, as the obtained values are extremely low. To clarify the role of the diamagnetic term, Fig.~\ref{fig: supp - diamagnetic term} shows the polariton eigenfrequencies in three different conditions. The left panel shows the eigenfrequencies using the same parameters of the fit obtained in the main text, that are: $\omega_c/2 \pi = 8.65$ GHz, $\Delta/2 \pi = 2.05$ GHz, $\lambda/2 \pi = 2.002$ GHz, $\alpha/\sqrt{2 \pi} = 3 \times 10^{-3} \ \text{GHz}^{\frac{3}{2}}$. The coupling strength is still too small to achieve a superradiant phase transition, but a slightly higher value would be enough to achieve it. Indeed, the continuous blue lines of the right panel show the polariton eigenfrequencies in the absence of the diamagnetic term and a larger coupling ($\beta = 0$ and $\lambda / 2 \pi = 3$ GHz). With these parameters, we can see a critical point in the region of small magnetic fields. On the other hand, as shown by the red dashed lines, the superradiant phase transition is forbidden by including the standard diamagnetic term of the Hopfield model ($\beta = \lambda^2 / \omega_b$).

\end{appendix}

%

\clearpage
\onecolumngrid

\section{Supplemental material}

\renewcommand\thefigure{S\arabic{figure}} 
\setcounter{figure}{0}

\label{supplementary}

\subsection{Experimental details}

\begin{figure}[ht]
\centering
\includegraphics[width=0.8\textwidth]{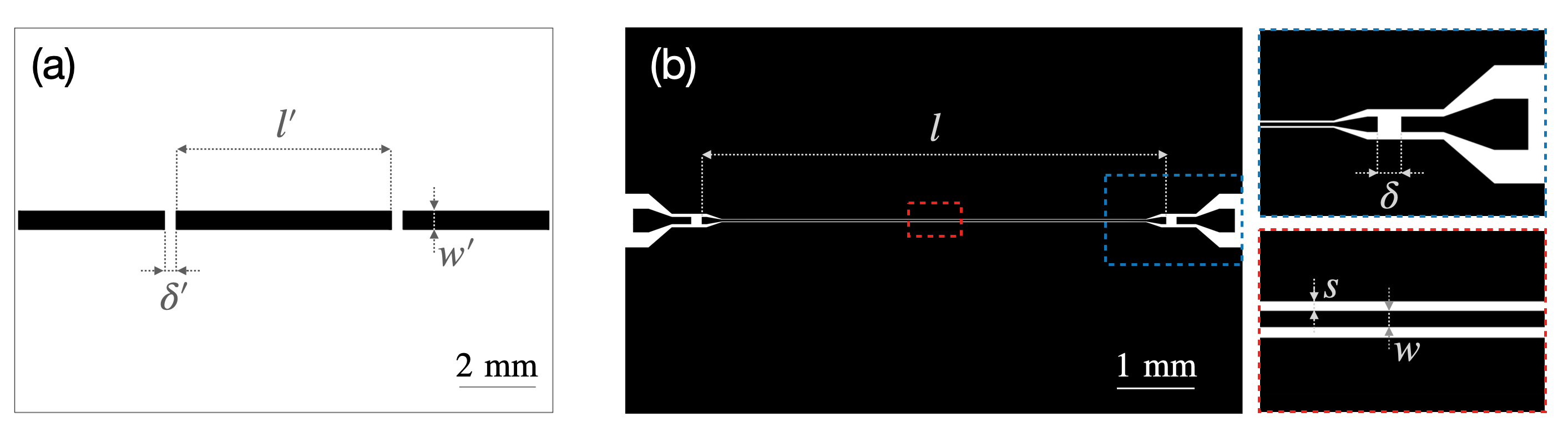}
\caption{Schematic representation of (a) microstrip resonator and (b) CPW resonator. In (a) the regions covered with Ag are black while the uncovered alumina substrate is white. In (b) YBCO regions are black while sapphire is white. The images on the right show blow-ups of launcher and central line of the CPW resonator as indicated by the different colors.}
\label{fig_suppl_mask}
\end{figure}

\begin{figure}[ht]
\centering
\includegraphics[width=0.8\textwidth]{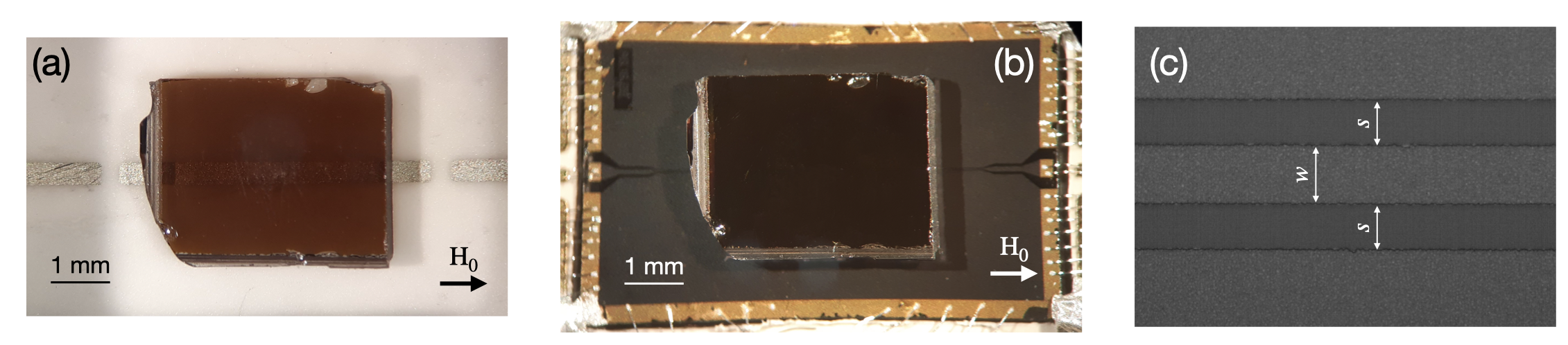}
\caption{Photographs of (a) Ag/alumina microstrip resonator and (b) YBCO/sapphire CPW transmission line, both with the YIG/GGG film positioned on top. (c) Blow-up of the central region of the CPW line. The lateral widths are $w=(17 \pm 1)~\mu$m and $s=(14 \pm 1)~\mu$m.}
\label{fig_suppl_exp}
\end{figure}

Microstrip lines were fabricated by optical lithography and wet etching of Ag films (thickness $\approx 3~\mu \mathrm{m}$), which were thermally evaporated onto alumina $14 \times 10 \times 0.63~\mathrm{mm}^3$ substrates \cite{GhirriAdvQTechnol20}. The central strip is $w^{\prime} =500~\mu$m wide, while two $\delta^{\prime} = 300~\mu$m coupling gaps were made to define the corresponding resonator (Fig.~\ref{fig_suppl_mask}(a)). Superconducting coplanar waveguide (CPW) broadband lines and resonators (Fig.~\ref{fig_suppl_mask}(b)) were fabricated from commercial  YBa$_2$Cu$_3$O$_7$ (YBCO) films (thickness 330~nm) deposited on a sapphire substrate and diced into $ 8 \times 5 \times 0.43~\mathrm{mm^3}$ blocks. Etching was carried out by Ar plasma in a reactive ion etching (RIE) chamber. The central conductor has typical width $w=(17 \pm 1)~\mu$m and is separated by $s=(14 \pm 1)~\mu$m from the lateral ground planes (Fig.~\ref{fig_suppl_exp}(c)). In the CPW resonator, the coupling gaps have width $\delta=140~\mu$m. 

We studied $5~\mu$m thick yttrium iron garnet (YIG) films which were grown by liquid-phase epitaxy on a gadolinium gallium garnet (GGG) substrate with (111) crystallographic orientation (Matesy GmbH). Reflection ($S_{11}$) and transmission ($S_{21}$) spectra were acquired in the 0.1-18~GHz range by means of a Vector Network Analyzer (VNA). The experiments were carried out in the 10-50~K temperature ($T$) range. Planar transmission lines and resonators were installed in a cryomagnetic set-up (Quantum Design PPMS), having variable temperature control and external magnetic field up to 7~T, by means of a cryogenic insert wired with a pair of silver-plated stainless steel coaxial cables connecting the low temperature stage to the external circuitry. The microstrip transmission line and resonator (Fig.~\ref{fig_suppl_exp}(a)) were mounted in a brass box and connected to the coaxial line by means of two metallic pins positioned on the microstrip. Coplanar waveguide (CPW) transmission line and resonator  (Fig.~\ref{fig_suppl_exp}(b)) were installed in a copper box and wire bonded (Al wire) to a printed-circuit board. Gold (thickness 200~nm) pads were patterned on the edges of the YBCO resonator to facilitate the bonding.

\begin{figure*}[ht]
\centering
\includegraphics[width=0.8\textwidth]{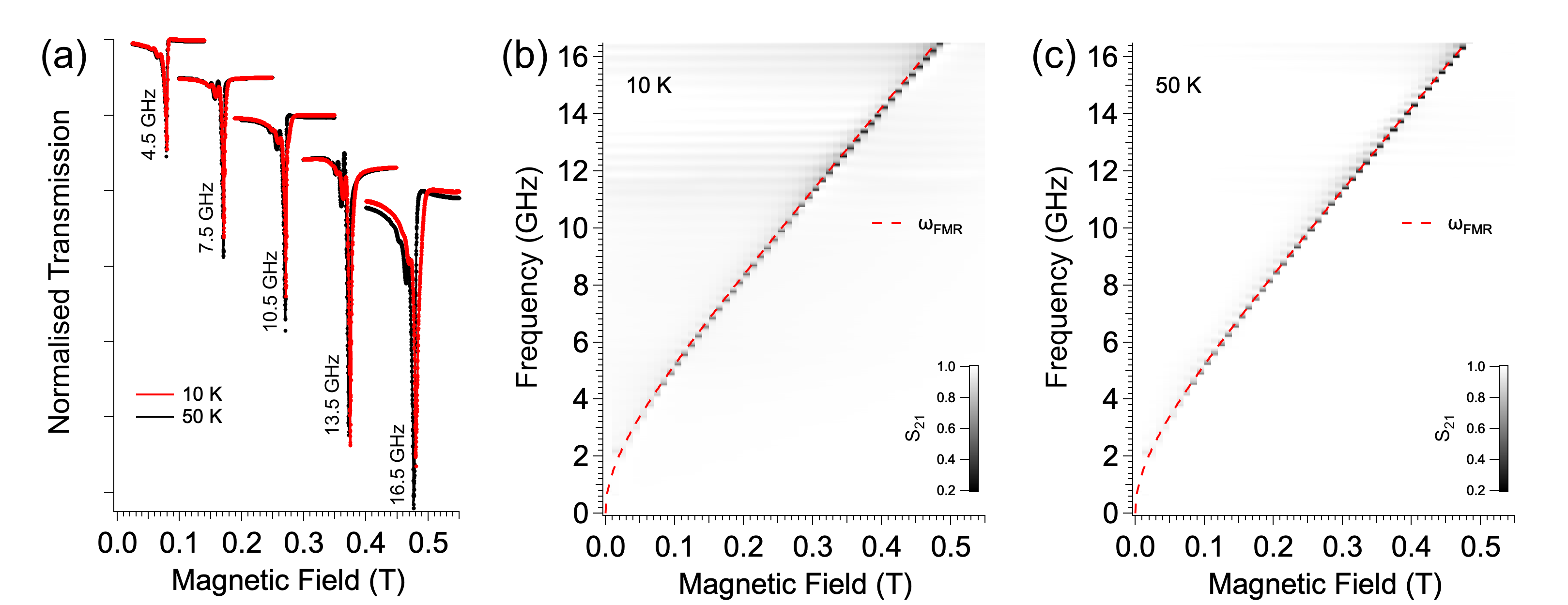}
\caption{Transmission spectra of the YIG film acquired by the microstrip broadband line at different temperatures. (a) Comparison between fixed-frequency field-swept spectra taken at 10~K and 50~K. (b,c) Spectral maps measured by sweeping the frequency at progressively increasing magnetic field. The temperature is (b) 10 K and (c) 50 K. }
\label{fig_suppl_temp}
\end{figure*}

In all the measurements, the incident power supplied to the planar transmission line is $P_{inc}\approx -8$~dBm. Spectra taken with $P_{inc}$ in the range between -28~dBm and 2~dBm showed no variations with the microwave power. The number of photons can be calculated as \cite{SageJAP11}
\begin{eqnarray}
n=\frac{P_{inc}Q_L 10^{-IL/20}}{\pi h f_0^2}, 
\label{nphotons}
\end{eqnarray}
where $f_0=\omega_0/2 \pi$ is the fundamental mode frequency, $Q_L$ is the loaded quality factor, $IL$ is the insertion loss and $h$ is the Planck constant. Depending on the resonator's parameters (see below), the estimated number of photons results $n_{mic} \approx 7 \times 10^{8}$ for the microstrip resonator and $n_{CPW} \approx 3 \times 10^{10}$ for the CPW resonators. In all the experiments the number of photons resulted much smaller than the estimated number of spins.

\subsection{Additional transmission spectroscopy data}

The YIG film has been initially characterized by means of the broadband microstrip line, which allowed the acquisition of ferromagnetic resonance spectra at different excitation frequencies in the 4.5-16.5~GHz range (Fig.~\ref{fig_suppl_temp}). Fixed-frequency field-swept transmission spectra taken at 10 and 50~K show similar features below $\approx 13.5$~GHz while little differences in absorption intensity and line position can be observed for higher frequencies (Fig.~\ref{fig_suppl_temp}(a)). Spectra taken with the microstrip line are characterized by maximum absorption amplitude in correspondence to the Kittel mode, while additional higher modes have lower amplitude. The measured spectral maps  (Fig.~\ref{fig_suppl_temp}(b,c)) show that the dependence on the lowest mode can be fitted with the Kittel formula (Eq.~1 in the main text) by considering the saturation magnetization $\mu_0 M_s=0.245$~T both for 10 and 50~K.

\begin{figure}[ht]
\centering
\includegraphics[width=0.7\linewidth]{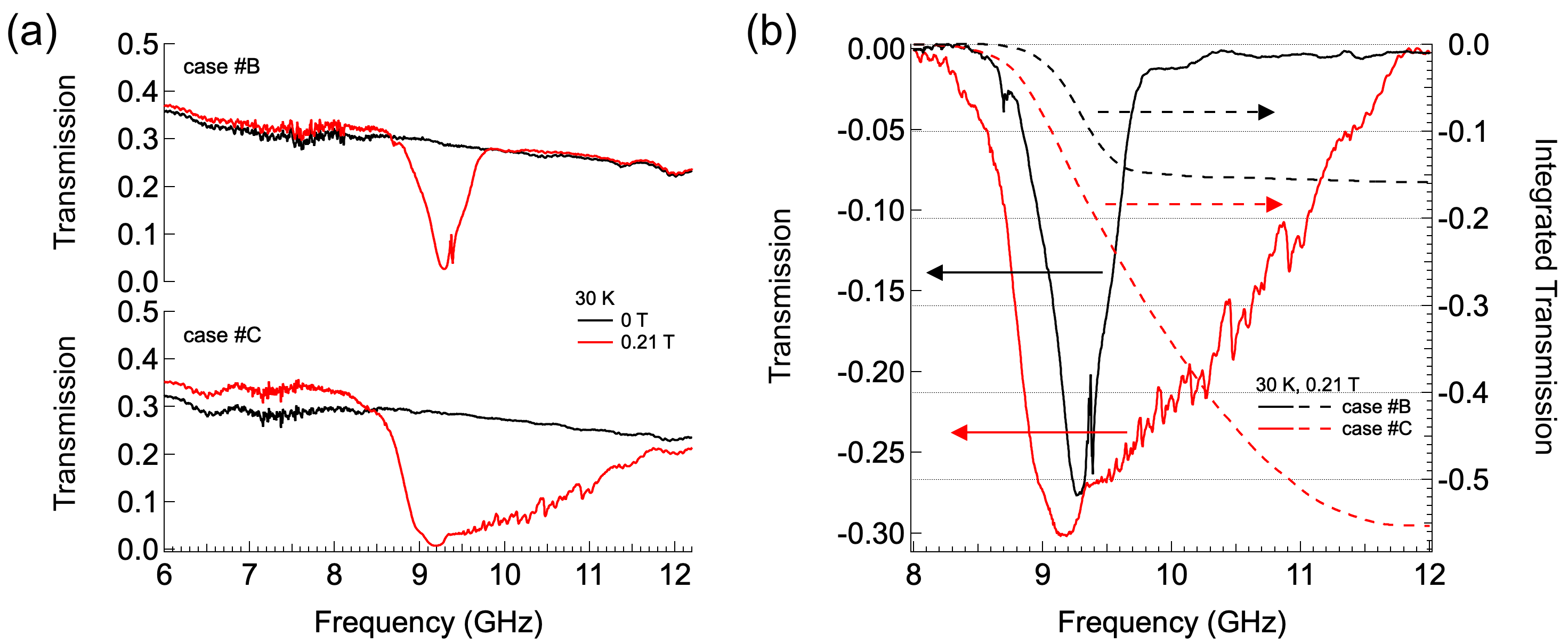}
\caption{Transmission spectra of the YIG film acquired by the CPW broadband line. (a) Comparison between frequency spectra taken for cases $\#B$ and $\#C$. (b) Direct comparison between the spectra  at 0.21~T after background subtraction. Dashed lines show the integral calculated between 8 and 12~GHz.}
\label{fig_suppl_intensity}
\end{figure}

Fig.~\ref{fig_suppl_intensity}(a) shows broadband transmission-vs-frequency spectra acquired by YBCO CPW lines (case $\#B$ and $\#C$, as defined in the main text) for zero and 0.21~T applied magnetic field. In panel (b) the spectra taken at 0.21~T are directly compared to evidence the increased depth and width of the $S_{21}$ spectrum when YIG and YBCO CPW line are put in direct contact. To quantify this point, we calculate the total absorption related to the transitions in Fig.~3(b) and (c) in the main text, which can be obtained by integrating $S_{21}(\omega)$ with respect to the frequency. For $H_0=0.21$~T, the ratio of the integrals calculated between 8 and 12~GHz (Fig.~\ref{fig_suppl_intensity}) gives $I_C/I_B= 3.49$. The total number of magnon excitations in this frequency range thus results much larger when the YIG film is positioned in direct contact with the YBCO resonator.

\begin{figure*}[ht]
\centering
\includegraphics[width=0.97\textwidth]{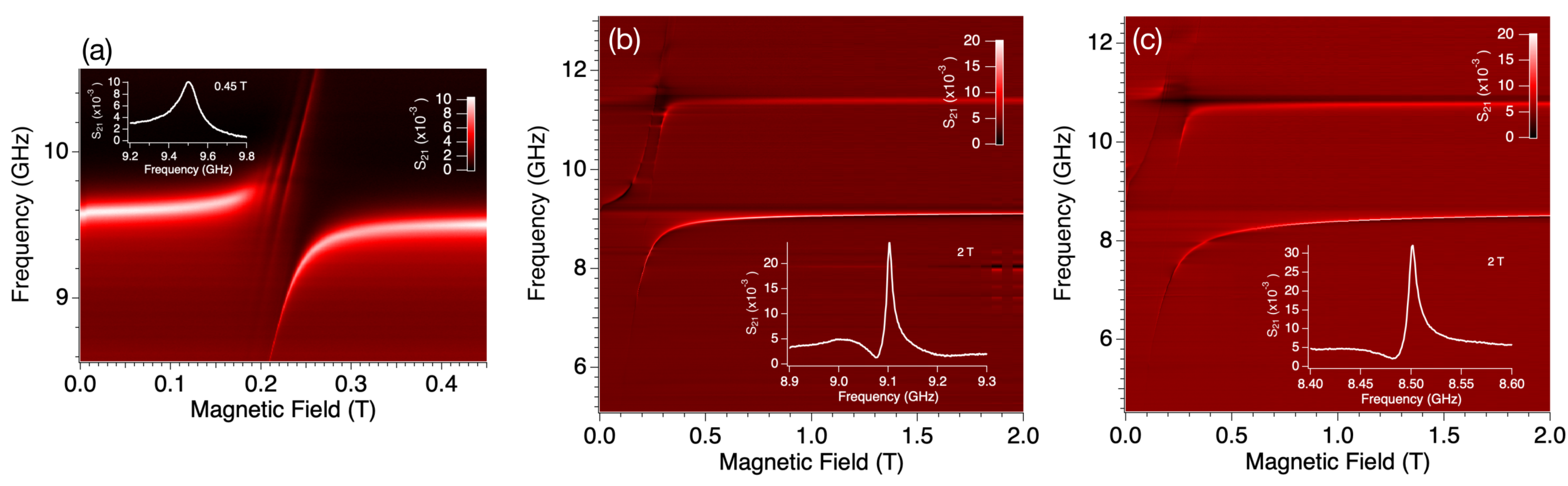}
\caption{Transmission spectral maps obtained with the same YIG film and different resonator. The background transmission curve has been subtracted from the data. (a) Metallic microstrip resonator ($T=50$~K, case $\#$A). Inset: transmission-vs-frequency spectrum taken at $H_0=0.45$~T. (b) Superconducting CPW resonator ($T=30$~K, case $\#$B). (c) CPW resonator with the YIG film pressed against the superconductor ($T=30$~K, case $\#$C). In (b) and (c) the insets show the transmission-vs-frequency spectrum taken at $H_0=2$~T. }
\label{fig_suppl_maps}
\end{figure*}
\begin{figure}[ht]
\centering
\includegraphics[width=0.35\textwidth]{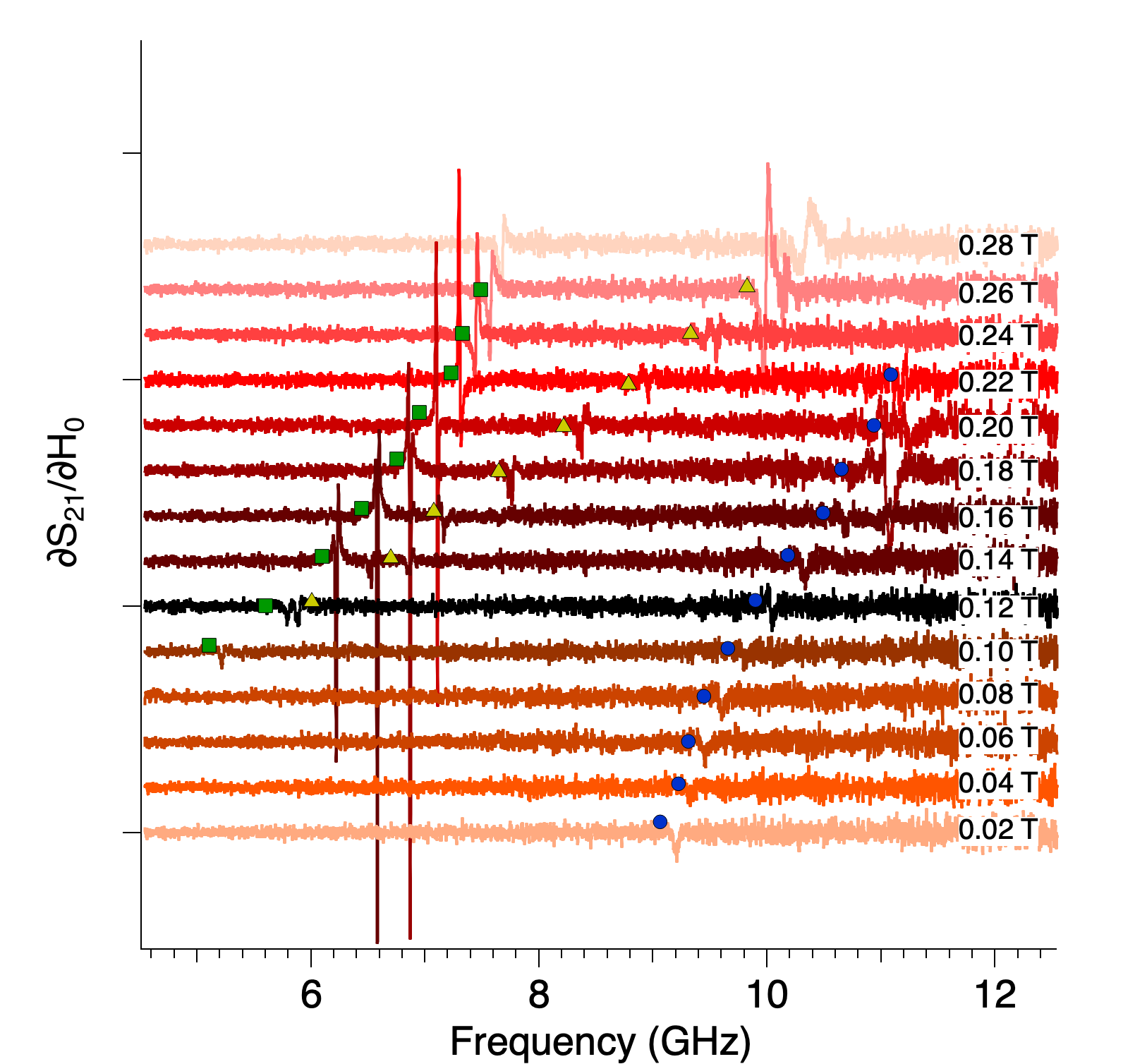}
\caption{Evolution of the derivative spectra  (case $\#$C) plotted as a function of frequency for fixed applied magnetic fields. Green squares and blue circles indicate respectively the position of the lower and upper polaritonic branches, yellow triangles shows the evolution of the FMR mode.}
\label{fig_suppl_spectra}
\end{figure}

Figure~\ref{fig_suppl_maps} shows $S_{21}(\omega, H_0)$  transmission spectral maps acquired with microstrip and CPW resonators corresponding to the same dataset reported in Fig.~4 in the main text. Since resonator and YIG are already coupled in zero magnetic field, we exploited the resilience of YBCO in high magnetic field \cite{GhirriAPL15} to characterize the uncoupled CPW resonators. From transmission-vs-frequency spectra acquired at $H_0=2$~T,  we obtain the resonator frequencies $\omega_{0, B}/2 \pi=9.1$~GHz and $\omega_{0, C}/2 \pi=8.5$~GHz, respectively for case $\#$B and $\#$C (Fig.~\ref{fig_suppl_maps}(b) and (c)). The discrepancy between the latter and $\omega_c/2\pi=8.65$~GHz obtained from the fit (Fig.~5 in the main text), indicates that at 2~T resonator and magnon modes are still weakly coupled, due to their large collective coupling strength. Since CPW resonators are fabricated from the same lithographic mask, the differences between $\omega_{0, B}$ and $\omega_{0, C}$ originate from specific experimental details, including sample installation in diverse ways. The insertion loss is $IL_B=32$~dB and $IL_C=30$~dB. In both cases, the loaded quality factor is $Q_L\approx 1000$, which is dominated by losses introduced by the GGG substrate (permittivity $\epsilon_r=11.99$ and $\tan{\delta}=5.2 \times 10^{-3}$ at room temperature \cite{ConnellyIEEE21}). The corresponding decay rate of the resonator ($\kappa_0=\omega_0/Q_L$) amounts to $\kappa_{0,B} /(2 \pi) = 9$~MHz and $\kappa_{0,C} /(2 \pi) =8$~MHz. For comparison, the microstrip resonator (case $\#$A) displays $\omega_{0, A}/2\pi=9.5$~GHz,  $IL_A=42$~dB and $Q_L\approx 120$ (Inset in Fig.~\ref{fig_suppl_maps}(a)), resulting in the decay rate $\kappa_{0,A} /(2 \pi) = 79$~MHz.

\begin{figure}[ht]
\centering
\includegraphics[width=0.8\textwidth]{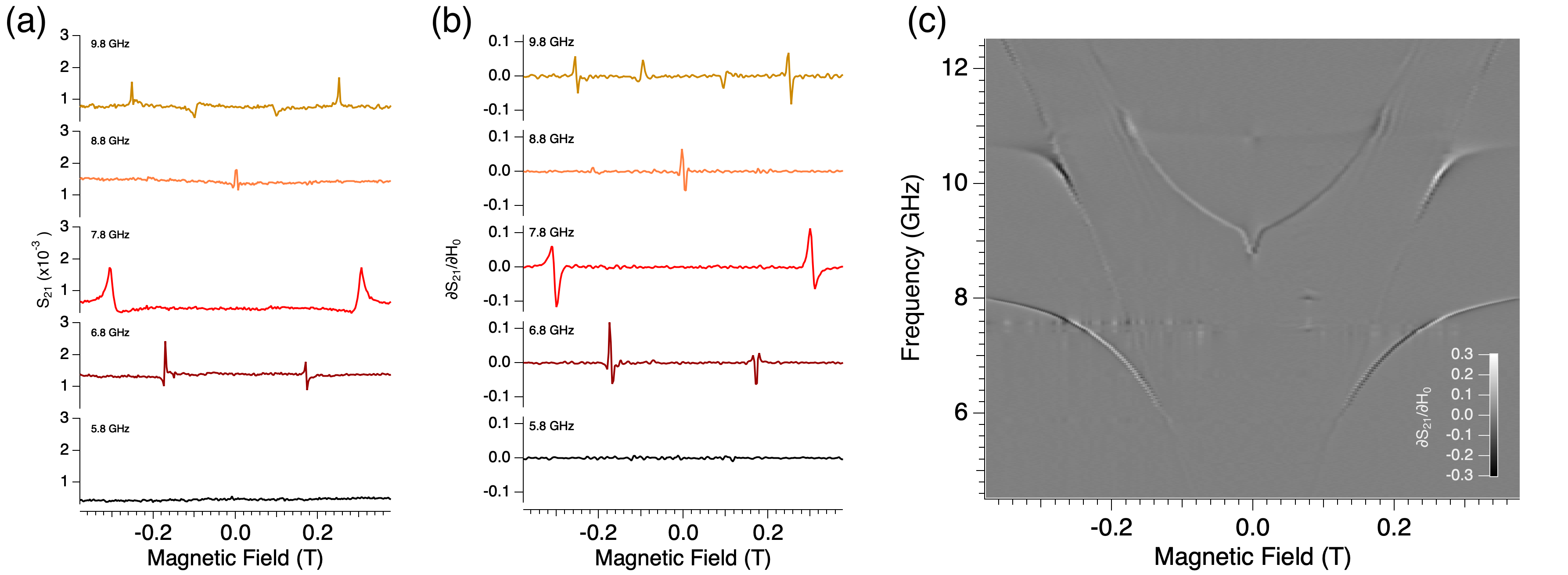}
\caption{Transmission spectral data taken with the YIG film in contact with the CPW resonator. $S_{21}$ (a) and $\partial S_{21}/\partial H_0$ (b)  spectra plotted as a function of the magnetic field for different frequencies, as indicated. (c) Spectral map acquired for $H_0$ spanning between 0.38 to -0.38~T.}
\label{fig_suppl_fullH}
\end{figure}

Figure~\ref{fig_suppl_spectra} shows the evolution of derivative spectra corresponding to the dataset in Fig.~4(c) of the main text and plotted as a function of the frequency for fixed fields. The lower polaritonic branch typically shows narrower and more marked derivative peaks with respect to the upper one. This behavior is evident also in the spectra plotted in Fig.~\ref{fig_suppl_fullH}(a,b) as a function of the magnetic field. Transmission $S_{21}$ spectra taken at positive and negative field shows a quite similar behavior (panel (c)). 

\begin{figure*}[ht]
\centering
\includegraphics[width=0.9\textwidth]{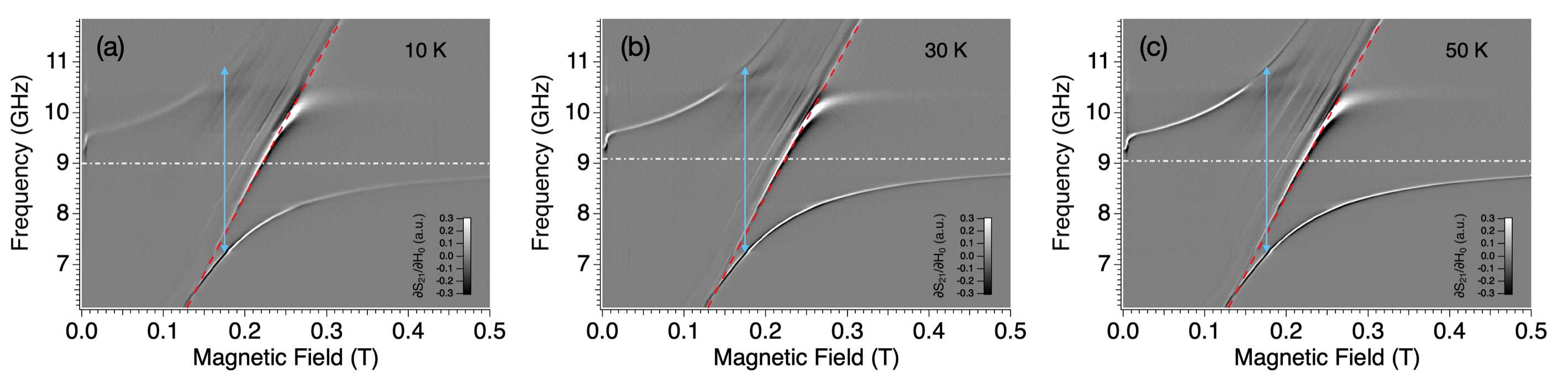}
\caption{Comparison between derivative maps taken at different temperatures, as indicated. The YIG/GGG film had an area of $4 \times 2~\mathrm{mm}^2$ and a thickness of $5~\mu$m. The red dashed lines show the Kittel equation ($\mu_0 M_s=0.245$~T) while the white dash-dot lines indicate the frequency of the resonator determined at $H_0=2$~T. Cyan arrows have the same length and indicate the splitting of the polaritonic branches.}
\label{fig_suppl_maps_temp}
\end{figure*}

The coupling between YIG film and superconducting CPW resonator was studied by means of spectral maps acquired at $T=10,~30$ and 50~K (Fig.~\ref{fig_suppl_maps_temp}). In this range of temperatures, the coupling strength remains essentially constant, while it is possible to observe a broadening of the polaritonic modes for decreasing temperatures. Since the quality factor of the YBCO resonator typically increases from 50 to 10~K \cite{GhirriAPL15}, this behavior can be ascribed to the effect of the GGG substrate, which is known to exhibit a paramagnetic behavior below 70~K that is reported to increase the damping of magnetization precession in the YIG film grown on its surface \cite{DanilovSovPhysJ89,MihalceanuPRB18, KosenAPLMater19}. We just mention here that narrow linewidths, lower than those measured at 300~K, were reported for temperatures as low as mK with  substrate-free YIG films obtained by detaching the GGG in different manners \cite{KosenAPLMater19, TremplerAPL20}.

\begin{figure*}[ht]
\centering
\includegraphics[width=0.35\textwidth]{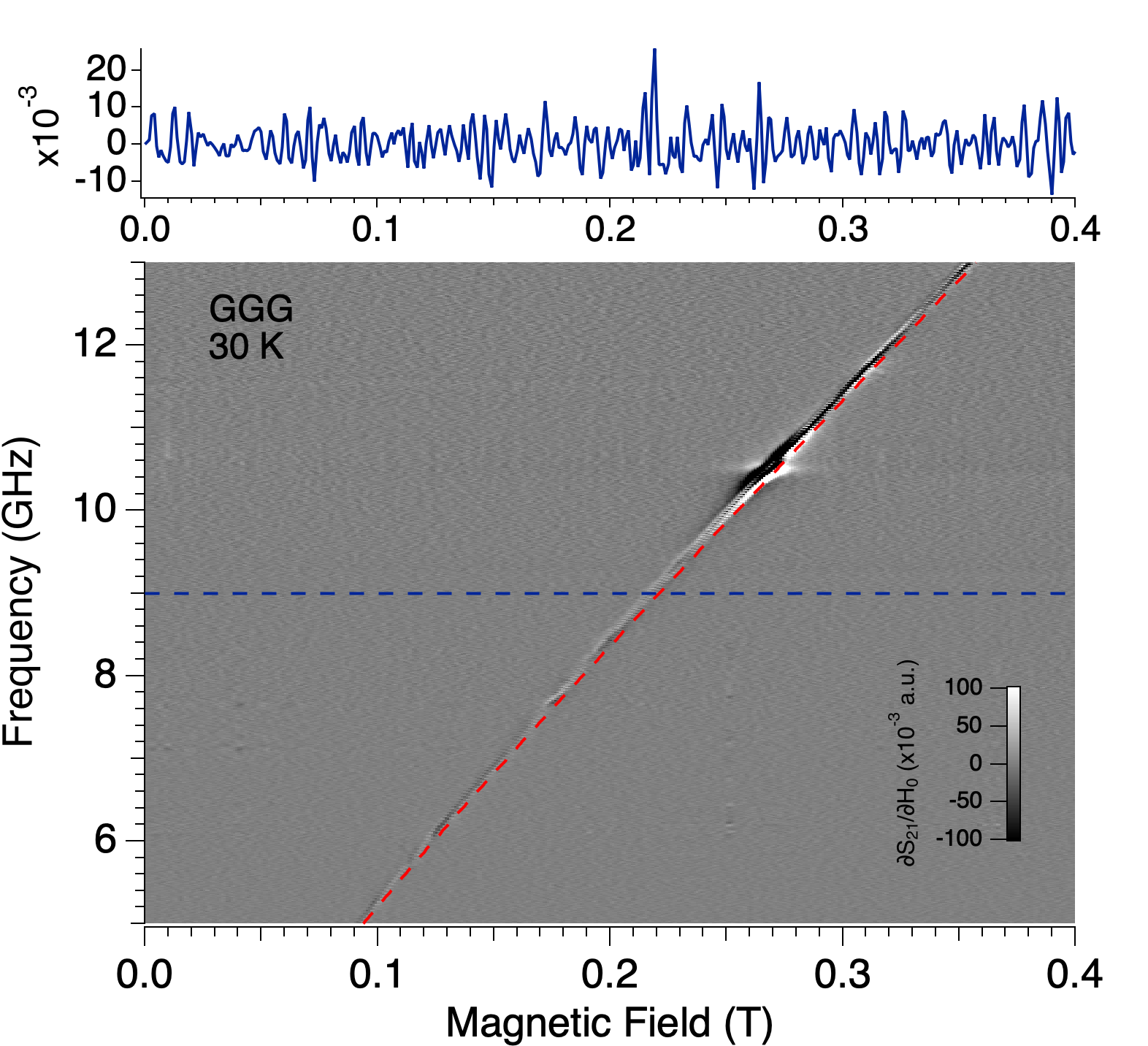}
\caption{Transmission spectral map measured with the GGG substrate (volume $4 \times 2 \times 0.5~\mathrm{mm}^3$) positioned in contact with the CPW resonator. The red dashed line shows the Kittel equation calculated with $\mu_0 M_s=0.245$~T. The upper panel shows the $\partial S_{21}/\partial H_0$-vs-$H_0$ spectrum taken at 9~GHz.}
\label{fig_suppl_GGG}
\end{figure*}

Additional effects reported for YIG films grown on paramagnetic GGG are the deviation of the resonance field from that of a free thin YIG plate \cite{DanilovSovPhysJ89} and the coupling between the magnetizations of ferrimagnetic YIG and paramagnetic GGG \cite{WangPRB20}. Fig.~\ref{fig_suppl_GGG} shows the transmission map of a YIG/GGG film measured with the GGG side in contact with the CPW resonator. In contrast with Fig.~\ref{fig_suppl_maps_temp}(b), only a weak absorption line is observed in this case. From such spectra, as well as from data in Figs.~\ref{fig_suppl_temp} and~\ref{fig_suppl_GGG} or in Fig.~3 of the main text, the presence of additional coupled modes~\cite{WangPRB20} was not observed. This likely follows from the different conditions in our experiment, particularly lower microwave power and narrower (and superconducting) CPW line with respect to Ref.~\cite{WangPRB20}. Thus, the magnetization coupling between YIG and GGG cannot be inferred from our data. However, the presence of the paramagnetic GGG substrate probably influences the value of the saturation magnetization 
$\mu_0 M_s=0.245$~T entering in the Kittel equation (Eq. 1 in the main text), that well reproduces the field dependence of the FMR line for temperatures between 50 and 10~K (Fig.~\ref{fig_suppl_temp}). We stress that little changes in the choice of $\mu_0 M_s$ by no means affect the interpretation of the results reported in our work.

\subsection{Additional finite-element electromagnetic simulations}

\begin{figure}[ht]
\centering
\includegraphics[width=0.7\textwidth]{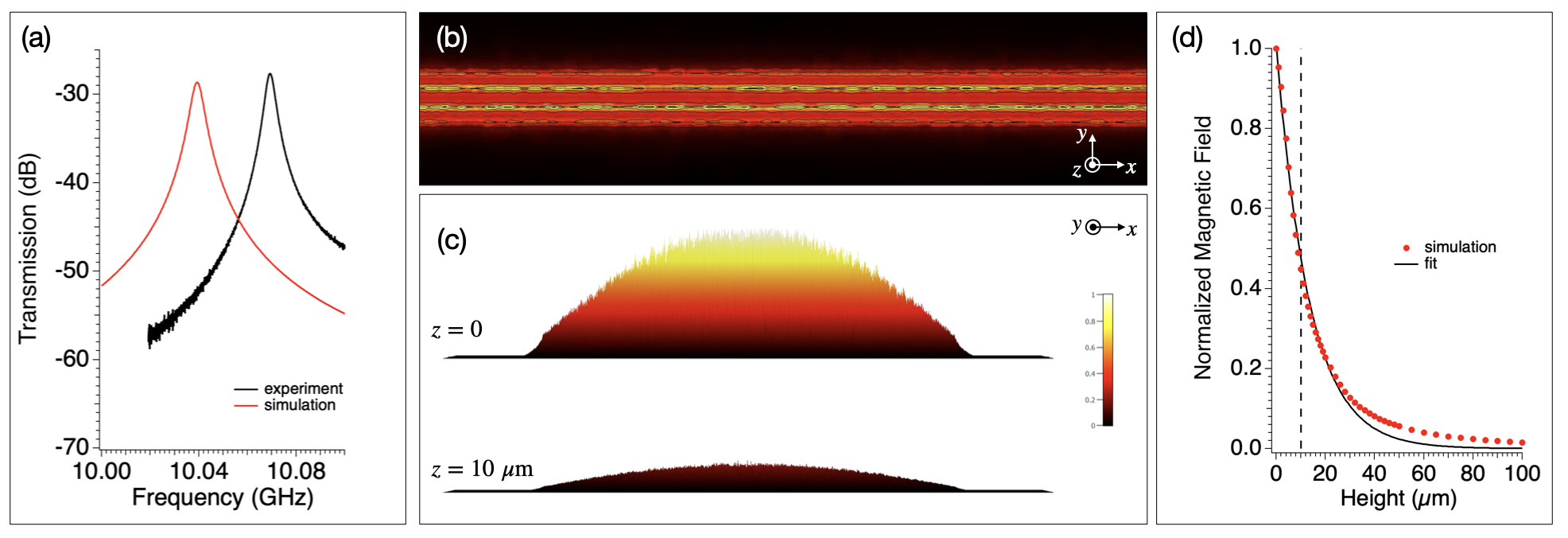}
\caption{Electromagnetic simulation of the bare CPW resonator. (a) Comparison between experimental ($T=30$~K) and simulated spectra. (b) Spectral map showing the normalized value of $h_{ac}$ around the central conductor. (c) $x$-dependence of the $y$-averaged value of $h_{ac}$ calculated for $z$ heights of 0 and $10~\mu$m. (d) Calculated dependence of the microwave magnetic field along the $z$-axis. The solid line show the fit with Eq.~\ref{hac_z}.}
\label{fig_suppl_sim_bare}
\end{figure}

Preliminary simulations of the electromagnetic field profile with commercial codes (CST Microwave Studio) have been used as a guide to evaluate and the distribution of the microwave field and to optimize the effective volume of interaction. The CPW resonator was modelled with realistic geometry, dimensions and materials parameters while the superconducting film was approximated as a perfect electric conductor (PEC). 

We initially calculated the transmission spectrum of the bare resonator . To obtain a substantial correspondence with the measured spectrum (Fig.~\ref{fig_suppl_sim_bare}(a)), we modelled the sapphire substrate (volume $8 \times 5 \times 0.43~\mathrm{mm}^3$) with the permittivity $\epsilon_r=10.74$, while losses where included by forcing $\tan \delta$ to values higher than those expected for sapphire. The frequency of the fundamental mode results in good agreement with the value obtained from $\omega_{c,bare}/2\pi=c/2\sqrt{\epsilon_{eff}} l$ being $l=6.02$~mm the length of the central strip (Fig.~\ref{fig_suppl_mask}(b)) and $\epsilon_{eff}=5.77$ the effective permittivity calculated for the CPW line.

The simulated distribution of the magnetic component of the fundamental mode of the resonator ($h_{ac}$) shows that the field is localized along $y$ in a region of approximate width $w+2s=45~\mu\mathrm{m}$ around the central conductor (panel (b)). The value of $h_{ac}$ shows a pronounced decay with increasing $z$ distance from the upper surface of the resonator ($z=0$), as it can be seen in the profiles calculated for $z=0$ and $10~\mu$m and plotted as a function of $x$ (Fig.~\ref{fig_suppl_sim_bare}(c)). To quantify this trend, we calculated the average of $\mathbf{h_{ac}}$ in regions centered in the middle of the CPW resonator (area $4~\mathrm{mm} \times 45~\mu \mathrm{m}$) with progressively increasing $z$ height. The obtained points ($\Bar{h}_{ac}(z)$) show a quasi exponential dependence from $z$ (Fig.~\ref{fig_suppl_sim_bare}(d)) that roughly follows 
\begin{eqnarray}
    \Bar{h}_{ac}(z)= h_{max}\exp(-z /\eta)
    \label{hac_z}
\end{eqnarray}
being $h_{max}=\Bar{h}_{ac}(z=0)$ and $\eta=13.5~\mu$m. Since $\Bar{h}_{ac}(10~\mathrm{\mu m})/h_{max}=0.45$ and by assuming that the normalized spin-photon coupling  $g_s(z)/g_s(z=0)$ scales as $\Bar{h}_{ac}(z)/h_{max}$, we obtain the ratio $g_s(z=0)/g_s(\mathrm{z=10~\mu m}) = 2.2$. This value is in good agreement with the ratio of the anticrossing splittings in configurations $\#$C and $\#$B derived from Fig.~4 in the main text.

The absolute value of $h_{max}$ obtained from the simulation can be used to estimate the amplitude of the zero-point vacuum fluctuation, $b_{vac}$. Being $b_{max}=\mu_0 h_{max}=26$~mT the value obtained for $P_{inc}=0.5$~W, we obtain $b_{vac}=b_{max}\sqrt{P_{vac}/P_{inc}}\approx3$~nT by considering the threshold power $P_{vac} \approx 5$~fW for single-photon operation (Eq.~\ref{nphotons}). On the other hand, the vacuum magnetic fluctuation can be calculated from \cite{TosiAIPAdv14}
\begin{eqnarray}
   b_{vac}\approx \frac{\mu_0 \omega_c}{4 w} \sqrt{\frac{h}{Z_0}}
   \label{bvac}
\end{eqnarray}
with $h=6.626 \times 10^{-34}$~J~s and $Z_0= 58~\mathrm{\Omega}$. From Eq.~\ref{bvac} we get $b_{vac} \approx 3$~nT in excellent agreement with the value resulting from finite-element simulations.

As a next step we modeled the effect of a GGG substrate having volume of $4\times 3 \times 0.5~\mathrm{mm}^3$ (Fig.~\ref{fig_suppl_sim_bare}). By using as permittivity $\epsilon_r=11.99$ and dielectric loss tangent $\tan{\delta}=5.2 \times 10^{-3}$ for GGG \cite{ConnellyIEEE21}, the simulated fundamental mode frequency $\omega_{c,sim}/2\pi=8.62$~GHz results in good agreement with $\omega_c/2\pi=8.65$~GHz obtained from the fit of the experimental data (Fig.~5 in the main text). With respect to the bare resonator, the presence of the GGG substrate determines the shift of the fundamental mode towards lower frequency and the broadening of resonant peaks. The latter gives rise to a higher decay rate $\kappa_c$. The distribution of $\mathbf{h_{ac}}=h_x \hat{x}+h_y \hat{y}+h_z \hat{z}$ simulated with the GGG substrate above the resonator is in line with that of the bare resonator (Fig.~2 in the main text): the magnetic antinode is located in the middle of the resonator, the microwave field is localized around the central conductor (panel (c)) and $h_x$ is negligible as expected for a quasi-transverse electromagnetic (TEM) mode (panel (d)).

\begin{figure}[ht]
\centering
\includegraphics[width=0.9\textwidth]{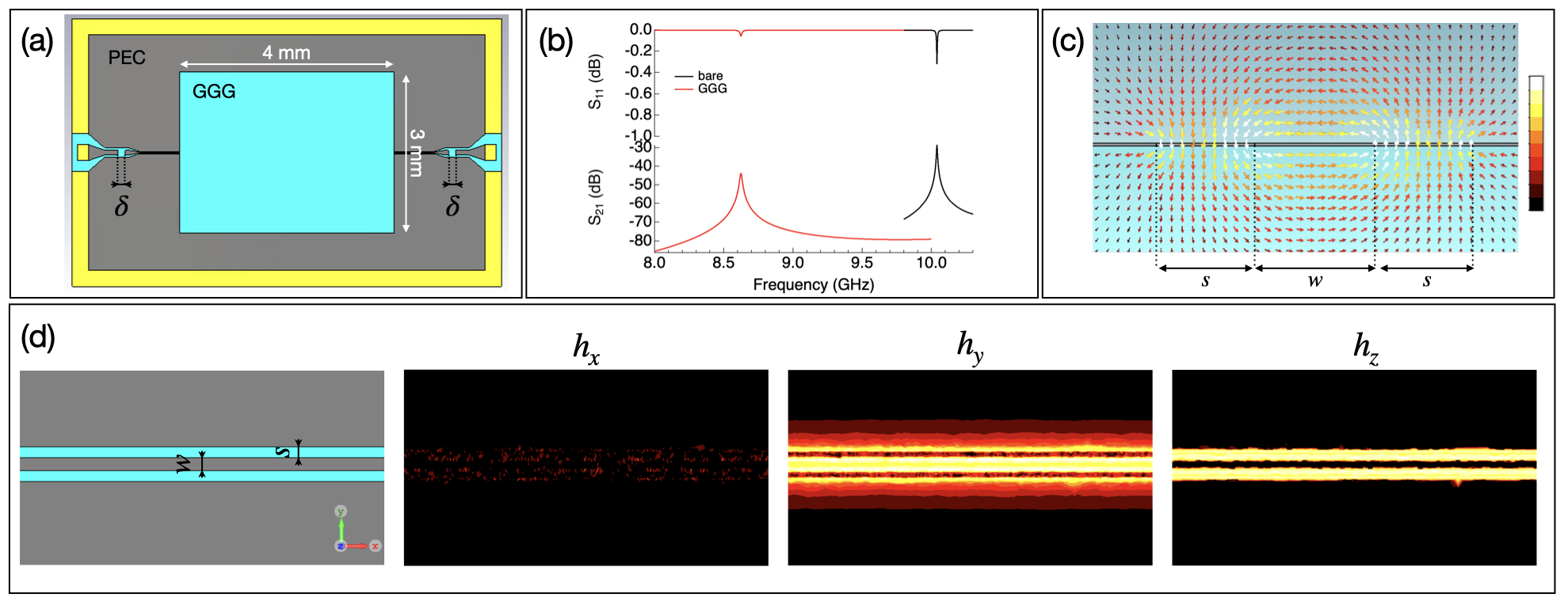}
\caption{Electromagnetic simulations of the CPW resonator with the GGG substrate positioned on top. (a) Representation of the model used for the simulations. Colors indicate different materials: PEC, grey; Au, yellow; sapphire and GGG, cyan. (b) Simulated reflection $S_{11}$ and transmission $S_{21}$ spectra showing the comparison between bare resonator and GGG cases. (c) Vertical section of the simulated $\mathbf{h_{ac}}$ field in the middle of the resonator (phase $\phi=0$). (d) Blow-up of the CPW line showing the simulated root-mean square cartesian components of $\mathbf{h_{ac}}$. Letters in (a,c,d) correspond to $\delta=140~\mu$m, $w=17~\mu$m and $s=14~\mu$m.}
\label{fig_suppl_sim_GGG}
\end{figure}

\subsection{Modelling of spin systems in a cavity}

Magnetic excitations, i.e.~spin waves in magnetically ordered materials, can be described by considering the precession of the macroscopic magnetic vector $\textbf{M}$ around the effective local magnetic field $\mathbf{H_{eff}}$ through the Landau-Lifshizt-Gilbert (LLG) equation, that also accounts for damping by a phenomenological term characterised by the Gilbert parameter. Exchange and dipolar interactions between different local values of $\textbf{M}$ can be included in the LLG equation as effective local fields,  $\mathbf{H_{exc}}$ and $\mathbf{H_{dip}}$ respectively, that contribute to $\mathbf{H_{eff}}$ along with the externally applied static field $\mathbf{H_0}$ \cite{GuerevichMelkov}. Both exchange and dipolar interactions are naturally included in a microscopic description of a regular lattice of interacting spins, the relative spectrum of excitation shows spin wave modes having characteristic dispersion between energy ($\hbar\omega$) and propagation wavevector ($k$) \cite{RameshtiPhysRep22}. For an isotropic infinite lattice, the low lying states assume parabolic dispersion law $\omega \approx k^2$ with a characteristic stationary ($k$=0) ground state (Kittel mode). Yet, for a real specimen with a specific shape, the finite size and the boundary conditions, as well as the way the antenna excites them, can make the spectrum of magnetic excitations much richer with the appearance of characteristic bands.

The collective coupling $\lambda$ between magnetic and microwave modes is the key parameter that needs to be compared, in the first instance, with the damping of both the magnetic system (i.e. the dissipation rate $\kappa_m$) and the photon losses ($\kappa_c$) of the cavity. Since $\lambda$ determines the rate at which the two systems exchange energy and information at resonance, when $\lambda\geq \kappa_m, \kappa_c$ we have coherent dynamics of the two systems, and the so-called \textit{strong coupling regime}. The dipolar coupling of single spins to the oscillating magnetic component of radiation $\mathbf{b_{vac}}$ has the relative coupling strength $g_s=\frac{1}{4}\gamma b_{vac}$
which turns out to be very weak, typically on the order of 0.1 to 50 Hz for $b_{vac} \sim$~nT. 
For spin ensembles, either paramagnets or magnetically ordered systems, the collective coupling strength $\lambda=g_s\sqrt{ N_s}$ is enhanced by a factor scaling as $\sqrt{N_s}$ \cite{ImamogluPRL2009}, being $N_s$ the number of spin transitions in the ensemble interacting with the electromagnetic radiation. For instance, with $N_s \sim 10^{12}-10^{14}$ we can get $\lambda=g_s\sqrt{N_s}=1$ to 10 MHz. For YIG, $N_s=2 s_{\mathrm{Fe}} N$, where $N$ is the number of spins and $s_{\mathrm{Fe}}=5/2$ is the ground state spin of Fe$^{3+}$. In a YIG sphere, $\kappa_m$ is of the order of MHz, thus for experiments with cavities having high quality factor, that is small $\kappa_c$, the strong coupling regime is achieved (see references in the main text).


\begin{thebibliography}{71}%
\makeatletter
\providecommand \@ifxundefined [1]{%
 \@ifx{#1\undefined}
}%
\providecommand \@ifnum [1]{%
 \ifnum #1\expandafter \@firstoftwo
 \else \expandafter \@secondoftwo
 \fi
}%
\providecommand \@ifx [1]{%
 \ifx #1\expandafter \@firstoftwo
 \else \expandafter \@secondoftwo
 \fi
}%
\providecommand \natexlab [1]{#1}%
\providecommand \enquote  [1]{``#1''}%
\providecommand \bibnamefont  [1]{#1}%
\providecommand \bibfnamefont [1]{#1}%
\providecommand \citenamefont [1]{#1}%
\providecommand \href@noop [0]{\@secondoftwo}%
\providecommand \href [0]{\begingroup \@sanitize@url \@href}%
\providecommand \@href[1]{\@@startlink{#1}\@@href}%
\providecommand \@@href[1]{\endgroup#1\@@endlink}%
\providecommand \@sanitize@url [0]{\catcode `\\12\catcode `\$12\catcode
  `\&12\catcode `\#12\catcode `\^12\catcode `\_12\catcode `\%12\relax}%
\providecommand \@@startlink[1]{}%
\providecommand \@@endlink[0]{}%
\providecommand \url  [0]{\begingroup\@sanitize@url \@url }%
\providecommand \@url [1]{\endgroup\@href {#1}{\urlprefix }}%
\providecommand \urlprefix  [0]{URL }%
\providecommand \Eprint [0]{\href }%
\providecommand \doibase [0]{https://doi.org/}%
\providecommand \selectlanguage [0]{\@gobble}%
\providecommand \bibinfo  [0]{\@secondoftwo}%
\providecommand \bibfield  [0]{\@secondoftwo}%
\providecommand \translation [1]{[#1]}%
\providecommand \BibitemOpen [0]{}%
\providecommand \bibitemStop [0]{}%
\providecommand \bibitemNoStop [0]{.\EOS\space}%
\providecommand \EOS [0]{\spacefactor3000\relax}%
\providecommand \BibitemShut  [1]{\csname bibitem#1\endcsname}%
\let\auto@bib@innerbib\@empty
\bibitem [{\citenamefont {Pirro}\ \emph {et~al.}(2021)\citenamefont {Pirro},
  \citenamefont {Vasyuchka}, \citenamefont {Serga},\ and\ \citenamefont
  {Hillebrands}}]{PirroNatRevMater21}%
  \BibitemOpen
  \bibfield  {author} {\bibinfo {author} {\bibfnamefont {P.}~\bibnamefont
  {Pirro}}, \bibinfo {author} {\bibfnamefont {V.~I.}\ \bibnamefont
  {Vasyuchka}}, \bibinfo {author} {\bibfnamefont {A.~A.}\ \bibnamefont
  {Serga}},\ and\ \bibinfo {author} {\bibfnamefont {B.}~\bibnamefont
  {Hillebrands}},\ }\bibfield  {title} {\bibinfo {title} {Advances in coherent
  magnonics},\ }\href {https://doi.org/10.1038/s41578-021-00332-w} {\bibfield
  {journal} {\bibinfo  {journal} {Nature Reviews Materials}\ }\textbf {\bibinfo
  {volume} {6}},\ \bibinfo {pages} {1114} (\bibinfo {year} {2021})}\BibitemShut
  {NoStop}%
\bibitem [{\citenamefont {Lachance-Quirion}\ \emph {et~al.}(2019)\citenamefont
  {Lachance-Quirion}, \citenamefont {Tabuchi}, \citenamefont {Gloppe},
  \citenamefont {Usami},\ and\ \citenamefont
  {Nakamura}}]{Lachance-QuirionApplPhysExpr19}%
  \BibitemOpen
  \bibfield  {author} {\bibinfo {author} {\bibfnamefont {D.}~\bibnamefont
  {Lachance-Quirion}}, \bibinfo {author} {\bibfnamefont {Y.}~\bibnamefont
  {Tabuchi}}, \bibinfo {author} {\bibfnamefont {A.}~\bibnamefont {Gloppe}},
  \bibinfo {author} {\bibfnamefont {K.}~\bibnamefont {Usami}},\ and\ \bibinfo
  {author} {\bibfnamefont {Y.}~\bibnamefont {Nakamura}},\ }\bibfield  {title}
  {\bibinfo {title} {Hybrid quantum systems based on magnonics},\ }\href
  {https://doi.org/10.7567/1882-0786/ab248d} {\bibfield  {journal} {\bibinfo
  {journal} {Applied Physics Express}\ }\textbf {\bibinfo {volume} {12}},\
  \bibinfo {pages} {070101} (\bibinfo {year} {2019})}\BibitemShut {NoStop}%
\bibitem [{\citenamefont {{Zare Rameshti}}\ \emph {et~al.}(2022)\citenamefont
  {{Zare Rameshti}}, \citenamefont {{Viola Kusminskiy}}, \citenamefont {Haigh},
  \citenamefont {Usami}, \citenamefont {Lachance-Quirion}, \citenamefont
  {Nakamura}, \citenamefont {Hu}, \citenamefont {Tang}, \citenamefont {Bauer},\
  and\ \citenamefont {Blanter}}]{RameshtiPhysRep22}%
  \BibitemOpen
  \bibfield  {author} {\bibinfo {author} {\bibfnamefont {B.}~\bibnamefont
  {{Zare Rameshti}}}, \bibinfo {author} {\bibfnamefont {S.}~\bibnamefont
  {{Viola Kusminskiy}}}, \bibinfo {author} {\bibfnamefont {J.~A.}\ \bibnamefont
  {Haigh}}, \bibinfo {author} {\bibfnamefont {K.}~\bibnamefont {Usami}},
  \bibinfo {author} {\bibfnamefont {D.}~\bibnamefont {Lachance-Quirion}},
  \bibinfo {author} {\bibfnamefont {Y.}~\bibnamefont {Nakamura}}, \bibinfo
  {author} {\bibfnamefont {C.-M.}\ \bibnamefont {Hu}}, \bibinfo {author}
  {\bibfnamefont {H.~X.}\ \bibnamefont {Tang}}, \bibinfo {author}
  {\bibfnamefont {G.~E.}\ \bibnamefont {Bauer}},\ and\ \bibinfo {author}
  {\bibfnamefont {Y.~M.}\ \bibnamefont {Blanter}},\ }\bibfield  {title}
  {\bibinfo {title} {Cavity magnonics},\ }\href
  {https://www.sciencedirect.com/science/article/pii/S0370157322002460}
  {\bibfield  {journal} {\bibinfo  {journal} {Physics Reports}\ }\textbf
  {\bibinfo {volume} {979}},\ \bibinfo {pages} {1} (\bibinfo {year}
  {2022})}\BibitemShut {NoStop}%
\bibitem [{\citenamefont {Li}\ \emph {et~al.}(2020)\citenamefont {Li},
  \citenamefont {Zhang}, \citenamefont {Tyberkevych}, \citenamefont {Kwok},
  \citenamefont {Hoffmann},\ and\ \citenamefont {Novosad}}]{HybMagnonics}%
  \BibitemOpen
  \bibfield  {author} {\bibinfo {author} {\bibfnamefont {Y.}~\bibnamefont
  {Li}}, \bibinfo {author} {\bibfnamefont {W.}~\bibnamefont {Zhang}}, \bibinfo
  {author} {\bibfnamefont {V.}~\bibnamefont {Tyberkevych}}, \bibinfo {author}
  {\bibfnamefont {W.-K.}\ \bibnamefont {Kwok}}, \bibinfo {author}
  {\bibfnamefont {A.}~\bibnamefont {Hoffmann}},\ and\ \bibinfo {author}
  {\bibfnamefont {V.}~\bibnamefont {Novosad}},\ }\bibfield  {title} {\bibinfo
  {title} {Hybrid magnonics: Physics, circuits, and applications for coherent
  information processing},\ }\href {https://doi.org/10.1063/5.0020277}
  {\bibfield  {journal} {\bibinfo  {journal} {Journal of Applied Physics}\
  }\textbf {\bibinfo {volume} {128}},\ \bibinfo {pages} {130902} (\bibinfo
  {year} {2020})}\BibitemShut {NoStop}%
\bibitem [{\citenamefont {Zhang}\ \emph {et~al.}(2015)\citenamefont {Zhang},
  \citenamefont {Zou}, \citenamefont {Zhu}, \citenamefont {Marquardt},
  \citenamefont {Jiang},\ and\ \citenamefont {Tang}}]{ZhangNatCommun15}%
  \BibitemOpen
  \bibfield  {author} {\bibinfo {author} {\bibfnamefont {X.}~\bibnamefont
  {Zhang}}, \bibinfo {author} {\bibfnamefont {C.-L.}\ \bibnamefont {Zou}},
  \bibinfo {author} {\bibfnamefont {N.}~\bibnamefont {Zhu}}, \bibinfo {author}
  {\bibfnamefont {F.}~\bibnamefont {Marquardt}}, \bibinfo {author}
  {\bibfnamefont {L.}~\bibnamefont {Jiang}},\ and\ \bibinfo {author}
  {\bibfnamefont {H.~X.}\ \bibnamefont {Tang}},\ }\bibfield  {title} {\bibinfo
  {title} {Magnon dark modes and gradient memory},\ }\href
  {https://doi.org/10.1038/ncomms9914} {\bibfield  {journal} {\bibinfo
  {journal} {Nature Communications}\ }\textbf {\bibinfo {volume} {6}},\
  \bibinfo {pages} {8914} (\bibinfo {year} {2015})}\BibitemShut {NoStop}%
\bibitem [{\citenamefont {Bai}\ \emph {et~al.}(2017)\citenamefont {Bai},
  \citenamefont {Harder}, \citenamefont {Hyde}, \citenamefont {Zhang},
  \citenamefont {Hu}, \citenamefont {Chen},\ and\ \citenamefont
  {Xiao}}]{BaiPRL17}%
  \BibitemOpen
  \bibfield  {author} {\bibinfo {author} {\bibfnamefont {L.}~\bibnamefont
  {Bai}}, \bibinfo {author} {\bibfnamefont {M.}~\bibnamefont {Harder}},
  \bibinfo {author} {\bibfnamefont {P.}~\bibnamefont {Hyde}}, \bibinfo {author}
  {\bibfnamefont {Z.}~\bibnamefont {Zhang}}, \bibinfo {author} {\bibfnamefont
  {C.-M.}\ \bibnamefont {Hu}}, \bibinfo {author} {\bibfnamefont {Y.~P.}\
  \bibnamefont {Chen}},\ and\ \bibinfo {author} {\bibfnamefont {J.~Q.}\
  \bibnamefont {Xiao}},\ }\bibfield  {title} {\bibinfo {title} {Cavity mediated
  manipulation of distant spin currents using a cavity-magnon-polariton},\
  }\href {https://doi.org/10.1103/PhysRevLett.118.217201} {\bibfield  {journal}
  {\bibinfo  {journal} {Phys. Rev. Lett.}\ }\textbf {\bibinfo {volume} {118}},\
  \bibinfo {pages} {217201} (\bibinfo {year} {2017})}\BibitemShut {NoStop}%
\bibitem [{\citenamefont {Crescini}\ \emph {et~al.}(2018)\citenamefont
  {Crescini}, \citenamefont {Alesini}, \citenamefont {Braggio}, \citenamefont
  {Carugno}, \citenamefont {Di~Gioacchino}, \citenamefont {Gallo},
  \citenamefont {Gambardella}, \citenamefont {Gatti}, \citenamefont {Iannone},
  \citenamefont {Lamanna}, \citenamefont {Ligi}, \citenamefont {Lombardi},
  \citenamefont {Ortolan}, \citenamefont {Pagano}, \citenamefont {Pengo},
  \citenamefont {Ruoso}, \citenamefont {Speake},\ and\ \citenamefont
  {Taffarello}}]{CresciniEurPhysJC18}%
  \BibitemOpen
  \bibfield  {author} {\bibinfo {author} {\bibfnamefont {N.}~\bibnamefont
  {Crescini}}, \bibinfo {author} {\bibfnamefont {D.}~\bibnamefont {Alesini}},
  \bibinfo {author} {\bibfnamefont {C.}~\bibnamefont {Braggio}}, \bibinfo
  {author} {\bibfnamefont {G.}~\bibnamefont {Carugno}}, \bibinfo {author}
  {\bibfnamefont {D.}~\bibnamefont {Di~Gioacchino}}, \bibinfo {author}
  {\bibfnamefont {C.~S.}\ \bibnamefont {Gallo}}, \bibinfo {author}
  {\bibfnamefont {U.}~\bibnamefont {Gambardella}}, \bibinfo {author}
  {\bibfnamefont {C.}~\bibnamefont {Gatti}}, \bibinfo {author} {\bibfnamefont
  {G.}~\bibnamefont {Iannone}}, \bibinfo {author} {\bibfnamefont
  {G.}~\bibnamefont {Lamanna}}, \bibinfo {author} {\bibfnamefont
  {C.}~\bibnamefont {Ligi}}, \bibinfo {author} {\bibfnamefont {A.}~\bibnamefont
  {Lombardi}}, \bibinfo {author} {\bibfnamefont {A.}~\bibnamefont {Ortolan}},
  \bibinfo {author} {\bibfnamefont {S.}~\bibnamefont {Pagano}}, \bibinfo
  {author} {\bibfnamefont {R.}~\bibnamefont {Pengo}}, \bibinfo {author}
  {\bibfnamefont {G.}~\bibnamefont {Ruoso}}, \bibinfo {author} {\bibfnamefont
  {C.~C.}\ \bibnamefont {Speake}},\ and\ \bibinfo {author} {\bibfnamefont
  {L.}~\bibnamefont {Taffarello}},\ }\bibfield  {title} {\bibinfo {title}
  {Operation of a ferromagnetic axion haloscope at
  {\$}{\$}m{\_}a=58{$\backslash$},{$\backslash$}upmu {$\backslash$}mathrm
  {\{}ev{\}}{\$}{\$}},\ }\href {https://doi.org/10.1140/epjc/s10052-018-6163-8}
  {\bibfield  {journal} {\bibinfo  {journal} {The European Physical Journal C}\
  }\textbf {\bibinfo {volume} {78}},\ \bibinfo {pages} {703} (\bibinfo {year}
  {2018})}\BibitemShut {NoStop}%
\bibitem [{\citenamefont {Hisatomi}\ \emph {et~al.}(2016)\citenamefont
  {Hisatomi}, \citenamefont {Osada}, \citenamefont {Tabuchi}, \citenamefont
  {Ishikawa}, \citenamefont {Noguchi}, \citenamefont {Yamazaki}, \citenamefont
  {Usami},\ and\ \citenamefont {Nakamura}}]{HisatomiPRB16}%
  \BibitemOpen
  \bibfield  {author} {\bibinfo {author} {\bibfnamefont {R.}~\bibnamefont
  {Hisatomi}}, \bibinfo {author} {\bibfnamefont {A.}~\bibnamefont {Osada}},
  \bibinfo {author} {\bibfnamefont {Y.}~\bibnamefont {Tabuchi}}, \bibinfo
  {author} {\bibfnamefont {T.}~\bibnamefont {Ishikawa}}, \bibinfo {author}
  {\bibfnamefont {A.}~\bibnamefont {Noguchi}}, \bibinfo {author} {\bibfnamefont
  {R.}~\bibnamefont {Yamazaki}}, \bibinfo {author} {\bibfnamefont
  {K.}~\bibnamefont {Usami}},\ and\ \bibinfo {author} {\bibfnamefont
  {Y.}~\bibnamefont {Nakamura}},\ }\bibfield  {title} {\bibinfo {title}
  {Bidirectional conversion between microwave and light via ferromagnetic
  magnons},\ }\href {https://doi.org/10.1103/PhysRevB.93.174427} {\bibfield
  {journal} {\bibinfo  {journal} {Phys. Rev. B}\ }\textbf {\bibinfo {volume}
  {93}},\ \bibinfo {pages} {174427} (\bibinfo {year} {2016})}\BibitemShut
  {NoStop}%
\bibitem [{\citenamefont {Yao}\ \emph {et~al.}(2023)\citenamefont {Yao},
  \citenamefont {Gui}, \citenamefont {Rao}, \citenamefont {Zhang},
  \citenamefont {Lu},\ and\ \citenamefont {Hu}}]{YaoPRL23}%
  \BibitemOpen
  \bibfield  {author} {\bibinfo {author} {\bibfnamefont {B.}~\bibnamefont
  {Yao}}, \bibinfo {author} {\bibfnamefont {Y.~S.}\ \bibnamefont {Gui}},
  \bibinfo {author} {\bibfnamefont {J.~W.}\ \bibnamefont {Rao}}, \bibinfo
  {author} {\bibfnamefont {Y.~H.}\ \bibnamefont {Zhang}}, \bibinfo {author}
  {\bibfnamefont {W.}~\bibnamefont {Lu}},\ and\ \bibinfo {author}
  {\bibfnamefont {C.-M.}\ \bibnamefont {Hu}},\ }\bibfield  {title} {\bibinfo
  {title} {Coherent microwave emission of gain-driven polaritons},\ }\href
  {https://doi.org/10.1103/PhysRevLett.130.146702} {\bibfield  {journal}
  {\bibinfo  {journal} {Phys. Rev. Lett.}\ }\textbf {\bibinfo {volume} {130}},\
  \bibinfo {pages} {146702} (\bibinfo {year} {2023})}\BibitemShut {NoStop}%
\bibitem [{\citenamefont {Roberts}\ \emph {et~al.}(2004)\citenamefont
  {Roberts}, \citenamefont {Auld},\ and\ \citenamefont
  {Schell}}]{RobertsJAP62}%
  \BibitemOpen
  \bibfield  {author} {\bibinfo {author} {\bibfnamefont {R.~W.}\ \bibnamefont
  {Roberts}}, \bibinfo {author} {\bibfnamefont {B.~A.}\ \bibnamefont {Auld}},\
  and\ \bibinfo {author} {\bibfnamefont {R.~R.}\ \bibnamefont {Schell}},\
  }\bibfield  {title} {\bibinfo {title} {{Magnetodynamic Mode Ferrite
  Amplifier}},\ }\href {https://doi.org/10.1063/1.1728685} {\bibfield
  {journal} {\bibinfo  {journal} {Journal of Applied Physics}\ }\textbf
  {\bibinfo {volume} {33}},\ \bibinfo {pages} {1267} (\bibinfo {year}
  {2004})}\BibitemShut {NoStop}%
\bibitem [{\citenamefont {Auld}(1963)}]{Auld1963}%
  \BibitemOpen
  \bibfield  {author} {\bibinfo {author} {\bibfnamefont {B.~A.}\ \bibnamefont
  {Auld}},\ }\bibfield  {title} {\bibinfo {title} {Coupling of electromagnetic
  and magnetostatic modes in ferrite loaded cavity resonators},\ }\href
  {https://doi.org/10.1063/1.1702646} {\bibfield  {journal} {\bibinfo
  {journal} {Journal of Applied Physics}\ }\textbf {\bibinfo {volume} {34}},\
  \bibinfo {pages} {1629} (\bibinfo {year} {1963})}\BibitemShut {NoStop}%
\bibitem [{\citenamefont {Gurevich}\ and\ \citenamefont
  {Melkov}(1996)}]{GuerevichMelkov}%
  \BibitemOpen
  \bibfield  {author} {\bibinfo {author} {\bibfnamefont {A.}~\bibnamefont
  {Gurevich}}\ and\ \bibinfo {author} {\bibfnamefont {G.}~\bibnamefont
  {Melkov}},\ }\href@noop {} {\emph {\bibinfo {title} {Magnetization
  Oscillations and Waves (1st ed.)}}}\ (\bibinfo  {publisher} {CRC Press.},\
  \bibinfo {year} {1996})\BibitemShut {NoStop}%
\bibitem [{\citenamefont {Maier-Flaig}\ \emph {et~al.}(2017)\citenamefont
  {Maier-Flaig}, \citenamefont {Klingler}, \citenamefont {Dubs}, \citenamefont
  {Surzhenko}, \citenamefont {Gross}, \citenamefont {Weiler}, \citenamefont
  {Huebl},\ and\ \citenamefont {Goennenwein}}]{MaierFlaigPRB17}%
  \BibitemOpen
  \bibfield  {author} {\bibinfo {author} {\bibfnamefont {H.}~\bibnamefont
  {Maier-Flaig}}, \bibinfo {author} {\bibfnamefont {S.}~\bibnamefont
  {Klingler}}, \bibinfo {author} {\bibfnamefont {C.}~\bibnamefont {Dubs}},
  \bibinfo {author} {\bibfnamefont {O.}~\bibnamefont {Surzhenko}}, \bibinfo
  {author} {\bibfnamefont {R.}~\bibnamefont {Gross}}, \bibinfo {author}
  {\bibfnamefont {M.}~\bibnamefont {Weiler}}, \bibinfo {author} {\bibfnamefont
  {H.}~\bibnamefont {Huebl}},\ and\ \bibinfo {author} {\bibfnamefont
  {S.~T.~B.}\ \bibnamefont {Goennenwein}},\ }\bibfield  {title} {\bibinfo
  {title} {Temperature-dependent magnetic damping of yttrium iron garnet
  spheres},\ }\href {https://link.aps.org/doi/10.1103/PhysRevB.95.214423}
  {\bibfield  {journal} {\bibinfo  {journal} {Phys. Rev. B}\ }\textbf {\bibinfo
  {volume} {95}},\ \bibinfo {pages} {214423} (\bibinfo {year}
  {2017})}\BibitemShut {NoStop}%
\bibitem [{\citenamefont {Imamo\ifmmode~\breve{g}\else
  \u{g}\fi{}lu}(2009)}]{ImamogluPRL2009}%
  \BibitemOpen
  \bibfield  {author} {\bibinfo {author} {\bibfnamefont {A.}~\bibnamefont
  {Imamo\ifmmode~\breve{g}\else \u{g}\fi{}lu}},\ }\bibfield  {title} {\bibinfo
  {title} {Cavity qed based on collective magnetic dipole coupling: Spin
  ensembles as hybrid two-level systems},\ }\href
  {https://link.aps.org/doi/10.1103/PhysRevLett.102.083602} {\bibfield
  {journal} {\bibinfo  {journal} {Phys. Rev. Lett.}\ }\textbf {\bibinfo
  {volume} {102}},\ \bibinfo {pages} {083602} (\bibinfo {year}
  {2009})}\BibitemShut {NoStop}%
\bibitem [{\citenamefont {Soykal}\ and\ \citenamefont
  {Flatt\'e}(2010)}]{SoykalPRL2010}%
  \BibitemOpen
  \bibfield  {author} {\bibinfo {author} {\bibfnamefont {O.~O.}\ \bibnamefont
  {Soykal}}\ and\ \bibinfo {author} {\bibfnamefont {M.~E.}\ \bibnamefont
  {Flatt\'e}},\ }\bibfield  {title} {\bibinfo {title} {Strong field
  interactions between a nanomagnet and a photonic cavity},\ }\href
  {https://link.aps.org/doi/10.1103/PhysRevLett.104.077202} {\bibfield
  {journal} {\bibinfo  {journal} {Phys. Rev. Lett.}\ }\textbf {\bibinfo
  {volume} {104}},\ \bibinfo {pages} {077202} (\bibinfo {year}
  {2010})}\BibitemShut {NoStop}%
\bibitem [{\citenamefont {Frisk~Kockum}\ \emph {et~al.}(2019)\citenamefont
  {Frisk~Kockum}, \citenamefont {Miranowicz}, \citenamefont {De~Liberato},
  \citenamefont {Savasta},\ and\ \citenamefont {Nori}}]{Kockum2019}%
  \BibitemOpen
  \bibfield  {author} {\bibinfo {author} {\bibfnamefont {A.}~\bibnamefont
  {Frisk~Kockum}}, \bibinfo {author} {\bibfnamefont {A.}~\bibnamefont
  {Miranowicz}}, \bibinfo {author} {\bibfnamefont {S.}~\bibnamefont
  {De~Liberato}}, \bibinfo {author} {\bibfnamefont {S.}~\bibnamefont
  {Savasta}},\ and\ \bibinfo {author} {\bibfnamefont {F.}~\bibnamefont
  {Nori}},\ }\bibfield  {title} {\bibinfo {title} {Ultrastrong coupling between
  light and matter},\ }\href {https://doi.org/10.1038/s42254-018-0006-2}
  {\bibfield  {journal} {\bibinfo  {journal} {Nature Reviews Physics}\ }\textbf
  {\bibinfo {volume} {1}},\ \bibinfo {pages} {19} (\bibinfo {year}
  {2019})}\BibitemShut {NoStop}%
\bibitem [{\citenamefont {Nataf}\ and\ \citenamefont
  {Ciuti}(2010)}]{nataf2010nogo}%
  \BibitemOpen
  \bibfield  {author} {\bibinfo {author} {\bibfnamefont {P.}~\bibnamefont
  {Nataf}}\ and\ \bibinfo {author} {\bibfnamefont {C.}~\bibnamefont {Ciuti}},\
  }\bibfield  {title} {\bibinfo {title} {No-go theorem for superradiant quantum
  phase transitions in cavity qed and counter-example in circuit qed},\ }\href
  {https://doi.org/10.1038/ncomms1069} {\bibfield  {journal} {\bibinfo
  {journal} {Nature Communications}\ }\textbf {\bibinfo {volume} {1}},\
  \bibinfo {pages} {72} (\bibinfo {year} {2010})}\BibitemShut {NoStop}%
\bibitem [{\citenamefont {Mazza}\ and\ \citenamefont
  {Georges}(2019)}]{mazza2019superradiant}%
  \BibitemOpen
  \bibfield  {author} {\bibinfo {author} {\bibfnamefont {G.}~\bibnamefont
  {Mazza}}\ and\ \bibinfo {author} {\bibfnamefont {A.}~\bibnamefont
  {Georges}},\ }\bibfield  {title} {\bibinfo {title} {Superradiant quantum
  materials},\ }\href {https://doi.org/10.1103/PhysRevLett.122.017401}
  {\bibfield  {journal} {\bibinfo  {journal} {Phys. Rev. Lett.}\ }\textbf
  {\bibinfo {volume} {122}},\ \bibinfo {pages} {017401} (\bibinfo {year}
  {2019})}\BibitemShut {NoStop}%
\bibitem [{\citenamefont {Andolina}\ \emph {et~al.}(2020)\citenamefont
  {Andolina}, \citenamefont {Pellegrino}, \citenamefont {Giovannetti},
  \citenamefont {MacDonald},\ and\ \citenamefont
  {Polini}}]{andolina2020condensation}%
  \BibitemOpen
  \bibfield  {author} {\bibinfo {author} {\bibfnamefont {G.~M.}\ \bibnamefont
  {Andolina}}, \bibinfo {author} {\bibfnamefont {F.~M.~D.}\ \bibnamefont
  {Pellegrino}}, \bibinfo {author} {\bibfnamefont {V.}~\bibnamefont
  {Giovannetti}}, \bibinfo {author} {\bibfnamefont {A.~H.}\ \bibnamefont
  {MacDonald}},\ and\ \bibinfo {author} {\bibfnamefont {M.}~\bibnamefont
  {Polini}},\ }\bibfield  {title} {\bibinfo {title} {Theory of photon
  condensation in a spatially varying electromagnetic field},\ }\href
  {https://doi.org/10.1103/PhysRevB.102.125137} {\bibfield  {journal} {\bibinfo
   {journal} {Phys. Rev. B}\ }\textbf {\bibinfo {volume} {102}},\ \bibinfo
  {pages} {125137} (\bibinfo {year} {2020})}\BibitemShut {NoStop}%
\bibitem [{\citenamefont {Rom\'an-Roche}\ \emph {et~al.}(2021)\citenamefont
  {Rom\'an-Roche}, \citenamefont {Luis},\ and\ \citenamefont
  {Zueco}}]{zueco2021condensation}%
  \BibitemOpen
  \bibfield  {author} {\bibinfo {author} {\bibfnamefont {J.}~\bibnamefont
  {Rom\'an-Roche}}, \bibinfo {author} {\bibfnamefont {F.}~\bibnamefont
  {Luis}},\ and\ \bibinfo {author} {\bibfnamefont {D.}~\bibnamefont {Zueco}},\
  }\bibfield  {title} {\bibinfo {title} {Photon condensation and enhanced
  magnetism in cavity qed},\ }\href
  {https://doi.org/10.1103/PhysRevLett.127.167201} {\bibfield  {journal}
  {\bibinfo  {journal} {Phys. Rev. Lett.}\ }\textbf {\bibinfo {volume} {127}},\
  \bibinfo {pages} {167201} (\bibinfo {year} {2021})}\BibitemShut {NoStop}%
\bibitem [{\citenamefont {Goryachev}\ \emph {et~al.}(2014)\citenamefont
  {Goryachev}, \citenamefont {Farr}, \citenamefont {Creedon}, \citenamefont
  {Fan}, \citenamefont {Kostylev},\ and\ \citenamefont
  {Tobar}}]{PhysRevApplied.2.054002}%
  \BibitemOpen
  \bibfield  {author} {\bibinfo {author} {\bibfnamefont {M.}~\bibnamefont
  {Goryachev}}, \bibinfo {author} {\bibfnamefont {W.~G.}\ \bibnamefont {Farr}},
  \bibinfo {author} {\bibfnamefont {D.~L.}\ \bibnamefont {Creedon}}, \bibinfo
  {author} {\bibfnamefont {Y.}~\bibnamefont {Fan}}, \bibinfo {author}
  {\bibfnamefont {M.}~\bibnamefont {Kostylev}},\ and\ \bibinfo {author}
  {\bibfnamefont {M.~E.}\ \bibnamefont {Tobar}},\ }\bibfield  {title} {\bibinfo
  {title} {High-cooperativity cavity qed with magnons at microwave
  frequencies},\ }\href
  {https://link.aps.org/doi/10.1103/PhysRevApplied.2.054002} {\bibfield
  {journal} {\bibinfo  {journal} {Phys. Rev. Applied}\ }\textbf {\bibinfo
  {volume} {2}},\ \bibinfo {pages} {054002} (\bibinfo {year}
  {2014})}\BibitemShut {NoStop}%
\bibitem [{\citenamefont {Bourhill}\ \emph {et~al.}(2016)\citenamefont
  {Bourhill}, \citenamefont {Kostylev}, \citenamefont {Goryachev},
  \citenamefont {Creedon},\ and\ \citenamefont {Tobar}}]{BourhillPRB16}%
  \BibitemOpen
  \bibfield  {author} {\bibinfo {author} {\bibfnamefont {J.}~\bibnamefont
  {Bourhill}}, \bibinfo {author} {\bibfnamefont {N.}~\bibnamefont {Kostylev}},
  \bibinfo {author} {\bibfnamefont {M.}~\bibnamefont {Goryachev}}, \bibinfo
  {author} {\bibfnamefont {D.~L.}\ \bibnamefont {Creedon}},\ and\ \bibinfo
  {author} {\bibfnamefont {M.~E.}\ \bibnamefont {Tobar}},\ }\bibfield  {title}
  {\bibinfo {title} {Ultrahigh cooperativity interactions between magnons and
  resonant photons in a yig sphere},\ }\href
  {https://doi.org/10.1103/PhysRevB.93.144420} {\bibfield  {journal} {\bibinfo
  {journal} {Phys. Rev. B}\ }\textbf {\bibinfo {volume} {93}},\ \bibinfo
  {pages} {144420} (\bibinfo {year} {2016})}\BibitemShut {NoStop}%
\bibitem [{\citenamefont {Kostylev}\ \emph {et~al.}(2016)\citenamefont
  {Kostylev}, \citenamefont {Goryachev},\ and\ \citenamefont
  {Tobar}}]{KostylevAPL2016}%
  \BibitemOpen
  \bibfield  {author} {\bibinfo {author} {\bibfnamefont {N.}~\bibnamefont
  {Kostylev}}, \bibinfo {author} {\bibfnamefont {M.}~\bibnamefont
  {Goryachev}},\ and\ \bibinfo {author} {\bibfnamefont {M.~E.}\ \bibnamefont
  {Tobar}},\ }\bibfield  {title} {\bibinfo {title} {Superstrong coupling of a
  microwave cavity to yttrium iron garnet magnons},\ }\href
  {https://doi.org/10.1063/1.4941730} {\bibfield  {journal} {\bibinfo
  {journal} {Applied Physics Letters}\ }\textbf {\bibinfo {volume} {108}},\
  \bibinfo {pages} {062402} (\bibinfo {year} {2016})}\BibitemShut {NoStop}%
\bibitem [{\citenamefont {Zare~Rameshti}\ \emph {et~al.}(2015)\citenamefont
  {Zare~Rameshti}, \citenamefont {Cao},\ and\ \citenamefont
  {Bauer}}]{RameshtiPRB15}%
  \BibitemOpen
  \bibfield  {author} {\bibinfo {author} {\bibfnamefont {B.}~\bibnamefont
  {Zare~Rameshti}}, \bibinfo {author} {\bibfnamefont {Y.}~\bibnamefont {Cao}},\
  and\ \bibinfo {author} {\bibfnamefont {G.~E.~W.}\ \bibnamefont {Bauer}},\
  }\bibfield  {title} {\bibinfo {title} {Magnetic spheres in microwave
  cavities},\ }\href {https://doi.org/10.1103/PhysRevB.91.214430} {\bibfield
  {journal} {\bibinfo  {journal} {Phys. Rev. B}\ }\textbf {\bibinfo {volume}
  {91}},\ \bibinfo {pages} {214430} (\bibinfo {year} {2015})}\BibitemShut
  {NoStop}%
\bibitem [{\citenamefont {Bourcin}\ \emph {et~al.}(2023)\citenamefont
  {Bourcin}, \citenamefont {Bourhill}, \citenamefont {Vlaminck},\ and\
  \citenamefont {Castel}}]{BourcinPRB23}%
  \BibitemOpen
  \bibfield  {author} {\bibinfo {author} {\bibfnamefont {G.}~\bibnamefont
  {Bourcin}}, \bibinfo {author} {\bibfnamefont {J.}~\bibnamefont {Bourhill}},
  \bibinfo {author} {\bibfnamefont {V.}~\bibnamefont {Vlaminck}},\ and\
  \bibinfo {author} {\bibfnamefont {V.}~\bibnamefont {Castel}},\ }\bibfield
  {title} {\bibinfo {title} {Strong to ultrastrong coherent coupling
  measurements in a yig/cavity system at room temperature},\ }\href
  {https://doi.org/10.1103/PhysRevB.107.214423} {\bibfield  {journal} {\bibinfo
   {journal} {Phys. Rev. B}\ }\textbf {\bibinfo {volume} {107}},\ \bibinfo
  {pages} {214423} (\bibinfo {year} {2023})}\BibitemShut {NoStop}%
\bibitem [{\citenamefont {Flower}\ \emph {et~al.}(2019)\citenamefont {Flower},
  \citenamefont {Goryachev}, \citenamefont {Bourhill},\ and\ \citenamefont
  {Tobar}}]{Flower_2019}%
  \BibitemOpen
  \bibfield  {author} {\bibinfo {author} {\bibfnamefont {G.}~\bibnamefont
  {Flower}}, \bibinfo {author} {\bibfnamefont {M.}~\bibnamefont {Goryachev}},
  \bibinfo {author} {\bibfnamefont {J.}~\bibnamefont {Bourhill}},\ and\
  \bibinfo {author} {\bibfnamefont {M.~E.}\ \bibnamefont {Tobar}},\ }\bibfield
  {title} {\bibinfo {title} {Experimental implementations of cavity-magnon
  systems: from ultra strong coupling to applications in precision
  measurement},\ }\href {https://doi.org/10.1088/1367-2630/ab3e1c} {\bibfield
  {journal} {\bibinfo  {journal} {New Journal of Physics}\ }\textbf {\bibinfo
  {volume} {21}},\ \bibinfo {pages} {095004} (\bibinfo {year}
  {2019})}\BibitemShut {NoStop}%
\bibitem [{\citenamefont {Liensberger}\ \emph {et~al.}(2019)\citenamefont
  {Liensberger}, \citenamefont {Kamra}, \citenamefont {Maier-Flaig},
  \citenamefont {Gepr\"ags}, \citenamefont {Erb}, \citenamefont {Goennenwein},
  \citenamefont {Gross}, \citenamefont {Belzig}, \citenamefont {Huebl},\ and\
  \citenamefont {Weiler}}]{PhysRevLett.123.117204}%
  \BibitemOpen
  \bibfield  {author} {\bibinfo {author} {\bibfnamefont {L.}~\bibnamefont
  {Liensberger}}, \bibinfo {author} {\bibfnamefont {A.}~\bibnamefont {Kamra}},
  \bibinfo {author} {\bibfnamefont {H.}~\bibnamefont {Maier-Flaig}}, \bibinfo
  {author} {\bibfnamefont {S.}~\bibnamefont {Gepr\"ags}}, \bibinfo {author}
  {\bibfnamefont {A.}~\bibnamefont {Erb}}, \bibinfo {author} {\bibfnamefont
  {S.~T.~B.}\ \bibnamefont {Goennenwein}}, \bibinfo {author} {\bibfnamefont
  {R.}~\bibnamefont {Gross}}, \bibinfo {author} {\bibfnamefont
  {W.}~\bibnamefont {Belzig}}, \bibinfo {author} {\bibfnamefont
  {H.}~\bibnamefont {Huebl}},\ and\ \bibinfo {author} {\bibfnamefont
  {M.}~\bibnamefont {Weiler}},\ }\bibfield  {title} {\bibinfo {title}
  {Exchange-enhanced ultrastrong magnon-magnon coupling in a compensated
  ferrimagnet},\ }\href
  {https://link.aps.org/doi/10.1103/PhysRevLett.123.117204} {\bibfield
  {journal} {\bibinfo  {journal} {Phys. Rev. Lett.}\ }\textbf {\bibinfo
  {volume} {123}},\ \bibinfo {pages} {117204} (\bibinfo {year}
  {2019})}\BibitemShut {NoStop}%
\bibitem [{\citenamefont {Bia\l{}ek}\ \emph {et~al.}(2021)\citenamefont
  {Bia\l{}ek}, \citenamefont {Zhang}, \citenamefont {Yu},\ and\ \citenamefont
  {Ansermet}}]{PhysRevApplied.15.044018}%
  \BibitemOpen
  \bibfield  {author} {\bibinfo {author} {\bibfnamefont {M.}~\bibnamefont
  {Bia\l{}ek}}, \bibinfo {author} {\bibfnamefont {J.}~\bibnamefont {Zhang}},
  \bibinfo {author} {\bibfnamefont {H.}~\bibnamefont {Yu}},\ and\ \bibinfo
  {author} {\bibfnamefont {J.-P.}\ \bibnamefont {Ansermet}},\ }\bibfield
  {title} {\bibinfo {title} {Strong coupling of antiferromagnetic resonance
  with subterahertz cavity fields},\ }\href
  {https://link.aps.org/doi/10.1103/PhysRevApplied.15.044018} {\bibfield
  {journal} {\bibinfo  {journal} {Phys. Rev. Applied}\ }\textbf {\bibinfo
  {volume} {15}},\ \bibinfo {pages} {044018} (\bibinfo {year}
  {2021})}\BibitemShut {NoStop}%
\bibitem [{\citenamefont {Everts}\ \emph {et~al.}(2020)\citenamefont {Everts},
  \citenamefont {King}, \citenamefont {Lambert}, \citenamefont {Kocsis},
  \citenamefont {Rogge},\ and\ \citenamefont {Longdell}}]{PhysRevB.101.214414}%
  \BibitemOpen
  \bibfield  {author} {\bibinfo {author} {\bibfnamefont {J.~R.}\ \bibnamefont
  {Everts}}, \bibinfo {author} {\bibfnamefont {G.~G.~G.}\ \bibnamefont {King}},
  \bibinfo {author} {\bibfnamefont {N.~J.}\ \bibnamefont {Lambert}}, \bibinfo
  {author} {\bibfnamefont {S.}~\bibnamefont {Kocsis}}, \bibinfo {author}
  {\bibfnamefont {S.}~\bibnamefont {Rogge}},\ and\ \bibinfo {author}
  {\bibfnamefont {J.~J.}\ \bibnamefont {Longdell}},\ }\bibfield  {title}
  {\bibinfo {title} {Ultrastrong coupling between a microwave resonator and
  antiferromagnetic resonances of rare-earth ion spins},\ }\href
  {https://link.aps.org/doi/10.1103/PhysRevB.101.214414} {\bibfield  {journal}
  {\bibinfo  {journal} {Phys. Rev. B}\ }\textbf {\bibinfo {volume} {101}},\
  \bibinfo {pages} {214414} (\bibinfo {year} {2020})}\BibitemShut {NoStop}%
\bibitem [{\citenamefont {Huebl}\ \emph {et~al.}(2013)\citenamefont {Huebl},
  \citenamefont {Zollitsch}, \citenamefont {Lotze}, \citenamefont {Hocke},
  \citenamefont {Greifenstein}, \citenamefont {Marx}, \citenamefont {Gross},\
  and\ \citenamefont {Goennenwein}}]{HueblPRL13}%
  \BibitemOpen
  \bibfield  {author} {\bibinfo {author} {\bibfnamefont {H.}~\bibnamefont
  {Huebl}}, \bibinfo {author} {\bibfnamefont {C.~W.}\ \bibnamefont
  {Zollitsch}}, \bibinfo {author} {\bibfnamefont {J.}~\bibnamefont {Lotze}},
  \bibinfo {author} {\bibfnamefont {F.}~\bibnamefont {Hocke}}, \bibinfo
  {author} {\bibfnamefont {M.}~\bibnamefont {Greifenstein}}, \bibinfo {author}
  {\bibfnamefont {A.}~\bibnamefont {Marx}}, \bibinfo {author} {\bibfnamefont
  {R.}~\bibnamefont {Gross}},\ and\ \bibinfo {author} {\bibfnamefont
  {S.~T.~B.}\ \bibnamefont {Goennenwein}},\ }\bibfield  {title} {\bibinfo
  {title} {High cooperativity in coupled microwave resonator ferrimagnetic
  insulator hybrids},\ }\href
  {https://link.aps.org/doi/10.1103/PhysRevLett.111.127003} {\bibfield
  {journal} {\bibinfo  {journal} {Phys. Rev. Lett.}\ }\textbf {\bibinfo
  {volume} {111}},\ \bibinfo {pages} {127003} (\bibinfo {year}
  {2013})}\BibitemShut {NoStop}%
\bibitem [{\citenamefont {Morris}\ \emph {et~al.}(2017)\citenamefont {Morris},
  \citenamefont {van Loo}, \citenamefont {Kosen},\ and\ \citenamefont
  {Karenowska}}]{morris2017}%
  \BibitemOpen
  \bibfield  {author} {\bibinfo {author} {\bibfnamefont {R.~G.~E.}\
  \bibnamefont {Morris}}, \bibinfo {author} {\bibfnamefont {A.~F.}\
  \bibnamefont {van Loo}}, \bibinfo {author} {\bibfnamefont {S.}~\bibnamefont
  {Kosen}},\ and\ \bibinfo {author} {\bibfnamefont {A.~D.}\ \bibnamefont
  {Karenowska}},\ }\bibfield  {title} {\bibinfo {title} {Strong coupling of
  magnons in a {YIG} sphere to photons in a planar superconducting resonator in
  the quantum limit},\ }\href {https://doi.org/10.1038/s41598-017-11835-4}
  {\bibfield  {journal} {\bibinfo  {journal} {Scientific Reports}\ }\textbf
  {\bibinfo {volume} {7}},\ \bibinfo {pages} {11511} (\bibinfo {year}
  {2017})}\BibitemShut {NoStop}%
\bibitem [{\citenamefont {Hou}\ and\ \citenamefont {Liu}(2019)}]{HouPRL2019}%
  \BibitemOpen
  \bibfield  {author} {\bibinfo {author} {\bibfnamefont {J.~T.}\ \bibnamefont
  {Hou}}\ and\ \bibinfo {author} {\bibfnamefont {L.}~\bibnamefont {Liu}},\
  }\bibfield  {title} {\bibinfo {title} {Strong coupling between microwave
  photons and nanomagnet magnons},\ }\href
  {https://link.aps.org/doi/10.1103/PhysRevLett.123.107702} {\bibfield
  {journal} {\bibinfo  {journal} {Phys. Rev. Lett.}\ }\textbf {\bibinfo
  {volume} {123}},\ \bibinfo {pages} {107702} (\bibinfo {year}
  {2019})}\BibitemShut {NoStop}%
\bibitem [{\citenamefont {Golovchanskiy}\ \emph
  {et~al.}(2021{\natexlab{a}})\citenamefont {Golovchanskiy}, \citenamefont
  {Abramov}, \citenamefont {Stolyarov}, \citenamefont {Weides}, \citenamefont
  {Ryazanov}, \citenamefont {Golubov}, \citenamefont {Ustinov},\ and\
  \citenamefont {Kupriyanov}}]{UstinovScAdv2021}%
  \BibitemOpen
  \bibfield  {author} {\bibinfo {author} {\bibfnamefont {I.~A.}\ \bibnamefont
  {Golovchanskiy}}, \bibinfo {author} {\bibfnamefont {N.~N.}\ \bibnamefont
  {Abramov}}, \bibinfo {author} {\bibfnamefont {V.~S.}\ \bibnamefont
  {Stolyarov}}, \bibinfo {author} {\bibfnamefont {M.}~\bibnamefont {Weides}},
  \bibinfo {author} {\bibfnamefont {V.~V.}\ \bibnamefont {Ryazanov}}, \bibinfo
  {author} {\bibfnamefont {A.~A.}\ \bibnamefont {Golubov}}, \bibinfo {author}
  {\bibfnamefont {A.~V.}\ \bibnamefont {Ustinov}},\ and\ \bibinfo {author}
  {\bibfnamefont {M.~Y.}\ \bibnamefont {Kupriyanov}},\ }\bibfield  {title}
  {\bibinfo {title} {Ultrastrong photon-to-magnon coupling in multilayered
  heterostructures involving superconducting coherence via ferromagnetic
  layers},\ }\href {https://www.science.org/doi/abs/10.1126/sciadv.abe8638}
  {\bibfield  {journal} {\bibinfo  {journal} {Science Advances}\ }\textbf
  {\bibinfo {volume} {7}},\ \bibinfo {pages} {eabe8638} (\bibinfo {year}
  {2021}{\natexlab{a}})}\BibitemShut {NoStop}%
\bibitem [{\citenamefont {Li}\ \emph {et~al.}(2022)\citenamefont {Li},
  \citenamefont {Yefremenko}, \citenamefont {Lisovenko}, \citenamefont
  {Trevillian}, \citenamefont {Polakovic}, \citenamefont {Cecil}, \citenamefont
  {Barry}, \citenamefont {Pearson}, \citenamefont {Divan}, \citenamefont
  {Tyberkevych}, \citenamefont {Chang}, \citenamefont {Welp}, \citenamefont
  {Kwok},\ and\ \citenamefont {Novosad}}]{PhysRevLett.128.047701}%
  \BibitemOpen
  \bibfield  {author} {\bibinfo {author} {\bibfnamefont {Y.}~\bibnamefont
  {Li}}, \bibinfo {author} {\bibfnamefont {V.~G.}\ \bibnamefont {Yefremenko}},
  \bibinfo {author} {\bibfnamefont {M.}~\bibnamefont {Lisovenko}}, \bibinfo
  {author} {\bibfnamefont {C.}~\bibnamefont {Trevillian}}, \bibinfo {author}
  {\bibfnamefont {T.}~\bibnamefont {Polakovic}}, \bibinfo {author}
  {\bibfnamefont {T.~W.}\ \bibnamefont {Cecil}}, \bibinfo {author}
  {\bibfnamefont {P.~S.}\ \bibnamefont {Barry}}, \bibinfo {author}
  {\bibfnamefont {J.}~\bibnamefont {Pearson}}, \bibinfo {author} {\bibfnamefont
  {R.}~\bibnamefont {Divan}}, \bibinfo {author} {\bibfnamefont
  {V.}~\bibnamefont {Tyberkevych}}, \bibinfo {author} {\bibfnamefont {C.~L.}\
  \bibnamefont {Chang}}, \bibinfo {author} {\bibfnamefont {U.}~\bibnamefont
  {Welp}}, \bibinfo {author} {\bibfnamefont {W.-K.}\ \bibnamefont {Kwok}},\
  and\ \bibinfo {author} {\bibfnamefont {V.}~\bibnamefont {Novosad}},\
  }\bibfield  {title} {\bibinfo {title} {Coherent coupling of two remote
  magnonic resonators mediated by superconducting circuits},\ }\href
  {https://link.aps.org/doi/10.1103/PhysRevLett.128.047701} {\bibfield
  {journal} {\bibinfo  {journal} {Phys. Rev. Lett.}\ }\textbf {\bibinfo
  {volume} {128}},\ \bibinfo {pages} {047701} (\bibinfo {year}
  {2022})}\BibitemShut {NoStop}%
\bibitem [{\citenamefont {Golovchanskiy}\ \emph
  {et~al.}(2021{\natexlab{b}})\citenamefont {Golovchanskiy}, \citenamefont
  {Abramov}, \citenamefont {Stolyarov}, \citenamefont {Golubov}, \citenamefont
  {Kupriyanov}, \citenamefont {Ryazanov},\ and\ \citenamefont
  {Ustinov}}]{UstinovPRA2021}%
  \BibitemOpen
  \bibfield  {author} {\bibinfo {author} {\bibfnamefont {I.~A.}\ \bibnamefont
  {Golovchanskiy}}, \bibinfo {author} {\bibfnamefont {N.~N.}\ \bibnamefont
  {Abramov}}, \bibinfo {author} {\bibfnamefont {V.~S.}\ \bibnamefont
  {Stolyarov}}, \bibinfo {author} {\bibfnamefont {A.~A.}\ \bibnamefont
  {Golubov}}, \bibinfo {author} {\bibfnamefont {M.~Y.}\ \bibnamefont
  {Kupriyanov}}, \bibinfo {author} {\bibfnamefont {V.~V.}\ \bibnamefont
  {Ryazanov}},\ and\ \bibinfo {author} {\bibfnamefont {A.~V.}\ \bibnamefont
  {Ustinov}},\ }\bibfield  {title} {\bibinfo {title} {Approaching deep-strong
  on-chip photon-to-magnon coupling},\ }\href
  {https://link.aps.org/doi/10.1103/PhysRevApplied.16.034029} {\bibfield
  {journal} {\bibinfo  {journal} {Phys. Rev. Applied}\ }\textbf {\bibinfo
  {volume} {16}},\ \bibinfo {pages} {034029} (\bibinfo {year}
  {2021}{\natexlab{b}})}\BibitemShut {NoStop}%
\bibitem [{\citenamefont {Mac\^edo}\ \emph {et~al.}(2021)\citenamefont
  {Mac\^edo}, \citenamefont {Holland}, \citenamefont {Baity}, \citenamefont
  {McLellan}, \citenamefont {Livesey}, \citenamefont {Stamps}, \citenamefont
  {Weides},\ and\ \citenamefont {Bozhko}}]{MacedoPRAppl21}%
  \BibitemOpen
  \bibfield  {author} {\bibinfo {author} {\bibfnamefont {R.}~\bibnamefont
  {Mac\^edo}}, \bibinfo {author} {\bibfnamefont {R.~C.}\ \bibnamefont
  {Holland}}, \bibinfo {author} {\bibfnamefont {P.~G.}\ \bibnamefont {Baity}},
  \bibinfo {author} {\bibfnamefont {L.~J.}\ \bibnamefont {McLellan}}, \bibinfo
  {author} {\bibfnamefont {K.~L.}\ \bibnamefont {Livesey}}, \bibinfo {author}
  {\bibfnamefont {R.~L.}\ \bibnamefont {Stamps}}, \bibinfo {author}
  {\bibfnamefont {M.~P.}\ \bibnamefont {Weides}},\ and\ \bibinfo {author}
  {\bibfnamefont {D.~A.}\ \bibnamefont {Bozhko}},\ }\bibfield  {title}
  {\bibinfo {title} {Electromagnetic approach to cavity spintronics},\ }\href
  {https://doi.org/10.1103/PhysRevApplied.15.024065} {\bibfield  {journal}
  {\bibinfo  {journal} {Phys. Rev. Appl.}\ }\textbf {\bibinfo {volume} {15}},\
  \bibinfo {pages} {024065} (\bibinfo {year} {2021})}\BibitemShut {NoStop}%
\bibitem [{\citenamefont {Mart\'{\i}nez-Losa~del Rinc\'on}\ \emph
  {et~al.}(2023)\citenamefont {Mart\'{\i}nez-Losa~del Rinc\'on}, \citenamefont
  {Gimeno}, \citenamefont {P\'erez-Bail\'on}, \citenamefont {Rollano},
  \citenamefont {Luis}, \citenamefont {Zueco},\ and\ \citenamefont
  {Mart\'{\i}nez-P\'erez}}]{RinconPRAppl23}%
  \BibitemOpen
  \bibfield  {author} {\bibinfo {author} {\bibfnamefont {S.}~\bibnamefont
  {Mart\'{\i}nez-Losa~del Rinc\'on}}, \bibinfo {author} {\bibfnamefont
  {I.}~\bibnamefont {Gimeno}}, \bibinfo {author} {\bibfnamefont
  {J.}~\bibnamefont {P\'erez-Bail\'on}}, \bibinfo {author} {\bibfnamefont
  {V.}~\bibnamefont {Rollano}}, \bibinfo {author} {\bibfnamefont
  {F.}~\bibnamefont {Luis}}, \bibinfo {author} {\bibfnamefont {D.}~\bibnamefont
  {Zueco}},\ and\ \bibinfo {author} {\bibfnamefont {M.~J.}\ \bibnamefont
  {Mart\'{\i}nez-P\'erez}},\ }\bibfield  {title} {\bibinfo {title} {Measuring
  the magnon-photon coupling in shaped ferromagnets: Tuning of the resonance
  frequency},\ }\href {https://doi.org/10.1103/PhysRevApplied.19.014002}
  {\bibfield  {journal} {\bibinfo  {journal} {Phys. Rev. Appl.}\ }\textbf
  {\bibinfo {volume} {19}},\ \bibinfo {pages} {014002} (\bibinfo {year}
  {2023})}\BibitemShut {NoStop}%
\bibitem [{\citenamefont {Ghirri}\ \emph {et~al.}(2015)\citenamefont {Ghirri},
  \citenamefont {Bonizzoni}, \citenamefont {Gerace}, \citenamefont {Sanna},
  \citenamefont {Cassinese},\ and\ \citenamefont {Affronte}}]{GhirriAPL15}%
  \BibitemOpen
  \bibfield  {author} {\bibinfo {author} {\bibfnamefont {A.}~\bibnamefont
  {Ghirri}}, \bibinfo {author} {\bibfnamefont {C.}~\bibnamefont {Bonizzoni}},
  \bibinfo {author} {\bibfnamefont {D.}~\bibnamefont {Gerace}}, \bibinfo
  {author} {\bibfnamefont {S.}~\bibnamefont {Sanna}}, \bibinfo {author}
  {\bibfnamefont {A.}~\bibnamefont {Cassinese}},\ and\ \bibinfo {author}
  {\bibfnamefont {M.}~\bibnamefont {Affronte}},\ }\bibfield  {title} {\bibinfo
  {title} {Yba2cu3o7 microwave resonators for strong collective coupling with
  spin ensembles},\ }\href {https://doi.org/10.1063/1.4920930} {\bibfield
  {journal} {\bibinfo  {journal} {Applied Physics Letters}\ }\textbf {\bibinfo
  {volume} {106}},\ \bibinfo {pages} {184101} (\bibinfo {year}
  {2015})}\BibitemShut {NoStop}%
\bibitem [{\citenamefont {Niedzielski}\ \emph {et~al.}(2023)\citenamefont
  {Niedzielski}, \citenamefont {Jia},\ and\ \citenamefont
  {Berakdar}}]{NiedzielskiPRAppl23}%
  \BibitemOpen
  \bibfield  {author} {\bibinfo {author} {\bibfnamefont {B.}~\bibnamefont
  {Niedzielski}}, \bibinfo {author} {\bibfnamefont {C.~L.}\ \bibnamefont
  {Jia}},\ and\ \bibinfo {author} {\bibfnamefont {J.}~\bibnamefont
  {Berakdar}},\ }\bibfield  {title} {\bibinfo {title} {Magnon-fluxon
  interaction in coupled superconductor/ferromagnet hybrid periodic
  structures},\ }\href {https://doi.org/10.1103/PhysRevApplied.19.024073}
  {\bibfield  {journal} {\bibinfo  {journal} {Phys. Rev. Appl.}\ }\textbf
  {\bibinfo {volume} {19}},\ \bibinfo {pages} {024073} (\bibinfo {year}
  {2023})}\BibitemShut {NoStop}%
\bibitem [{\citenamefont {Bonizzoni}\ \emph {et~al.}(2022)\citenamefont
  {Bonizzoni}, \citenamefont {Maksutoglu}, \citenamefont {Ghirri},
  \citenamefont {van Tol}, \citenamefont {Rameev},\ and\ \citenamefont
  {Affronte}}]{bonizzoni_coupling_2022}%
  \BibitemOpen
  \bibfield  {author} {\bibinfo {author} {\bibfnamefont {C.}~\bibnamefont
  {Bonizzoni}}, \bibinfo {author} {\bibfnamefont {M.}~\bibnamefont
  {Maksutoglu}}, \bibinfo {author} {\bibfnamefont {A.}~\bibnamefont {Ghirri}},
  \bibinfo {author} {\bibfnamefont {J.}~\bibnamefont {van Tol}}, \bibinfo
  {author} {\bibfnamefont {B.}~\bibnamefont {Rameev}},\ and\ \bibinfo {author}
  {\bibfnamefont {M.}~\bibnamefont {Affronte}},\ }\bibfield  {title} {\bibinfo
  {title} {Coupling {Sub}-nanoliter {BDPA} {Organic} {Radical} {Spin}
  {Ensembles} with {YBCO} {Inverse} {Anapole} {Resonators}},\ }\href
  {https://doi.org/10.1007/s00723-022-01505-8} {\bibfield  {journal} {\bibinfo
  {journal} {Applied Magnetic Resonance}\ } (\bibinfo {year}
  {2022})}\BibitemShut {NoStop}%
\bibitem [{\citenamefont {Ghirri}\ \emph {et~al.}(2020)\citenamefont {Ghirri},
  \citenamefont {Herrero}, \citenamefont {Mazerat}, \citenamefont {Mallah},
  \citenamefont {Moze},\ and\ \citenamefont {Affronte}}]{GhirriAdvQTechnol20}%
  \BibitemOpen
  \bibfield  {author} {\bibinfo {author} {\bibfnamefont {A.}~\bibnamefont
  {Ghirri}}, \bibinfo {author} {\bibfnamefont {C.}~\bibnamefont {Herrero}},
  \bibinfo {author} {\bibfnamefont {S.}~\bibnamefont {Mazerat}}, \bibinfo
  {author} {\bibfnamefont {T.}~\bibnamefont {Mallah}}, \bibinfo {author}
  {\bibfnamefont {O.}~\bibnamefont {Moze}},\ and\ \bibinfo {author}
  {\bibfnamefont {M.}~\bibnamefont {Affronte}},\ }\bibfield  {title} {\bibinfo
  {title} {Coupling nanostructured csnicr prussian blue analogue to resonant
  microwave fields},\ }\href
  {https://doi.org/https://doi.org/10.1002/qute.201900101} {\bibfield
  {journal} {\bibinfo  {journal} {Advanced Quantum Technologies}\ }\textbf
  {\bibinfo {volume} {3}},\ \bibinfo {pages} {1900101} (\bibinfo {year}
  {2020})}\BibitemShut {NoStop}%
\bibitem [{sup()}]{supplementary}%
  \BibitemOpen
  \href@noop {} {\bibinfo  {journal} {See Supplemental Material at [URL will be
  inserted by publisher] for experimental details, additional transmission
  spectroscopy data and simulations}\ }\BibitemShut {NoStop}%
\bibitem [{\citenamefont {Tosi}\ \emph {et~al.}(2014)\citenamefont {Tosi},
  \citenamefont {Mohiyaddin}, \citenamefont {Huebl},\ and\ \citenamefont
  {Morello}}]{TosiAIPAdv14}%
  \BibitemOpen
\bibfield  {journal} {  }\bibfield  {author} {\bibinfo {author} {\bibfnamefont
  {G.}~\bibnamefont {Tosi}}, \bibinfo {author} {\bibfnamefont {F.~A.}\
  \bibnamefont {Mohiyaddin}}, \bibinfo {author} {\bibfnamefont
  {H.}~\bibnamefont {Huebl}},\ and\ \bibinfo {author} {\bibfnamefont
  {A.}~\bibnamefont {Morello}},\ }\bibfield  {title} {\bibinfo {title}
  {Circuit-quantum electrodynamics with direct magnetic coupling to single-atom
  spin qubits in isotopically enriched 28si},\ }\href
  {https://doi.org/10.1063/1.4893242} {\bibfield  {journal} {\bibinfo
  {journal} {AIP Advances}\ }\textbf {\bibinfo {volume} {4}},\ \bibinfo {pages}
  {087122} (\bibinfo {year} {2014})}\BibitemShut {NoStop}%
\bibitem [{\citenamefont {Kittel}(1948)}]{KittelPR48}%
  \BibitemOpen
  \bibfield  {author} {\bibinfo {author} {\bibfnamefont {C.}~\bibnamefont
  {Kittel}},\ }\bibfield  {title} {\bibinfo {title} {On the theory of
  ferromagnetic resonance absorption},\ }\href
  {https://link.aps.org/doi/10.1103/PhysRev.73.155} {\bibfield  {journal}
  {\bibinfo  {journal} {Phys. Rev.}\ }\textbf {\bibinfo {volume} {73}},\
  \bibinfo {pages} {155} (\bibinfo {year} {1948})}\BibitemShut {NoStop}%
\bibitem [{\citenamefont {Kittel}(2004)}]{Kittel}%
  \BibitemOpen
  \bibfield  {author} {\bibinfo {author} {\bibfnamefont {C.}~\bibnamefont
  {Kittel}},\ }\href
  {http://www.amazon.com/Introduction-Solid-Physics-Charles-Kittel/dp/047141526X/ref=dp_ob_title_bk}
  {\emph {\bibinfo {title} {Introduction to Solid State Physics}}},\ \bibinfo
  {edition} {8th}\ ed.\ (\bibinfo  {publisher} {Wiley},\ \bibinfo {year}
  {2004})\BibitemShut {NoStop}%
\bibitem [{\citenamefont {Kajiwara}\ \emph {et~al.}(2010)\citenamefont
  {Kajiwara}, \citenamefont {Harii}, \citenamefont {Takahashi}, \citenamefont
  {Ohe}, \citenamefont {Uchida}, \citenamefont {Mizuguchi}, \citenamefont
  {Umezawa}, \citenamefont {Kawai}, \citenamefont {Ando}, \citenamefont
  {Takanashi}, \citenamefont {Maekawa},\ and\ \citenamefont
  {Saitoh}}]{KajiwaraNature10}%
  \BibitemOpen
  \bibfield  {author} {\bibinfo {author} {\bibfnamefont {Y.}~\bibnamefont
  {Kajiwara}}, \bibinfo {author} {\bibfnamefont {K.}~\bibnamefont {Harii}},
  \bibinfo {author} {\bibfnamefont {S.}~\bibnamefont {Takahashi}}, \bibinfo
  {author} {\bibfnamefont {J.}~\bibnamefont {Ohe}}, \bibinfo {author}
  {\bibfnamefont {K.}~\bibnamefont {Uchida}}, \bibinfo {author} {\bibfnamefont
  {M.}~\bibnamefont {Mizuguchi}}, \bibinfo {author} {\bibfnamefont
  {H.}~\bibnamefont {Umezawa}}, \bibinfo {author} {\bibfnamefont
  {H.}~\bibnamefont {Kawai}}, \bibinfo {author} {\bibfnamefont
  {K.}~\bibnamefont {Ando}}, \bibinfo {author} {\bibfnamefont {K.}~\bibnamefont
  {Takanashi}}, \bibinfo {author} {\bibfnamefont {S.}~\bibnamefont {Maekawa}},\
  and\ \bibinfo {author} {\bibfnamefont {E.}~\bibnamefont {Saitoh}},\
  }\bibfield  {title} {\bibinfo {title} {Transmission of electrical signals by
  spin-wave interconversion in a magnetic insulator},\ }\href
  {https://doi.org/10.1038/nature08876} {\bibfield  {journal} {\bibinfo
  {journal} {Nature}\ }\textbf {\bibinfo {volume} {464}},\ \bibinfo {pages}
  {262} (\bibinfo {year} {2010})}\BibitemShut {NoStop}%
\bibitem [{\citenamefont {Maksymov}\ and\ \citenamefont
  {Kostylev}(2015)}]{MaksymovPhysE15}%
  \BibitemOpen
  \bibfield  {author} {\bibinfo {author} {\bibfnamefont {I.~S.}\ \bibnamefont
  {Maksymov}}\ and\ \bibinfo {author} {\bibfnamefont {M.}~\bibnamefont
  {Kostylev}},\ }\bibfield  {title} {\bibinfo {title} {Broadband stripline
  ferromagnetic resonance spectroscopy of ferromagnetic films, multilayers and
  nanostructures},\ }\href
  {https://www.sciencedirect.com/science/article/pii/S1386947714004664}
  {\bibfield  {journal} {\bibinfo  {journal} {Physica E: Low-dimensional
  Systems and Nanostructures}\ }\textbf {\bibinfo {volume} {69}},\ \bibinfo
  {pages} {253} (\bibinfo {year} {2015})}\BibitemShut {NoStop}%
\bibitem [{\citenamefont {Kennewell}\ \emph {et~al.}(2007)\citenamefont
  {Kennewell}, \citenamefont {Kostylev},\ and\ \citenamefont
  {Stamps}}]{Kennewell2007}%
  \BibitemOpen
  \bibfield  {author} {\bibinfo {author} {\bibfnamefont {K.~J.}\ \bibnamefont
  {Kennewell}}, \bibinfo {author} {\bibfnamefont {M.}~\bibnamefont
  {Kostylev}},\ and\ \bibinfo {author} {\bibfnamefont {R.~L.}\ \bibnamefont
  {Stamps}},\ }\bibfield  {title} {\bibinfo {title} {Calculation of spin wave
  mode response induced by a coplanar microwave line},\ }\href
  {https://doi.org/10.1063/1.2710068} {\bibfield  {journal} {\bibinfo
  {journal} {Journal of Applied Physics}\ }\textbf {\bibinfo {volume} {101}},\
  \bibinfo {pages} {09D107} (\bibinfo {year} {2007})}\BibitemShut {NoStop}%
\bibitem [{\citenamefont {Wang}\ \emph {et~al.}(2020)\citenamefont {Wang},
  \citenamefont {Lu}, \citenamefont {Zhao}, \citenamefont {Zhang},
  \citenamefont {Chen}, \citenamefont {Tian}, \citenamefont {Yan},
  \citenamefont {Bai},\ and\ \citenamefont {Harder}}]{WangPRB20}%
  \BibitemOpen
  \bibfield  {author} {\bibinfo {author} {\bibfnamefont {L.}~\bibnamefont
  {Wang}}, \bibinfo {author} {\bibfnamefont {Z.}~\bibnamefont {Lu}}, \bibinfo
  {author} {\bibfnamefont {X.}~\bibnamefont {Zhao}}, \bibinfo {author}
  {\bibfnamefont {W.}~\bibnamefont {Zhang}}, \bibinfo {author} {\bibfnamefont
  {Y.}~\bibnamefont {Chen}}, \bibinfo {author} {\bibfnamefont {Y.}~\bibnamefont
  {Tian}}, \bibinfo {author} {\bibfnamefont {S.}~\bibnamefont {Yan}}, \bibinfo
  {author} {\bibfnamefont {L.}~\bibnamefont {Bai}},\ and\ \bibinfo {author}
  {\bibfnamefont {M.}~\bibnamefont {Harder}},\ }\bibfield  {title} {\bibinfo
  {title} {Magnetization coupling in a yig/ggg structure},\ }\href
  {https://doi.org/10.1103/PhysRevB.102.144428} {\bibfield  {journal} {\bibinfo
   {journal} {Phys. Rev. B}\ }\textbf {\bibinfo {volume} {102}},\ \bibinfo
  {pages} {144428} (\bibinfo {year} {2020})}\BibitemShut {NoStop}%
\bibitem [{\citenamefont {Tsutsumi}\ \emph {et~al.}(1996)\citenamefont
  {Tsutsumi}, \citenamefont {Fukusako},\ and\ \citenamefont
  {Yoshida}}]{TsutsumiIEEE97}%
  \BibitemOpen
  \bibfield  {author} {\bibinfo {author} {\bibfnamefont {M.}~\bibnamefont
  {Tsutsumi}}, \bibinfo {author} {\bibfnamefont {T.}~\bibnamefont {Fukusako}},\
  and\ \bibinfo {author} {\bibfnamefont {S.}~\bibnamefont {Yoshida}},\
  }\bibfield  {title} {\bibinfo {title} {Propagation characteristics of the
  magnetostatic surface wave in the ybco-yig film-layered structure},\
  }\href@noop {} {\bibfield  {journal} {\bibinfo  {journal} {IEEE Transactions
  on Microwave Theory and Techniques}\ }\textbf {\bibinfo {volume} {44}},\
  \bibinfo {pages} {1410} (\bibinfo {year} {1996})}\BibitemShut {NoStop}%
\bibitem [{\citenamefont {Oates}\ \emph {et~al.}(1997)\citenamefont {Oates},
  \citenamefont {Pique}, \citenamefont {Harshavardhan}, \citenamefont {Moses},
  \citenamefont {Yang},\ and\ \citenamefont {Dionne}}]{OatesIEEE97}%
  \BibitemOpen
  \bibfield  {author} {\bibinfo {author} {\bibfnamefont {D.}~\bibnamefont
  {Oates}}, \bibinfo {author} {\bibfnamefont {A.}~\bibnamefont {Pique}},
  \bibinfo {author} {\bibfnamefont {K.}~\bibnamefont {Harshavardhan}}, \bibinfo
  {author} {\bibfnamefont {J.}~\bibnamefont {Moses}}, \bibinfo {author}
  {\bibfnamefont {F.}~\bibnamefont {Yang}},\ and\ \bibinfo {author}
  {\bibfnamefont {G.}~\bibnamefont {Dionne}},\ }\bibfield  {title} {\bibinfo
  {title} {Tunable ybco resonators on yig substrates},\ }\href@noop {}
  {\bibfield  {journal} {\bibinfo  {journal} {IEEE Transactions on Applied
  Superconductivity}\ }\textbf {\bibinfo {volume} {7}},\ \bibinfo {pages}
  {2338} (\bibinfo {year} {1997})}\BibitemShut {NoStop}%
\bibitem [{\citenamefont {Diniz}\ \emph {et~al.}(2011)\citenamefont {Diniz},
  \citenamefont {Portolan}, \citenamefont {Ferreira}, \citenamefont {G\'erard},
  \citenamefont {Bertet},\ and\ \citenamefont {Auff\`eves}}]{DinizPRA11}%
  \BibitemOpen
  \bibfield  {author} {\bibinfo {author} {\bibfnamefont {I.}~\bibnamefont
  {Diniz}}, \bibinfo {author} {\bibfnamefont {S.}~\bibnamefont {Portolan}},
  \bibinfo {author} {\bibfnamefont {R.}~\bibnamefont {Ferreira}}, \bibinfo
  {author} {\bibfnamefont {J.~M.}\ \bibnamefont {G\'erard}}, \bibinfo {author}
  {\bibfnamefont {P.}~\bibnamefont {Bertet}},\ and\ \bibinfo {author}
  {\bibfnamefont {A.}~\bibnamefont {Auff\`eves}},\ }\bibfield  {title}
  {\bibinfo {title} {Strongly coupling a cavity to inhomogeneous ensembles of
  emitters: Potential for long-lived solid-state quantum memories},\ }\href
  {https://link.aps.org/doi/10.1103/PhysRevA.84.063810} {\bibfield  {journal}
  {\bibinfo  {journal} {Phys. Rev. A}\ }\textbf {\bibinfo {volume} {84}},\
  \bibinfo {pages} {063810} (\bibinfo {year} {2011})}\BibitemShut {NoStop}%
\bibitem [{\citenamefont {Ghirri}\ \emph {et~al.}(2016)\citenamefont {Ghirri},
  \citenamefont {Bonizzoni}, \citenamefont {Troiani}, \citenamefont {Buccheri},
  \citenamefont {Beverina}, \citenamefont {Cassinese},\ and\ \citenamefont
  {Affronte}}]{GhirriPRA16}%
  \BibitemOpen
  \bibfield  {author} {\bibinfo {author} {\bibfnamefont {A.}~\bibnamefont
  {Ghirri}}, \bibinfo {author} {\bibfnamefont {C.}~\bibnamefont {Bonizzoni}},
  \bibinfo {author} {\bibfnamefont {F.}~\bibnamefont {Troiani}}, \bibinfo
  {author} {\bibfnamefont {N.}~\bibnamefont {Buccheri}}, \bibinfo {author}
  {\bibfnamefont {L.}~\bibnamefont {Beverina}}, \bibinfo {author}
  {\bibfnamefont {A.}~\bibnamefont {Cassinese}},\ and\ \bibinfo {author}
  {\bibfnamefont {M.}~\bibnamefont {Affronte}},\ }\bibfield  {title} {\bibinfo
  {title} {Coherently coupling distinct spin ensembles through a high-${T}_{c}$
  superconducting resonator},\ }\href
  {https://link.aps.org/doi/10.1103/PhysRevA.93.063855} {\bibfield  {journal}
  {\bibinfo  {journal} {Phys. Rev. A}\ }\textbf {\bibinfo {volume} {93}},\
  \bibinfo {pages} {063855} (\bibinfo {year} {2016})}\BibitemShut {NoStop}%
\bibitem [{\citenamefont {Forn-D\'{\i}az}\ \emph
  {et~al.}(2019{\natexlab{b}})\citenamefont {Forn-D\'{\i}az}, \citenamefont
  {Lamata}, \citenamefont {Rico}, \citenamefont {Kono},\ and\ \citenamefont
  {Solano}}]{FornDiazRMP19}%
  \BibitemOpen
  \bibfield  {author} {\bibinfo {author} {\bibfnamefont {P.}~\bibnamefont
  {Forn-D\'{\i}az}}, \bibinfo {author} {\bibfnamefont {L.}~\bibnamefont
  {Lamata}}, \bibinfo {author} {\bibfnamefont {E.}~\bibnamefont {Rico}},
  \bibinfo {author} {\bibfnamefont {J.}~\bibnamefont {Kono}},\ and\ \bibinfo
  {author} {\bibfnamefont {E.}~\bibnamefont {Solano}},\ }\bibfield  {title}
  {\bibinfo {title} {Ultrastrong coupling regimes of light-matter
  interaction},\ }\href {https://link.aps.org/doi/10.1103/RevModPhys.91.025005}
  {\bibfield  {journal} {\bibinfo  {journal} {Rev. Mod. Phys.}\ }\textbf
  {\bibinfo {volume} {91}},\ \bibinfo {pages} {025005} (\bibinfo {year}
  {2019}{\natexlab{b}})}\BibitemShut {NoStop}%
\bibitem [{\citenamefont {Holstein}\ and\ \citenamefont
  {Primakoff}(1940)}]{holstein1940field}%
  \BibitemOpen
  \bibfield  {author} {\bibinfo {author} {\bibfnamefont {T.}~\bibnamefont
  {Holstein}}\ and\ \bibinfo {author} {\bibfnamefont {H.}~\bibnamefont
  {Primakoff}},\ }\bibfield  {title} {\bibinfo {title} {Field dependence of the
  intrinsic domain magnetization of a ferromagnet},\ }\href
  {https://link.aps.org/doi/10.1103/PhysRev.58.1098} {\bibfield  {journal}
  {\bibinfo  {journal} {Phys. Rev.}\ }\textbf {\bibinfo {volume} {58}},\
  \bibinfo {pages} {1098} (\bibinfo {year} {1940})}\BibitemShut {NoStop}%
\bibitem [{\citenamefont {Hopfield}(1958)}]{hopfield1958theory}%
  \BibitemOpen
  \bibfield  {author} {\bibinfo {author} {\bibfnamefont {J.~J.}\ \bibnamefont
  {Hopfield}},\ }\bibfield  {title} {\bibinfo {title} {Theory of the
  contribution of excitons to the complex dielectric constant of crystals},\
  }\href {https://link.aps.org/doi/10.1103/PhysRev.112.1555} {\bibfield
  {journal} {\bibinfo  {journal} {Phys. Rev.}\ }\textbf {\bibinfo {volume}
  {112}},\ \bibinfo {pages} {1555} (\bibinfo {year} {1958})}\BibitemShut
  {NoStop}%
\bibitem [{\citenamefont {Kalinikos}\ and\ \citenamefont
  {Slavin}(1986)}]{KalinikosJPhysC86}%
  \BibitemOpen
  \bibfield  {author} {\bibinfo {author} {\bibfnamefont {B.~A.}\ \bibnamefont
  {Kalinikos}}\ and\ \bibinfo {author} {\bibfnamefont {A.~N.}\ \bibnamefont
  {Slavin}},\ }\bibfield  {title} {\bibinfo {title} {Theory of dipole-exchange
  spin wave spectrum for ferromagnetic films with mixed exchange boundary
  conditions},\ }\href {https://doi.org/10.1088/0022-3719/19/35/014} {\bibfield
   {journal} {\bibinfo  {journal} {Journal of Physics C: Solid State Physics}\
  }\textbf {\bibinfo {volume} {19}},\ \bibinfo {pages} {7013} (\bibinfo {year}
  {1986})}\BibitemShut {NoStop}%
\bibitem [{\citenamefont {Demokritov}\ and\ \citenamefont
  {Slavin}(2021)}]{DemokritovSpringer21}%
  \BibitemOpen
  \bibfield  {author} {\bibinfo {author} {\bibfnamefont {S.~O.}\ \bibnamefont
  {Demokritov}}\ and\ \bibinfo {author} {\bibfnamefont {A.~N.}\ \bibnamefont
  {Slavin}},\ }\bibinfo {title} {Spin waves},\ in\ \href
  {https://doi.org/10.1007/978-3-030-63210-6_6} {\emph {\bibinfo {booktitle}
  {Handbook of Magnetism and Magnetic Materials}}},\ \bibinfo {editor} {edited
  by\ \bibinfo {editor} {\bibfnamefont {J.~M.~D.}\ \bibnamefont {Coey}}\ and\
  \bibinfo {editor} {\bibfnamefont {S.~S.}\ \bibnamefont {Parkin}}}\ (\bibinfo
  {publisher} {Springer International Publishing},\ \bibinfo {address} {Cham},\
  \bibinfo {year} {2021})\ pp.\ \bibinfo {pages} {281--346}\BibitemShut
  {NoStop}%
\bibitem [{\citenamefont {Golovchanskiy}\ \emph
  {et~al.}(2018{\natexlab{a}})\citenamefont {Golovchanskiy}, \citenamefont
  {Abramov}, \citenamefont {Stolyarov}, \citenamefont {Bolginov}, \citenamefont
  {Ryazanov}, \citenamefont {Golubov},\ and\ \citenamefont
  {Ustinov}}]{UstinovAFM2018}%
  \BibitemOpen
  \bibfield  {author} {\bibinfo {author} {\bibfnamefont {I.~A.}\ \bibnamefont
  {Golovchanskiy}}, \bibinfo {author} {\bibfnamefont {N.~N.}\ \bibnamefont
  {Abramov}}, \bibinfo {author} {\bibfnamefont {V.~S.}\ \bibnamefont
  {Stolyarov}}, \bibinfo {author} {\bibfnamefont {V.~V.}\ \bibnamefont
  {Bolginov}}, \bibinfo {author} {\bibfnamefont {V.~V.}\ \bibnamefont
  {Ryazanov}}, \bibinfo {author} {\bibfnamefont {A.~A.}\ \bibnamefont
  {Golubov}},\ and\ \bibinfo {author} {\bibfnamefont {A.~V.}\ \bibnamefont
  {Ustinov}},\ }\bibfield  {title} {\bibinfo {title}
  {Ferromagnet/superconductor hybridization for magnonic applications},\ }\href
  {https://onlinelibrary.wiley.com/doi/abs/10.1002/adfm.201802375} {\bibfield
  {journal} {\bibinfo  {journal} {Advanced Functional Materials}\ }\textbf
  {\bibinfo {volume} {28}},\ \bibinfo {pages} {1802375} (\bibinfo {year}
  {2018}{\natexlab{a}})}\BibitemShut {NoStop}%
\bibitem [{\citenamefont {Golovchanskiy}\ \emph
  {et~al.}(2018{\natexlab{b}})\citenamefont {Golovchanskiy}, \citenamefont
  {Abramov}, \citenamefont {Stolyarov}, \citenamefont {Ryazanov}, \citenamefont
  {Golubov},\ and\ \citenamefont {Ustinov}}]{UstinovJAP2018}%
  \BibitemOpen
  \bibfield  {author} {\bibinfo {author} {\bibfnamefont {I.~A.}\ \bibnamefont
  {Golovchanskiy}}, \bibinfo {author} {\bibfnamefont {N.~N.}\ \bibnamefont
  {Abramov}}, \bibinfo {author} {\bibfnamefont {V.~S.}\ \bibnamefont
  {Stolyarov}}, \bibinfo {author} {\bibfnamefont {V.~V.}\ \bibnamefont
  {Ryazanov}}, \bibinfo {author} {\bibfnamefont {A.~A.}\ \bibnamefont
  {Golubov}},\ and\ \bibinfo {author} {\bibfnamefont {A.~V.}\ \bibnamefont
  {Ustinov}},\ }\bibfield  {title} {\bibinfo {title} {Modified dispersion law
  for spin waves coupled to a superconductor},\ }\href
  {https://doi.org/10.1063/1.5077086} {\bibfield  {journal} {\bibinfo
  {journal} {Journal of Applied Physics}\ }\textbf {\bibinfo {volume} {124}},\
  \bibinfo {pages} {233903} (\bibinfo {year} {2018}{\natexlab{b}})}\BibitemShut
  {NoStop}%
\bibitem [{\citenamefont {Lee}\ \emph {et~al.}(2023)\citenamefont {Lee},
  \citenamefont {Fakhrul}, \citenamefont {Ross},\ and\ \citenamefont
  {Beach}}]{LeePRL20}%
  \BibitemOpen
  \bibfield  {author} {\bibinfo {author} {\bibfnamefont {B.~H.}\ \bibnamefont
  {Lee}}, \bibinfo {author} {\bibfnamefont {T.}~\bibnamefont {Fakhrul}},
  \bibinfo {author} {\bibfnamefont {C.~A.}\ \bibnamefont {Ross}},\ and\
  \bibinfo {author} {\bibfnamefont {G.~S.~D.}\ \bibnamefont {Beach}},\
  }\bibfield  {title} {\bibinfo {title} {Large anomalous frequency shift in
  perpendicular standing spin wave modes in biyig films induced by thin
  metallic overlayers},\ }\href
  {https://doi.org/10.1103/PhysRevLett.130.126703} {\bibfield  {journal}
  {\bibinfo  {journal} {Phys. Rev. Lett.}\ }\textbf {\bibinfo {volume} {130}},\
  \bibinfo {pages} {126703} (\bibinfo {year} {2023})}\BibitemShut {NoStop}%
\bibitem [{\citenamefont {Bayer}\ \emph {et~al.}(2006)\citenamefont {Bayer},
  \citenamefont {Jorzick}, \citenamefont {Demokritov}, \citenamefont {Slavin},
  \citenamefont {Guslienko}, \citenamefont {Berkov}, \citenamefont {Gorn},
  \citenamefont {Kostylev},\ and\ \citenamefont
  {Hillebrands}}]{BayerSpringer06}%
  \BibitemOpen
  \bibfield  {author} {\bibinfo {author} {\bibfnamefont {C.}~\bibnamefont
  {Bayer}}, \bibinfo {author} {\bibfnamefont {J.}~\bibnamefont {Jorzick}},
  \bibinfo {author} {\bibfnamefont {S.~O.}\ \bibnamefont {Demokritov}},
  \bibinfo {author} {\bibfnamefont {A.~N.}\ \bibnamefont {Slavin}}, \bibinfo
  {author} {\bibfnamefont {K.~Y.}\ \bibnamefont {Guslienko}}, \bibinfo {author}
  {\bibfnamefont {D.~V.}\ \bibnamefont {Berkov}}, \bibinfo {author}
  {\bibfnamefont {N.~L.}\ \bibnamefont {Gorn}}, \bibinfo {author}
  {\bibfnamefont {M.~P.}\ \bibnamefont {Kostylev}},\ and\ \bibinfo {author}
  {\bibfnamefont {B.}~\bibnamefont {Hillebrands}},\ }\bibinfo {title}
  {Spin-wave excitations in finite rectangular elements},\ in\ \href
  {https://doi.org/10.1007/10938171_2} {\emph {\bibinfo {booktitle} {Spin
  Dynamics in Confined Magnetic Structures III}}},\ \bibinfo {editor} {edited
  by\ \bibinfo {editor} {\bibfnamefont {B.}~\bibnamefont {Hillebrands}}\ and\
  \bibinfo {editor} {\bibfnamefont {A.}~\bibnamefont {Thiaville}}}\ (\bibinfo
  {publisher} {Springer Berlin Heidelberg},\ \bibinfo {address} {Berlin,
  Heidelberg},\ \bibinfo {year} {2006})\ pp.\ \bibinfo {pages}
  {57--103}\BibitemShut {NoStop}%
\bibitem [{\citenamefont {Sage}\ \emph {et~al.}(2011)\citenamefont {Sage},
  \citenamefont {Bolkhovsky}, \citenamefont {Oliver}, \citenamefont {Turek},\
  and\ \citenamefont {Welander}}]{SageJAP11}%
  \BibitemOpen
  \bibfield  {author} {\bibinfo {author} {\bibfnamefont {J.~M.}\ \bibnamefont
  {Sage}}, \bibinfo {author} {\bibfnamefont {V.}~\bibnamefont {Bolkhovsky}},
  \bibinfo {author} {\bibfnamefont {W.~D.}\ \bibnamefont {Oliver}}, \bibinfo
  {author} {\bibfnamefont {B.}~\bibnamefont {Turek}},\ and\ \bibinfo {author}
  {\bibfnamefont {P.~B.}\ \bibnamefont {Welander}},\ }\bibfield  {title}
  {\bibinfo {title} {Study of loss in superconducting coplanar waveguide
  resonators},\ }\href {https://doi.org/10.1063/1.3552890} {\bibfield
  {journal} {\bibinfo  {journal} {Journal of Applied Physics}\ }\textbf
  {\bibinfo {volume} {109}},\ \bibinfo {pages} {063915} (\bibinfo {year}
  {2011})}\BibitemShut {NoStop}%
\bibitem [{\citenamefont {Trempler}\ \emph {et~al.}(2020)\citenamefont
  {Trempler}, \citenamefont {Dreyer}, \citenamefont {Geyer}, \citenamefont
  {Hauser}, \citenamefont {Woltersdorf},\ and\ \citenamefont
  {Schmidt}}]{TremplerAPL20}%
  \BibitemOpen
  \bibfield  {author} {\bibinfo {author} {\bibfnamefont {P.}~\bibnamefont
  {Trempler}}, \bibinfo {author} {\bibfnamefont {R.}~\bibnamefont {Dreyer}},
  \bibinfo {author} {\bibfnamefont {P.}~\bibnamefont {Geyer}}, \bibinfo
  {author} {\bibfnamefont {C.}~\bibnamefont {Hauser}}, \bibinfo {author}
  {\bibfnamefont {G.}~\bibnamefont {Woltersdorf}},\ and\ \bibinfo {author}
  {\bibfnamefont {G.}~\bibnamefont {Schmidt}},\ }\bibfield  {title} {\bibinfo
  {title} {Integration and characterization of micron-sized yig structures with
  very low gilbert damping on arbitrary substrates},\ }\href
  {https://doi.org/10.1063/5.0026120} {\bibfield  {journal} {\bibinfo
  {journal} {Applied Physics Letters}\ }\textbf {\bibinfo {volume} {117}},\
  \bibinfo {pages} {232401} (\bibinfo {year} {2020})}\BibitemShut {NoStop}%
\bibitem [{\citenamefont {Kosen}\ \emph {et~al.}(2019)\citenamefont {Kosen},
  \citenamefont {van Loo}, \citenamefont {Bozhko}, \citenamefont {Mihalceanu},\
  and\ \citenamefont {Karenowska}}]{KosenAPLMater19}%
  \BibitemOpen
  \bibfield  {author} {\bibinfo {author} {\bibfnamefont {S.}~\bibnamefont
  {Kosen}}, \bibinfo {author} {\bibfnamefont {A.~F.}\ \bibnamefont {van Loo}},
  \bibinfo {author} {\bibfnamefont {D.~A.}\ \bibnamefont {Bozhko}}, \bibinfo
  {author} {\bibfnamefont {L.}~\bibnamefont {Mihalceanu}},\ and\ \bibinfo
  {author} {\bibfnamefont {A.~D.}\ \bibnamefont {Karenowska}},\ }\bibfield
  {title} {\bibinfo {title} {Microwave magnon damping in yig films at
  millikelvin temperatures},\ }\href {https://doi.org/10.1063/1.5115266}
  {\bibfield  {journal} {\bibinfo  {journal} {APL Materials}\ }\textbf
  {\bibinfo {volume} {7}},\ \bibinfo {pages} {101120} (\bibinfo {year}
  {2019})}\BibitemShut {NoStop}%
\bibitem [{\citenamefont {Zhang}\ \emph {et~al.}(2016)\citenamefont {Zhang},
  \citenamefont {Zou}, \citenamefont {Jiang},\ and\ \citenamefont
  {Tang}}]{ZhangJAP16}%
  \BibitemOpen
  \bibfield  {author} {\bibinfo {author} {\bibfnamefont {X.}~\bibnamefont
  {Zhang}}, \bibinfo {author} {\bibfnamefont {C.}~\bibnamefont {Zou}}, \bibinfo
  {author} {\bibfnamefont {L.}~\bibnamefont {Jiang}},\ and\ \bibinfo {author}
  {\bibfnamefont {H.~X.}\ \bibnamefont {Tang}},\ }\bibfield  {title} {\bibinfo
  {title} {Superstrong coupling of thin film magnetostatic waves with microwave
  cavity},\ }\href {https://doi.org/10.1063/1.4939134} {\bibfield  {journal}
  {\bibinfo  {journal} {Journal of Applied Physics}\ }\textbf {\bibinfo
  {volume} {119}},\ \bibinfo {pages} {023905} (\bibinfo {year}
  {2016})}\BibitemShut {NoStop}%
\bibitem [{\citenamefont {Danilov}\ \emph {et~al.}(1989)\citenamefont
  {Danilov}, \citenamefont {Lyfar'}, \citenamefont {Lyubon'ko}, \citenamefont
  {Nechiporuk},\ and\ \citenamefont {Ryabchenko}}]{DanilovSovPhysJ89}%
  \BibitemOpen
  \bibfield  {author} {\bibinfo {author} {\bibfnamefont {V.~V.}\ \bibnamefont
  {Danilov}}, \bibinfo {author} {\bibfnamefont {D.~L.}\ \bibnamefont {Lyfar'}},
  \bibinfo {author} {\bibfnamefont {Y.~V.}\ \bibnamefont {Lyubon'ko}}, \bibinfo
  {author} {\bibfnamefont {A.~Y.}\ \bibnamefont {Nechiporuk}},\ and\ \bibinfo
  {author} {\bibfnamefont {S.~M.}\ \bibnamefont {Ryabchenko}},\ }\bibfield
  {title} {\bibinfo {title} {Low-temperature ferromagnetic resonance in
  epitaxial garnet films on paramagnetic substrates},\ }\href
  {https://doi.org/10.1007/BF00897267} {\bibfield  {journal} {\bibinfo
  {journal} {Soviet Physics Journal}\ }\textbf {\bibinfo {volume} {32}},\
  \bibinfo {pages} {276} (\bibinfo {year} {1989})}\BibitemShut {NoStop}%
\bibitem [{\citenamefont {Mihalceanu}\ \emph {et~al.}(2018)\citenamefont
  {Mihalceanu}, \citenamefont {Vasyuchka}, \citenamefont {Bozhko},
  \citenamefont {Langner}, \citenamefont {Nechiporuk}, \citenamefont
  {Romanyuk}, \citenamefont {Hillebrands},\ and\ \citenamefont
  {Serga}}]{MihalceanuPRB18}%
  \BibitemOpen
  \bibfield  {author} {\bibinfo {author} {\bibfnamefont {L.}~\bibnamefont
  {Mihalceanu}}, \bibinfo {author} {\bibfnamefont {V.~I.}\ \bibnamefont
  {Vasyuchka}}, \bibinfo {author} {\bibfnamefont {D.~A.}\ \bibnamefont
  {Bozhko}}, \bibinfo {author} {\bibfnamefont {T.}~\bibnamefont {Langner}},
  \bibinfo {author} {\bibfnamefont {A.~Y.}\ \bibnamefont {Nechiporuk}},
  \bibinfo {author} {\bibfnamefont {V.~F.}\ \bibnamefont {Romanyuk}}, \bibinfo
  {author} {\bibfnamefont {B.}~\bibnamefont {Hillebrands}},\ and\ \bibinfo
  {author} {\bibfnamefont {A.~A.}\ \bibnamefont {Serga}},\ }\bibfield  {title}
  {\bibinfo {title} {Temperature-dependent relaxation of dipole-exchange
  magnons in yttrium iron garnet films},\ }\href
  {https://doi.org/10.1103/PhysRevB.97.214405} {\bibfield  {journal} {\bibinfo
  {journal} {Phys. Rev. B}\ }\textbf {\bibinfo {volume} {97}},\ \bibinfo
  {pages} {214405} (\bibinfo {year} {2018})}\BibitemShut {NoStop}%
\bibitem [{\citenamefont {Connelly}\ \emph {et~al.}(2021)\citenamefont
  {Connelly}, \citenamefont {Aquino}, \citenamefont {Robbins}, \citenamefont
  {Bernstein}, \citenamefont {Orlov}, \citenamefont {Porod},\ and\
  \citenamefont {Chisum}}]{ConnellyIEEE21}%
  \BibitemOpen
  \bibfield  {author} {\bibinfo {author} {\bibfnamefont {D.~A.}\ \bibnamefont
  {Connelly}}, \bibinfo {author} {\bibfnamefont {H.~R.~O.}\ \bibnamefont
  {Aquino}}, \bibinfo {author} {\bibfnamefont {M.}~\bibnamefont {Robbins}},
  \bibinfo {author} {\bibfnamefont {G.~H.}\ \bibnamefont {Bernstein}}, \bibinfo
  {author} {\bibfnamefont {A.}~\bibnamefont {Orlov}}, \bibinfo {author}
  {\bibfnamefont {W.}~\bibnamefont {Porod}},\ and\ \bibinfo {author}
  {\bibfnamefont {J.}~\bibnamefont {Chisum}},\ }\bibfield  {title} {\bibinfo
  {title} {Complex permittivity of gadolinium gallium garnet from 8.2 to 12.4
  ghz},\ }\href@noop {} {\bibfield  {journal} {\bibinfo  {journal} {IEEE
  Magnetics Letters}\ }\textbf {\bibinfo {volume} {12}},\ \bibinfo {pages} {1}
  (\bibinfo {year} {2021})}\BibitemShut {NoStop}%
\bibitem [{\citenamefont {Garziano}\ \emph {et~al.}(2020)\citenamefont
  {Garziano}, \citenamefont {Settineri}, \citenamefont {Di~Stefano},
  \citenamefont {Savasta},\ and\ \citenamefont {Nori}}]{garziano2020gauge}%
  \BibitemOpen
  \bibfield  {author} {\bibinfo {author} {\bibfnamefont {L.}~\bibnamefont
  {Garziano}}, \bibinfo {author} {\bibfnamefont {A.}~\bibnamefont {Settineri}},
  \bibinfo {author} {\bibfnamefont {O.}~\bibnamefont {Di~Stefano}}, \bibinfo
  {author} {\bibfnamefont {S.}~\bibnamefont {Savasta}},\ and\ \bibinfo {author}
  {\bibfnamefont {F.}~\bibnamefont {Nori}},\ }\bibfield  {title} {\bibinfo
  {title} {Gauge invariance of the dicke and hopfield models},\ }\href
  {https://link.aps.org/doi/10.1103/PhysRevA.102.023718} {\bibfield  {journal}
  {\bibinfo  {journal} {Phys. Rev. A}\ }\textbf {\bibinfo {volume} {102}},\
  \bibinfo {pages} {023718} (\bibinfo {year} {2020})}\BibitemShut {NoStop}%
\end{thebibliography}
\end{document}